\documentclass[aps,pra,twocolumn,floats]{revtex4-2} 
\usepackage{amsmath}
\usepackage{graphicx}
\usepackage{dcolumn}
\usepackage{bm}
\usepackage{amssymb}
\usepackage{latexsym}
\usepackage{color}
\usepackage{subfigure}

\definecolor{purple}{rgb}{0.5, 0.0, 0.5}

\newcommand{\veps}{\varepsilon}
\def\mbfe{\mathbf{e}}

\def\mbfr{\mathbf{r}}

\def\mbfS{\mathbf{S}}
\def\mbfL{\mathbf{L}}
\def\mbfJ{\mathbf{J}}

\def\1S0{${}^{1}$S$_{0}$}
\def\3P1{${}^{3}$P$_{1}$}

\begin{document}

\title{Hydrogen and hydrogen-like-ion bound states and hyperfine splittings: finite nuclear size effects}

\author{Igor Kuzmenko$^{1,2}$, Tetyana Kuzmenko$^{1}$, Y. Avishai$^{1,3}$, Y. B. Band$^{1,2,4}$}

\affiliation{
  ${}^{1}$Department of Physics,
  Ben-Gurion University of the Negev,
  Beer-Sheva 84105, Israel
  \\
  $^2$Department of Chemistry,
  Ben-Gurion University of the Negev,
  Beer-Sheva 84105, Israel
  \\
  $^3$ Yukawa Institute for Theoretical Physics, Kyoto, Japan
  \\
  $^4$The Ilse Katz Center for Nano-Science,
  Ben-Gurion University of the Negev,
  Beer-Sheva 84105, Israel}

\begin{abstract}
Using the Dirac equation, we study corrections to electron binding energies and hyperfine splittings of atomic hydrogen and hydrogen-like ions due to finite nuclear size (FNS) effects, relativistic QED radiative corrections and nuclear recoil corrections. Three models for the charge distribution and the magnetic moment distribution within the nucleus are considered.  Calculations are carried for light atoms (H, He and K) and heavy atoms (Rb, Cs, Pb, Bi, U).  The FNS corrections to the ground-state energy are shown to be smaller than the electron-nucleus reduced mass corrections, and comparable to the relativistic QED radiative corrections for the light nuclei, but much larger  than both these corrections for heavy nuclei. Comparison is made with an experiment on the $1s$-$2s$ transition frequency for hydrogen. FNS corrections to the ground state hyperfine splitting are comparable in size to the relativistic QED radiative corrections for light nuclei, but are larger for heavy nuclei.
\end{abstract}

\maketitle

\section{Introduction} \label{sec:Intro}

Hydrogen is the most abundant element in the universe, and atomic hydrogen 
has the simplest electronic structure.  If the proton in atomic hydrogen 
is taken as a point charge, analytic expressions for wave functions 
and bound-state energies exist within both the non-relativistic framework 
(using the Schr\"odinger equation) and the relativistic one (using the 
Dirac equation) \cite{Bethe_57, Akhiezer_65, Sakurai_67, Berestetskii_82, 
Mawhin_2010}.  This is also true for hydrogen-like ions (HLI) if the nucleus 
is taken as a point charge.  Moreover, the ground state hyperfine 
splitting of these systems can also be obtained analytically (if the 
nuclear $g$-factor is known) \cite{Essen_71, Dupays_03, Kanda_18}.
However, nuclei are not point-like; even the proton is a composite particle
with finite radius, $r_p = 0.8414(19) \times 10^{-15} \mbox{ m} = 1.5900 (36) \times 10^{-5}$
Bohr \cite{Bohr_radius}.  Here we consider the effects of a finite nuclear 
size (FNS) $r_N$ (and the nuclear charge distribution), nuclear recoil corrections
and QED radiative corrections on the electronic properties and 
hyperfine splittings of hydrogen and hydrogen-like ions (H\&HLI). 

Finite nuclear size (FNS) corrections (i.e., finite nuclear radius corrections) to the binding energies of H\&HLI have been extensively studied, see e.g., Refs.~\cite{proton_radius, CODATA-2002, Beiersdorfer_01, BW-1950, BW-2015, Adamu_18, Martensoon-Pendrill-2003, Deck_05, Essen_71, Dupays_03, Kanda_18, QED, Shabaev-JPhysB-1993, Shabaev-PRA-1997, Valuev_20} and references quoted therein. Theoretical investigations 
often use the Schr\"odinger equation with relativistic corrections, e.g., spin-orbit 
coupling, obtained using $1/c$ expansions \cite{Berestetskii_82}.  One can apply 
Rayleigh--Schr\"odinger perturbation theory to calculate nuclear structure 
corrections for the ground state hyperfine splitting as well as the 
electronic transition energies \cite{Adamu_18}.  However, the ground state binding 
energies, $\epsilon$, of H\&HLI with large nuclear charge, $Ze$, are not small compared to 
$m_e c^2$, where $m_e$ is the electron mass, hence fully relativistic calculations are 
necessary.  For example, for the uranium hydrogen-like ion, $\epsilon \approx 0.2588 \, m_e c^2$.
Hence, using the Schr\"odinger equation (even with spin-orbit interaction) is not 
satisfactory for heavy nuclei; one must use the Dirac equation.  Moreover, even for
light nuclei, one should use relativistic calculations to obtain high accuracy.
In Ref.~\cite{Adamu_18}, the authors solve the Schr\"odinger equation
and add relativistic corrections such as spin-orbit interaction. They assume 
that the nuclear charge is distributed homogeneously inside the nucleus 
and find the correction to the ground-state energy 
$(2/3) Z \epsilon_H r_{N}^{2}/a_0^{2} \ll Z \epsilon_H$, where in Gaussian units 
$\epsilon_H = e^2/a_0$ is the Hartree energy, $r_N$ is the nuclear charge radius, 
and $a_0$ is the Bohr radius.
However, for high $Z$ nuclei, the relativistic Dirac equation
should be applied.  In Refs.~\cite{Martensoon-Pendrill-2003, Deck_05},
the authors solve the Dirac equation for hydrogen having a FNS with the 
charge distributed on the surface of the nucleus, approximate the position dependence of 
the wave function inside the nucleus by a polynomial, and find the binding energies 
for the ground and excited states. Note that
perturbation theory for the FNS corrections using the Dirac equation is a good 
approximation for hydrogen and light {\color {red} nuclei}.  Such an approximation leads to a closed 
form expression for the FNS correction in the non-relativistic limit [see Eq.~(\ref{eq:EFNS})].
However, perturbation theory is not valid for high $Z$ nuclei.

H\&HLI systems are traditionally used in ultra-high precision experimental tests of quantum 
mechanics, as well as of nuclear structure, since complications due to many-electron effects are absent 
\cite{Parthey, Matveev}.  It will be crucial to compare the results of our calculations with these
experiments.  Furthermore, accurate experiments of muonic hydrogen
atoms (muonic hydrogen contains a negatively charged muon instead of an electron) have 
been reported, see e.g., \cite{Beiersdorfer_01, Kanda_18}.

Here, within a relativistic Dirac equation formulation, we calculate the FNS corrections to the ground ($1S_{1/2}$) and excited $S$ states energies, and to the ground state hyperfine energy splittings, for a number of elements and their isotopes for H\&HLI.  We use three different models for the spherically symmetric charge and magnetic moment distributions within the nucleus:  In model (a) the nuclear charge and magnetic moment are distributed homogeneously on the surface of a sphere with radius $r_a$.  In model (b) the charge and magnetic moment are distributed homogeneously inside the nucleus with radius $r_b$.  In model (c) we use the so called two-parameter Fermi model (TPFM) for the nuclear charge density and magnetic moment density distributions \cite{Shabaev-JPhysB-1993, Shabaev-PRA-1997, deVries_87}, which  fall off smoothly near the surface of nucleus.  We use the same distribution for both nuclear charge and nuclear magnetic moment because of lack of parameters for the magnetic moment density.

We also calculate FNS corrections to the ground state hyperfine splitting of H\&HLI due to (1) the correction for the charge distribution of the nucleus and (2) the correction for the magnetic moment distribution of the nucleus, i.e., the Bohr-Weisskopf correction \cite{BW-1950, BW-2015}.  Measurements of the ground state hyperfine splitting of the H\&HLI are among the most accurately measured quantities  \cite{Essen_71, Dupays_03, Kanda_18}.  These measurements have been used to determine both nuclear radii and nuclear $g$-factors. To compare theoretical calculations for FNS corrections with experimental measurements, one needs to incorporate relativistic quantum electrodynamics (QED) radiative corrections \cite{QED}. FNS corrections to the ground-state hyperfine splitting have been analyzed in Refs.~\cite{Essen_71, Dupays_03, Kanda_18, BW-1950, BW-2015, QED, Shabaev-JPhysB-1993, Shabaev-PRA-1997, Volotka-PhD-2006}.  In Ref.~\cite{Kanda_18}, the authors report measurements of the hyperfine splitting of the ground state of hydrogen used to estimate the proton radius.  In Refs.~\cite{Dupays_03, BW-1950, BW-2015, QED, Shabaev-JPhysB-1993},  the authors apply QED radiative corrections and calculate the FNS corrections to the ground-state hyperfine splitting of H\&HLI.  References~\cite{Shabaev-PRA-1997, Volotka-PhD-2006} studied the hyperfine splitting for high-$Z$ HLI.

The paper is organized as follows.  In Sec.~\ref{sec:Dirac}, we formulate the Dirac equation 
for the H\&HLI with potential energy of the electron determined
using models (a) and (b).  Electron bound states for H\&HLI are studied 
in Sec.~\ref{sec:energies-bound}.  The solution of the Dirac equation for a point-like nucleus
is presented in Sec.~\ref{subsec:Dirac-rp=0-energy}.  The analytic solution of the Dirac 
equation for model (a) is given in Sec~\ref{subsec:Dirac-finite-radius-model-a} and the numerical solution for model (b) is given in Sec~\ref{subsec:Dirac-finite-radius-model-b}.  FNS effects on the ground-state energy for both models (a) and (b) are calculated in Sec.~\ref{sec:corrections} within perturbation theory.  Section~\ref{Fermi_model} discusses the TPFM for the nuclear charge distribution.  Ground state hyperfine splittings are calculated in Sec.~\ref{sec:hyperfine}.  Section \ref{sec:uncertainty} estimates the uncertainty of the hyperfine splitting due to uncertainty in the nuclear $g$-factors, $\delta g_I$, and nuclear radii $\delta r_N$ and Sec.~\ref{sec:comparison} compares the results of models (a), (b) and (c) with experiments.  In Sec.~\ref{sec:conclude} we summarize our main results. Several technical details are presented in the Appendices.  Specifically, in Appendix \ref{append:nuclear-charge-radius} we discuss the root mean squared nuclear charge radius, $r_N$, in Appendix~\ref{append:er_alpha_matrix_elemets} we calculate the matrix elements of $\mbfe_r \times \boldsymbol\alpha$, where $\mbfe_r$ is the basis vector in the direction of the position vector $\mbfr$ from the nucleus to the electron and $\boldsymbol\alpha$ is the vector of the Dirac matrices. Appendix~\ref{append:magnetic-moment-distribution} discusses the magnetic moment distribution for nuclei with one nucleon outside a closed nuclear shell.

\section{Models for FNS potentials}  \label{sec:Dirac}

The stationary Dirac equation for H\&HLI is given by,
\begin{equation}   \label{eq:Dirac}
  \Big[
      -i \hbar c \boldsymbol\alpha \cdot \nabla +
      V(r) + \beta m_e c^2
  \Big] \psi(\mbfr) = \veps \psi(\mbfr),
\end{equation}
where $m_e$ is the electron mass, $\boldsymbol\alpha$ and $\beta$ are the 4$\times$4 
Dirac matrices, and $V(r) = V_C(r) = -Ze^2/r$ is the Coulomb potential wherein $Z$ is the 
nuclear charge. In model (a), the nuclear charge is distributed on the surface 
of the sphere of radius $r_a$ and therefore the potential is constant within the nucleus, 
and in model (b) the nuclear charge density $\varrho_b(r)$ is uniformly distributed within 
a sphere of radius $r_b$,
\begin{equation}  \label{Eq:rho_b}
  \varrho_b(r) = \frac{3Ze}{4 \pi r_{b}^{3}} \, \Theta(r_b - r) ,
\end{equation}
where $\Theta(\bullet)$ is the Heaviside theta function. Therefore the potential for model 
(b) is quadratic inside the nucleus.  Explicit expressions for the potentials in models (a) and (b) are:
\begin{eqnarray}  
&&  V_a(r) =
  \begin{cases}
    \displaystyle
    - \frac{Z e^2}{r_a}, \ \ r \le r_a,
    \\
    \displaystyle
    - \frac{Z e^2}{r}, \ \ r > r_a ,
  \end{cases} \label{Eq:Va} \\
&&  V_b(r) = 
  \left\{
    \begin{array}{cc}
    \displaystyle
    - \frac{Z e^2}{2 r_b} \Big( 3 - \frac{r^2}{r_{b}^{2}} \Big), & r \le r_b,
    \\
    \displaystyle
    - \frac{Z e^2}{r}, & r > r_b,
    \end{array}
  \right.   \label{Eq:Vb}
\end{eqnarray}
The nuclear radii $r_a$ and $r_b$ are expressed in terms of root mean square (RMS) 
nuclear charge radius $r_N$ as \cite{proton_radius, nuclear-radii-ADNDT-2013}:
\begin{equation}
  r_a = r_N,
  \quad
  r_b = \sqrt{ \frac{5}{3} } \, r_N.
  \label{eq:r_a-r_b-vs-r_N}
\end{equation}
See Appendix~\ref{append:nuclear-charge-radius} for details.

\begin{figure}
\centering
\includegraphics[width=0.9\linewidth,angle=0] {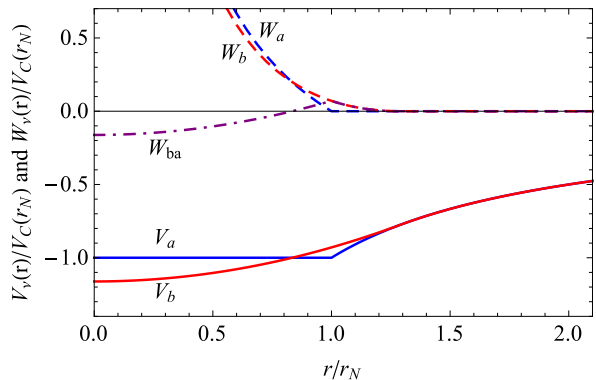}
\caption{\footnotesize  
  The potentials $V_{\nu=a}(r)$ and $V_{\nu=b}(r)$ versus $r$ for a hydrogen 
  atom defined in Eqs.~(\ref{Eq:Va}) and (\ref{Eq:Vb}) (solid blue and red curves respectively).
  The repulsive potentials $W_{\nu}(r) = V_{\nu}(r) - V_C(r)$, (blue and red dashed curves respectively),
  and the difference potentials $W_{ab}(r)$ defined in Eq.~(\ref{Eq:W_ba}) (dashed-dotted purple curve). 
  Note that for $r > r_{\nu}$, $W_{\nu}(r) = 0$.
  Here $r_a$ and $r_b$ are given by Eq.~(\ref{eq:r_a-r_b-vs-r_N}).}
\label{Fig:V_W_surface-volume}
\end{figure}

The difference between the FNS potential and the Coulomb potential is the {\it repulsive} potential
\begin{equation}    \label{eq:W}
  W_{\nu}(r) = V_{\nu}(r) - V_C(r)
  \quad
  \nu = a, b.
\end{equation}
Note that $V_{\nu}(r) = V_C(r) + W_{\nu}(r)$ is {\it less attractive} than $V_C(r)$
in both models.  We also define the difference potential
$W_{ba}(r) = V_b(r) - V_a(r)$,
\begin{equation}   \label{Eq:W_ba}
  W_{ba}(r) =
  \left\{
    \begin{array}{cc}
      \displaystyle
      \frac{Z e^2}{r_a} -
      \frac{Z e^2}{2 r_b} \Big( 3 - \frac{r^2}{r_{b}^{2}} \Big),
      &
      r \leq r_a,
      \\
      \displaystyle
      \frac{Z e^2}{r} -
      \frac{Z e^2}{2 r_b} \Big( 3 - \frac{r^2}{r_{b}^{2}} \Big),
      &
      r_a < r \leq r_b,
      \\
      0 & r > r_b,
    \end{array}
  \right \}
\end{equation}
which will be useful in our perturbation analysis below.  $W_{ba}(r)$ needs to be added to 
$V_a(r)$ to get $V_b(r)$.  Figure \ref{Fig:V_W_surface-volume} plots the potentials $V_a(r)$ and 
$V_b(r)$ given in Eqs.~(\ref{Eq:Va}) and (\ref{Eq:Vb}), the repulsive potentials $W_{a,b}(r)$, 
and the difference potential $W_{ba}(r)$ versus $r$.

\begin{figure}
\centering
\includegraphics[width=0.9\linewidth,angle=0] {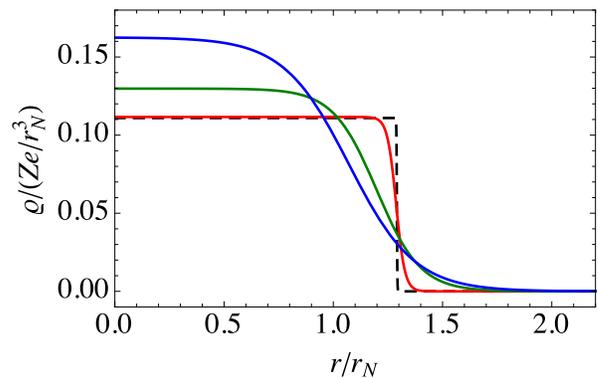}
\caption{\footnotesize  
  The charge density $\varrho$ versus $r/r_N$ for the TPFM 
  [see Eq.~(\ref{eq:Fermi-distribution})] for three different smoothing parameters 
  (solid colored curves) and the model (b) charge density [see Eq.~(\ref{Eq:rho_b})] (dashed 
  black curve).  The integral of $\varrho$ over all space equals $Ze$.}
\label{Fig_Fermi_2_parameter_model}
\end{figure}

The nuclear radii $r_a, r_b$ and their uncertainties $\delta r_a, \delta r_b$,
are expressed in terms of the RMS nuclear charge
radius $r_N$ and its uncertainty $\delta r_N$ as
\begin{eqnarray}   \label{eq:r_b-vs-r_N}
  &&r_a \pm \delta r_a = r_N \pm \delta r_N, \nonumber \\
  &&r_b \pm \delta r_b = \sqrt{ \frac{5}{3} } \, \big(  r_N \pm \delta r_N \big).
\end{eqnarray}
The uncertainties $\delta r_\nu$ give rise to the uncertainties in $V_\nu(r)$:
\begin{eqnarray}
  V_\nu(r) \pm \delta V_\nu(r) =
  V_\nu(r) \pm
  \bigg| \frac{\partial V_\nu(r)}{\partial r_{\nu}} \bigg| \,
  \delta r_\nu.
  \label{eq:V_b-pm-delts_V_nu}
\end{eqnarray}
Explicitly,
\begin{equation}   \label{eq:delta_V_a}
  \delta V_a(r) = \frac{Z e^2 \delta r_a}{r_a^2} \Theta \big( r_a - r \big),
\end{equation}
and
\begin{equation}
  \delta V_b(r) =
  \frac{3 Z e^2 \delta r_b}{2 r_{b}^{2}} \,
  \bigg( 1 - \frac{r^2}{r_{b}^{2}} \bigg) \,
  \Theta \big( r_b - r \big) .
  \label{eq:delta_V_b}
\end{equation}
In Sec.~\ref{Fermi_model} we will introduce the TPFM for the nuclear charge distribution \cite{deVries_87,Shabaev-JPhysB-1993, Shabaev-PRA-1997}, and we will refer to it as model (c) [see Eq.~(\ref{eq:Fermi-distribution})]. In this model the nuclear charge distribution near $r_N$ is smooth. Figure~\ref{Fig_Fermi_2_parameter_model} plots the charge distribution for three different smoothing parameters and the charge distribution of model (b) (see the dashed curve).  The shape of the TPFM charge distribution is sometimes called the Woods-Saxon shape.  The potential of the TPFM charge distribution is also referred to as the Woods-Saxon potential. It can be determined analytically and then numerical solution of the Dirac equation can be obtained.

The potential for this model can be determined analytically and then numerical solution of the Dirac equation eigenstates with this potential by numerically solving the Dirac equation for heavy nuclei.  

\ \\
\section{Dirac Equation Bound states}
  \label{sec:energies-bound}
The Dirac $4$-component spinor wave function $\psi(\mbfr)$
can be written in the form \cite{Landau-Lifshitz-4-hydrogen},
\begin{equation}
  \psi(\mbfr) =
  \left(
    \begin{array}{c}
      g(r) \, \Omega_{j l m}
      \\
      i f(r) \, \Omega_{j l' m}
    \end{array}
  \right),
  \label{eq:spinor-vs-spherical-harmonics}
\end{equation}
where $\Omega_{j l m}$ is a normalized spherical harmonic spinor
which is an eigenfunction of the total electronic angular momentum operator squared, $\mbfJ^2$, 
the electronic orbital angular momentum operator squared, $\mbfL^2$, the electronic spin angular 
momentum operator squared, $\mbfS^2$ and the $z$-projection of the total electronic angular momentum
operator, $J_z$.  These have eigenvalues $j (j+1)$, $l (l+1)$, $\frac{3}{4}$ and $m$, 
where $l = j \pm \frac{1}{2}$ and $l' = 2 j - l$.
The radial wave functions $g(r)$ and $f(r)$ satisfy the equations,
\begin{subequations}    \label{subeqs:radial}
\begin{eqnarray}
  \hbar c \,
  \Big[
      \big( r g(r) \big)' +
      \frac{\kappa}{r} \,
      \big( r g(r) \big)
  \Big]
  = \nonumber \\
  \big[ m_e c^2 - V(r) + \veps \big]
  \big( r f(r) \big),
  \label{eq1-radial}
  \\
  \hbar c \,
  \Big[
      \big( r f(r) \big)' -
      \frac{\kappa}{r} \,
      \big( r f(r) \big)
  \Big]
  = \nonumber \\
  \big[ m_e c^2 + V(r) - \veps \big]
  \big( r g(r) \big),
  \label{eq2-radial}
\end{eqnarray}
\end{subequations}
where $\kappa = -1$ for the $s$-state.
{\color {red} $S$ or $s$ ?}
\subsection{Solution of the Dirac equation (\ref{subeqs:radial}) for the point-like nucleus}
  \label{subsec:Dirac-rp=0-energy}

The solution of Eqs.~(\ref{subeqs:radial}) for the point-like nucleus
(i.e., for $r_N = 0$) is well known \cite{Landau-Lifshitz-4-hydrogen}; 
for an $s$-state,
\begin{subequations}    \label{subeqs:f-g-vs-Q1-Q2-point}
\begin{equation}  \label{eq1-f-vs-Q1-Q2-point}
  g(r) =
  \mathcal{N}
  \sqrt{m_e c^2 + \veps} \,
  \Big[ Q_1(\rho) + Q_2(\rho) \Big] \,
  \rho^{\gamma - 1}
  e^{-\rho/2},
\end{equation}
\begin{equation}   \label{eq2:g-vs-Q1-Q2-point}
  f(r) =
  -\mathcal{N}
  \sqrt{m_e c^2 - \veps} \,
  \Big[ Q_1(\rho) - Q_2(\rho) \Big] \,
  \rho^{\gamma - 1}
  e^{-\rho/2},
\end{equation}
\end{subequations}
where $\mathcal{N}$ is a normalization constant found from the equation,
$\int\limits_{0}^{\infty}
  \Big(
      \big| g(r) \big|^{2} +
      \big| f(r) \big|^{2}
  \Big)  r^2 dr  = 1$.
Here we define the quantities
\begin{eqnarray}
  \rho = \frac{2 \lambda r}{\hbar c},
  \quad
  \lambda = \sqrt{m_e^2 c^4 - \veps^2},
  \quad
  \gamma = \sqrt{\kappa^2 - Z^2 \alpha^2},
  \label{eq:rho-lambda-gamma-def}
\end{eqnarray}
and $\alpha = \frac{e^2}{\hbar c}$ is the fine structure constant (where the right hand side
of the equation is given in Gaussian units).  The functions $Q_1(\rho)$ and $Q_2(\rho)$ are
\cite{Landau-Lifshitz-4-hydrogen}
\begin{subequations}    \label{subeqs:Q1-Q2-hypergeometric}
\begin{eqnarray}
  Q_1(\rho) &=&
  A_1 \,
  F
  \bigg(
       \gamma - \frac{Z \alpha \veps}{\lambda}, \,
       2 \gamma + 1, \,
       \rho
  \bigg),
  \label{eq:Q1-hypergeometric}
  \\
  Q_2(\rho) &=&
  A_2 \,
  F
  \bigg(
       \gamma + 1 - \frac{Z \alpha \veps}{\lambda}, \,
       2 \gamma + 1, \,
       \rho
  \bigg),
  \label{eq:Q2-hypergeometric}
\end{eqnarray}
\end{subequations}
where $F(a, b, \rho)$ is the Kummer confluent hypergeometric function.
The normalization constants $A_1$ and $A_2$ are related as follows:
\begin{equation}
  A_2 =
  -\frac{\gamma \lambda - Z \alpha \veps}
        {\kappa \lambda - Z \alpha m_e c^2} \,
  A_1.
  \label{eq:B-vs-A-point}
\end{equation}
In the special case,
\begin{equation}   \label{eq:energy-r_n=0}
  \gamma - \frac{Z \alpha \veps}{\lambda} = - n_r,
\end{equation}
where $n_r$ is a positive integer, the Kummer function reduces to a Laguerre polynomial,
\begin{equation*}
  F \big( -n_r, 2 \gamma + 1, \rho \big) =
  \frac{\Gamma (2 \gamma + 1) \Gamma ( n_r + 1)}
       {\Gamma ( 2 \gamma + n_r + 1)} \,
  L_{n_r}^{(2 \gamma)}(\rho),
\end{equation*}
{\color {red} why there is no number?}\\
and thus, the functions $g(r)$ and $f(r)$ in
Eq.~(\ref{subeqs:f-g-vs-Q1-Q2-point}) decay exponentially
for $r \to \infty$.
In the special case,
\begin{equation*}
  \gamma - \frac{Z \alpha \veps}{\lambda} = 0, \ \ \mbox{and}  \quad \kappa < 0,
\end{equation*}
$A_2 = 0$ and $Q_1(\rho)$ decay exponentially for $r \to \infty$.
Otherwise, when
$$
  \gamma - \frac{Z \alpha \veps}{\lambda} \neq - n_r,
$$
$Q_1(\rho)$ and $Q_2(\rho)$ diverge as $e^{\rho}$ when
$\rho \to \infty$, and thus $g(r)$ and $f(r)$ in
Eq.~(\ref{subeqs:f-g-vs-Q1-Q2-point}) diverge as $e^{\rho/2}$.

Let us recall the expressions for the relativistic energies 
$\veps_{n}^{(0)}$ of a H\&HLI system calculated
for a point-like nucleus \cite{Landau-Lifshitz-4-hydrogen},
\begin{eqnarray}
  \veps_{n}^{(0)} = m_e c^2 +
  h c R_{\infty}  \frac{2 \mathcal{G}(n)}{1 + \mu_{ep}} ,  \label{eq:energy-hydrgen}
\end{eqnarray}
where, for arbitrary $Z$,  $\mu_{ep} \equiv m_e/M(Z,A)$ is the electron to nuclear mass 
ratio, $M(Z,A)$ is the mass of the nucleus comprising $Z$ protons and $A - Z$ neutrons, 
and $c R_{\infty}$ is the Rydberg frequency, which is related
to the Hartree energy $\epsilon_{H}$, $c R_{\infty} = \frac{\epsilon_{H}}{2 h} 
[= \frac{e^4 m_e}{4\pi \hbar^3}$ in Gaussian units].  In Eq.~(\ref{eq:energy-hydrgen})
$n = 1, 2, 3, \ldots$ is the principal quantum number, and
\begin{eqnarray}
  \mathcal{G}(n) &=&
  \frac{1}{Z^2 \alpha^2} \,
  \bigg[
    \frac{1}{\sqrt{1 + X_{n}^{2}}} -
    1
  \bigg]
  \nonumber \\ &=&
  - \frac{1}{Z^2 \alpha^2} \,
  \frac{X_{n}^{2}}{\sqrt{1 + X_{n}^{2}} \, \big[ \sqrt{1 + X_{n}^{2}} + 1 \big]} ,
  \label{eq:F_n}
\end{eqnarray}
where $X_n = \frac{Z \alpha}{\sqrt{1 - Z^2 \alpha^2} + n - 1}$.
We found that the first form on the right-hand side of Eq.~(\ref{eq:F_n}) can lead to 
significant numerical errors, so for numerical calculations, the second form is preferable.
As noted in Refs.~\cite{reduced-mass1, reduced-mass2, reduced-mass3, Eides-hydrogen-atom}, 
the concept of a reduced mass in the Dirac equation is ambiguous to some extent.
Nevertheless, the necessity of introducing the reduced mass as the substitution 
for electron mass {\em in the Dirac energy and the Dirac equation} in order to 
obtain theoretical results that are close to experimental results has been 
stressed in the literature (see Refs.~\cite{reduced-mass1, reduced-mass2, 
reduced-mass3, Eides-hydrogen-atom}).  The ground state energy in 
Eq.~(\ref{eq:energy-hydrgen}) ($n = 1$) is
\begin{equation}    \label{eq:energy-hydrgen-ground}
\veps_{1}^{(0)}(Z,A) = m_e c^2 + m c^2 ( \sqrt{1-Z^2 \alpha^2} - 1) ,
\end{equation}
where $m \equiv m(Z,A)$ is the electron-nucleus reduced mass, 
\begin{equation}   \label{eq:reduced-mass}
  m \equiv m(Z,A) = \frac{m_e}{1 + \mu_{ep}},
\end{equation}
and the ground-state wave functions in Eq.~(\ref{subeqs:f-g-vs-Q1-Q2-point}) are,
\begin{subequations}    \label{subeqs:f-g-ground-point}
\begin{eqnarray}
  g(r) &=&
  \frac{2 Z^{3/2} \sqrt{1 + \gamma}}{a_{B}^{3/2} \sqrt{\Gamma(2 \gamma + 1)}} \,
  \rho^{\gamma - 1} e^{-\rho/2},
  \label{eq:g-ground-point}
  \\
  f(r) &=&
  -\frac{2 Z^{3/2} \sqrt{1 - \gamma}}{a_{B}^{3/2} \sqrt{\Gamma(2 \gamma + 1)}} \,
  \rho^{\gamma - 1} e^{-\rho/2},
  \label{eq:f-ground-point}
\end{eqnarray}
\end{subequations}
where $a_B$ is taken to be the {\em effective} Bohr radius, including the 
reduced mass effect, and is given (in Gaussian units) by
\begin{equation}    \label{eq:a_Z}
  a_B = \frac{\hbar^2}{m e^2}.
\end{equation}
The nuclear radii $r_N$ are tabulated in Table~\ref{Table-atomic-nuclear-radii} 
for a number of nuclei.

\begin{table}
\caption{\footnotesize 
  RMS nuclear charge radii $r_N$ and their uncertainties $\delta r_N$ (in Bohr radius units $a_B$).
  Data for the proton and deuteron are taken from Ref.~\cite{proton_radius}, and data for the other
  isotopes are taken from Ref.~\cite{nuclear-radii-ADNDT-2013}.
  Here (and below) $E^{-n}$ is shorthand notation for $\times 10^{-n}$.}
\begin{center}
  \begin{tabular}{|c|c|c|}
    \hline
    Isotope ($Z,A$) &
    $r_N$ &
    $\delta r_N$
    \\
    \hline
    H(1, 1) &
    $1.5900 E^{-5}$ &
    $3.6 E^{-8}$
    \\
    \hline
    H(1, 2) &
    $4.02132 E^{-5}$ &
    $1.40 E^{-8}$
    \\
    \hline
    H(1, 3) &
    $3.3242 E^{-5}$ &
    $6.86 E^{-7}$
    \\
    \hline
    He(2, 3) &
    $3.7154 E^{-5}$ &
    $5.7 E^{-8}$
    \\
    \hline
    He(2, 4) &
    $3.1662 E^{-5}$ &
    $5.3 E^{-8}$
    \\
    \hline
    K(19, 39) &
    $6.4910 E^{-5}$ &
    $3.6 E^{-8}$
    \\
    \hline
    K(19, 40) &
    $6.4971 E^{-5}$ &
    $5.3 E^{-8}$
    \\
    \hline
    K(19, 41) &
    $6.5230 E^{-5}$ &
    $1.04 E^{-7}$
    \\
    \hline
    Rb(37, 85) &
    $7.9437 E^{-5}$ &
    $4.5 E^{-8}$
    \\
    \hline
    Rb(37, 87) &
    $7.9348 E^{-5}$ &
    $3.4 E^{-8}$
    \\
    \hline
    Cs(55, 133) &
    $9.0784 E^{-5}$ &
    $8.7 E^{-8}$
    \\
    \hline
    Cs(55, 135) &
    $9.0833 E^{-5}$ &
    $8.9 E^{-8}$
    \\
    \hline
    Pb(82, 204) &
    $1.03563 E^{-4}$ &
    $2.6 E^{-8}$
    \\
    \hline
    Pb(82, 206) &
    $1.03750 E^{-4}$ &
    $2.6 E^{-8}$
    \\
    \hline
    Pb(82, 207) &
    $1.03827 E^{-4}$ &
    $2.6 E^{-8}$
    \\
    \hline
    Pb(82, 208) &
    $1.03958 E^{-4}$ &
    $2.5 E^{-8}$
    \\
    \hline
    Bi(83, 209) &
    $1.04334 E^{-4}$ &
    $4.9 E^{-8}$
    \\
    \hline
    U(92, 235) &
    $1.10241 E^{-4}$ &
    $7.7 E^{-8}$
    \\
    \hline
    U(92, 236) &
    $1.10419 E^{-4}$ &
    $7.2 E^{-8}$
    \\ 
    \hline
    U(92, 238) &
    $1.10683 E^{-4}$ &
    $6.2 E^{-8}$
    \\
    \hline
  \end{tabular}
\end{center}
\label{Table-atomic-nuclear-radii}
\end{table}%

\begin{figure}
\centering 
 \includegraphics[width=0.9 \linewidth,angle=0]
   {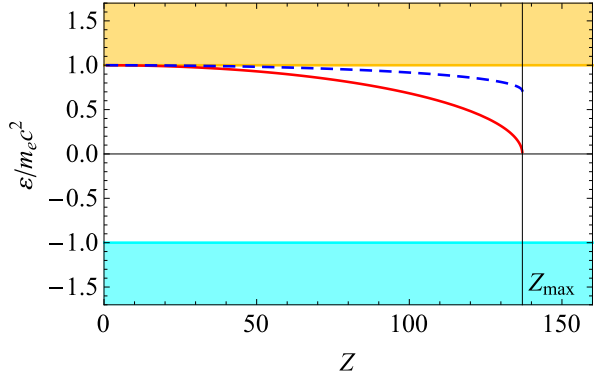}
\caption{\footnotesize 
  Energies of the $1s$ bound state $\varepsilon_1^{(0)}$ (solid red curve) and 
  the excited $2s$ state $\varepsilon_2^{(0)}$ (dashed blue curve) for the H\&HLI 
  isotopes defined in Eq.~(\ref{eq:energy-hydrgen}) as functions of the proton number $Z$.
  The gold and turquoise regions show the continuous energy spectrum
  with $|\veps| > m_e c^2$.  The horizontal line is $\veps = 0$, and
  the vertical line indicates $Z_{\max} = 137$.
  }
\label{Fig-energy-vs-Z}
\end{figure}

The energies $\veps_{1}^{(0)}$ and $\veps_{2}^{(0)}$
in Eq.~(\ref{eq:energy-hydrgen}) are plotted in 
Fig.~\ref{Fig-energy-vs-Z} as functions of $Z$.  For $Z \alpha < 1$, 
$\veps_{1}^{(0)}$ and $\veps_{2}^{(0)}$ satisfy the inequalities,
$$
  0 < \veps_{1}^{(0)} < \veps_{2}^{(0)} < m_e c^2.
$$
For  $Z = Z_{\max} = 137$, $\veps_{1}^{(0)} = 0.02292 \, m_e c^2$ and
$\veps_{2}^{(0)} = 0.71516 \, m_e c^2$.
For $Z > Z_{\max}$, the parameter $\gamma$ in Eq.~(\ref{eq:rho-lambda-gamma-def}) 
becomes imaginary, and the wave function in Eq.~(\ref{subeqs:f-g-ground-point}) 
is singular as $r \to 0$ \cite{Landau-Lifshitz-4-hydrogen}.
Indeed, the Dirac equation (\ref{subeqs:radial}) with the Coulomb potential 
of a point-like nucleus is meaningful only for $Z < 137$.

\subsection{Dirac equation (\ref{subeqs:radial}) solution for Model (a)}
\label{subsec:Dirac-finite-radius-model-a}

Equations~(\ref{subeqs:radial}) for model (a) potential are solved separately for 
$r > r_N$ and for $r < r_N$, to obtain the corresponding wave functions $f_{\pm }(r)$ and $g_{\pm}(r)$
up to as yet unknown multiplicative constants $A_{\pm}$.  The matching conditions at $r = r_N$ are,
\begin{equation}
  A_{-} g_{-}(r_N) = A_{+} g_{+}(r_N),
  \quad
  A_{-} f_{-}(r_N) = A_{+} f_{+}(r_N).
  \label{eq:boundary-conditions}
\end{equation}
The set of equations (\ref{eq:boundary-conditions})
has nontrivial solutions when the determinant 
\begin{equation}  \label{eq:det}
  \mathfrak{D}(\veps) \equiv
  g_{-}(r_N) f_{+}(r_N) -
  g_{+}(r_N) f_{-}(r_N) = 0 .
\end{equation}
Note that the condition above is necessary but not sufficient to assure that 
$g_-=g_+$ and $f_-=f_+$ at $r_N$.

\subsubsection{Solution of Dirac equation (\ref{subeqs:radial}) for $r > r_N$}
  \label{subsubsec:Dirac-r>rp-energy}

The solution of Eqs.~(\ref{subeqs:radial}) can be written in
the form \cite{Landau-Lifshitz-4-hydrogen}
\begin{subequations}   \label{subeqs:f-g-vs-Q1-Q2}
\begin{equation}
  f_{+}(r) =
  \sqrt{ \frac{m_e c^2 - \veps}{m_e c^2} } \,
  \Big[ Q_1(\rho) + Q_2(\rho) \Big] \,
  \rho^{\gamma - 1}
  e^{-\rho/2},
  \label{eq1-f-vs-Q1-Q2}
\end{equation}
\begin{equation}
  g_{+}(r) =
  \sqrt{ \frac{m_e c^2 + \veps}{m_e c^2} } \,
  \Big[ Q_2(\rho) - Q_1(\rho) \Big] \,
  \rho^{\gamma - 1}
  e^{-\rho/2},
  \label{eq2:g-vs-Q1-Q2}
\end{equation}
\end{subequations}
where $\rho$, $\lambda$ and $\gamma$ are given by Eq.~(\ref{eq:rho-lambda-gamma-def}).
The functions $Q_1(\rho)$ and $Q_2(\rho)$ satisfy the equations,
\begin{subequations}    \label{subeqs:Q1-Q2-1st-order}
\begin{eqnarray}
  \rho Q'_1(\rho) +
  \bigg(
       \gamma -
       \frac{Z \alpha \veps}{\lambda}
  \bigg)
  Q_1(\rho)
  \nonumber \\ -
  \bigg(
       \kappa -
       \frac{Z \alpha m_e c^2}{\lambda}
  \bigg)
  Q_2(\rho)
  &=& 0,
  \label{eq1:Q1-vs-Q2-1st-order}
  \\
  \rho Q'_2(\rho) +
  \bigg(
       \gamma +
       \frac{Z \alpha \veps}{\lambda} -
       \rho
  \bigg)
  Q_2(\rho)
  \nonumber \\ -
  \bigg(
       \kappa +
       \frac{Z \alpha m_e c^2}{\lambda}
  \bigg)
  Q_1(\rho)
  &=& 0,
  \label{eq2:Q2-vs-Q1-1st-order}
\end{eqnarray}
\end{subequations}
where the prime denotes differentiation with respect to $\rho$, and
$\kappa = -1$ for the $s$ states.
{\color {red} $S$ or $s$ ?}
Substituting  $\kappa = -1$ into Eq.~(\ref{eq1:Q1-vs-Q2-1st-order}),  we 
can express $Q_2(\rho)$ in terms of $Q_1(\rho)$ and $Q'_1(\rho)$ as follows,
\begin{eqnarray}
  Q_2(\rho) &=&
  -\frac{\lambda}{\lambda + Z \alpha m_e c^2} \,
  \Bigg[
       \rho Q'_1(\rho)
  \nonumber \\ && +
       \bigg( \gamma - \frac{Z \alpha \veps}{\lambda} \bigg) \, Q_1(\rho)
  \Bigg].
  \label{eq:Q2-vs-Q1-dQ1}
\end{eqnarray}
Substituting this equation into Eq.~(\ref{eq2:Q2-vs-Q1-1st-order}), we find
\begin{equation}    \label{eq-for-Q1-2nd-order}
  \rho Q''_1(\rho) + \big( 2 \gamma + 1 - \rho \big) Q'_1(\rho) -
  \bigg( \gamma - \frac{Z \alpha \veps}{\lambda} \bigg) Q_1(\rho) = 0.
\end{equation}
The solution of Eq.~(\ref{eq-for-Q1-2nd-order}) that vanishes at $\rho \to \infty$ is
\cite{Landau-Lifshitz-4-hydrogen}
\begin{equation}
  Q_1(\rho) =
  A U \bigg( \gamma - \frac{Z \alpha \veps}{\lambda}, \, 2 \gamma + 1, \, \rho \bigg),
  \label{eq:Q1-res}
\end{equation}
where $U(a, b, \rho)$ is the confluent hypergeometric function.

Similarly, substituting in Eq.~(\ref{eq2:Q2-vs-Q1-1st-order}) $\kappa = -1$, 
we can express $Q_1(\rho)$ in terms of $Q_2(\rho)$ and $Q'_2(\rho)$,
\begin{eqnarray}
  Q_1(\rho) &=&
  -\frac{\lambda}{\lambda - Z \alpha m_e c^2} \,
  \Bigg[
       \rho Q'_2(\rho)
  \nonumber \\ && +
       \bigg( \gamma + \frac{Z \alpha \veps}{\lambda} - \rho\bigg) \, Q_2(\rho)
  \Bigg].
  \label{eq:Q1-vs-Q2-dQ2}
\end{eqnarray}
Substituting Eq.~(\ref{eq:Q1-vs-Q2-dQ2}) 
into Eq.~(\ref{eq1:Q1-vs-Q2-1st-order}), we get
\begin{eqnarray}
  \rho Q''_2(\rho) +
  \big( 2 \gamma + 1 - \rho \big) Q'_2(\rho) &-&
  \nonumber \\
  \bigg( \gamma + 1 - \frac{Z \alpha \veps}{\lambda} \bigg) Q_2(\rho)
  &=& 0.
  \label{eq-for-Q2-2nd-order}
\end{eqnarray}
Solution of Eq.~(\ref{eq-for-Q2-2nd-order}) vanishing at $\rho \to \infty$ is
\cite{Landau-Lifshitz-4-hydrogen}
\begin{equation}
  Q_2(\rho) =
  B U \bigg( \gamma + 1 - \frac{Z \alpha \veps}{\lambda}, \, 2 \gamma + 1, \, \rho \bigg).
  \label{eq:Q2-res}
\end{equation}

Substituting Eqs.~(\ref{eq:Q1-res}) and (\ref{eq:Q2-res}) into
Eq.~(\ref{eq:Q2-vs-Q1-dQ1}),
we get
\begin{equation}
  A = \frac{\lambda}{m_e c^2},
  \quad
  B = \frac{\lambda - Z \alpha m_e c^2}{m_e c^2}.
  \label{eq:B-vs-A}
\end{equation}
Alternatively, we can substitute Eqs.~(\ref{eq:Q1-res}) and (\ref{eq:Q2-res})
into Eq.~(\ref{eq:Q1-vs-Q2-dQ2}), and get Eq.~(\ref{eq:B-vs-A}).
Finally, the solution of the Dirac equation (\ref{subeqs:radial})
for $r > r_N$ is,
\begin{subequations}
\begin{eqnarray}
  f_{+}(r) &=&
  \sqrt{ \frac{m_e c^2 - \veps}{m_e c^2} } \,
  \rho^{\gamma - 1}
  e^{-\rho/2}
  \times \nonumber \\ &&
  \Bigg[
       \frac{\lambda}{m_e c^2} \,
       U \bigg( \gamma - \frac{Z \alpha \veps}{\lambda}, 2 \gamma + 1, \rho \bigg) +
  \nonumber \\ &&
       \frac{\lambda - Z \alpha m_e c^2}{m_e c^2} \,
       U \bigg( \gamma + 1 - \frac{Z \alpha \veps}{\lambda}, 2 \gamma + 1, \rho \bigg)
  \Bigg],
  \nonumber \\
  \label{eq:f-res-plus}
  \\
  \nonumber
  \\
  g_{+}(r) &=&
  -\sqrt{ \frac{m_e c^2 + \veps}{m_e c^2} } \,
  \rho^{\gamma - 1}
  e^{-\rho/2}
  \times \nonumber \\ &&
  \Bigg[
       \frac{\lambda}{m_e c^2} \,
       U \bigg( \gamma - \frac{Z \alpha \veps}{\lambda}, 2 \gamma + 1, \rho \bigg) -
  \nonumber \\ &&
       \frac{\lambda - Z \alpha m_e c^2}{m_e c^2} \,
       U \bigg( \gamma + 1 - \frac{Z \alpha \veps}{\lambda}, 2 \gamma + 1, \rho \bigg)
  \Bigg].
  \nonumber \\
  \label{eq:g-res-plus}
\end{eqnarray}
  \label{subeqs:fg-res-plus}
\end{subequations}

\subsubsection{Dirac equation (\ref{subeqs:radial}) solution for $r < r_N$}
\label{subsubsec:Dirac-r<rp-energy}

An analytic solution for the case of constant potential inside the nucleus is well-known 
\cite{Akhiezer_65}.  Equation~(\ref{eq1-radial}) allows us to express $(r f_{-}(r))$
in terms of $(r g_{-}(r))$ and $(r g_{-}(r))'$,
\begin{eqnarray}
  \big( r f_{-}(r) \big) &=&
  \frac{\hbar c}{m_e c^2 - V(r_N) + \veps}
  \times \nonumber \\ &&
  \bigg[
       \big( r g_{-}(r) \big)' -
       \frac{1}{r} \,
       \big( r g_{-}(r) \big)
  \bigg].
  \label{eq:f-vs-g-r<rp}
\end{eqnarray}
Substituting Eq.~(\ref{eq:f-vs-g-r<rp}) into Eq.~(\ref{eq2-radial}),
we get
\begin{equation}
  \big( r g_{-}(r) \big)'' +
  k^2(\veps)
  \big( r g_{-}(r) \big)
  = 0,
  \label{eq-for-g-2nd-r<rp}
\end{equation}
where
\begin{equation}
  k(\veps) = \frac{1}{\hbar c} \sqrt{ ( \veps - V(r_N) )^{2} - m_e^2 c^4 }.
  \label{eq:k-epsilon}
\end{equation}
The nonsingular solution of Eq.~(\ref{eq-for-g-2nd-r<rp}) at $r = 0$  is,
\begin{equation}
  g_{-}(r) = C \, \frac{\sin ( k(\veps) r)}{k(\veps) r},
  \label{eq:g-res-r<rp}
\end{equation}
where $C$ is a normalization constant.

Equation~(\ref{eq2-radial}) allows us to express $(r g_{-}(r))$
in terms of $(r f_{-}(r))$ and $(r f_{-}(r))'$,
\begin{eqnarray}
  \big( r g_{-}(r) \big) &=&
  \frac{\hbar c}{m_e c^2 + V(r_N) - \veps}
  \times \nonumber \\ &&
  \bigg[
       \big( r f_{-}(r) \big)' +
       \frac{1}{r} \,
       \big( r f_{-}(r) \big)
  \bigg].
  \label{eq:g-vs-f-r<rp}
\end{eqnarray}
Substituting Eq.~(\ref{eq:g-vs-f-r<rp}) into Eq.~(\ref{eq1-radial}),
we get
\begin{equation}
  \big( r f_{-}(r) \big)'' +
  \bigg( k^2(\veps) - \frac{2}{r^2} \bigg)
  \big( r f_{-}(r) \big)
  = 0.
  \label{eq-for-f-2nd-r<rp}
\end{equation}
The nonsingular solution of Eq.~(\ref{eq-for-f-2nd-r<rp}) at $r = 0$ is,
\begin{equation}
  f_{-}(r) = D \, \frac{k(\veps) r \cos ( k(\veps) r ) - \sin ( k(\veps) r)}{k^2(\veps) r^2},
  \label{eq:f-res-r<rp}
\end{equation}
where $D$ is a normalization constant.

Substituting Eqs.~(\ref{eq:g-res-r<rp}) and (\ref{eq:f-res-r<rp}) into
Eq.~(\ref{eq:f-vs-g-r<rp}), we get
\begin{equation}
  \frac{D}{C} = \sqrt{ \frac{\veps - V(r_N) - m_e c^2}{\veps - V(r_N) + m_e c^2} }.
  \label{eq:D-vs-C}
\end{equation}
Alternatively, we can substitute Eqs.~(\ref{eq:g-res-r<rp}) and
(\ref{eq:f-res-r<rp}) into Eq.~(\ref{eq:g-vs-f-r<rp}), and get
Eq.~(\ref{eq:D-vs-C}).
Finally, the solution of the Dirac equation (\ref{subeqs:radial})
for $r < r_N$ is,
\begin{subequations}
\begin{eqnarray}
  f_{-}(r) &=&
  \sqrt{ \frac{\veps - V(r_N) - m_e c^2}{m_e c^2 } }
  \times \nonumber \\ &&
  \frac{k(\veps) r \cos ( k(\veps) r ) - \sin ( k(\veps) r)}{k^2(\veps) r^2},
  \label{eq:f-res-minus}
  \\
  g_{-}(r) &=&
  \sqrt{ \frac{\veps - V(r_N) + m_e c^2}{m_e c^2 } } \,
  \frac{\sin ( k(\veps) r)}{k(\veps) r}.
  \label{eq:g-res-minus}
\end{eqnarray}
  \label{subeqs:fg-res-minus}
\end{subequations}

\subsubsection{Boundary conditions at $r = r_N$}
  \label{subsubsec:Dirac-r=rp-boundary}

\begin{figure}
\centering
 \subfigure[]
 {\includegraphics[width=0.9 \linewidth,angle=0]
   {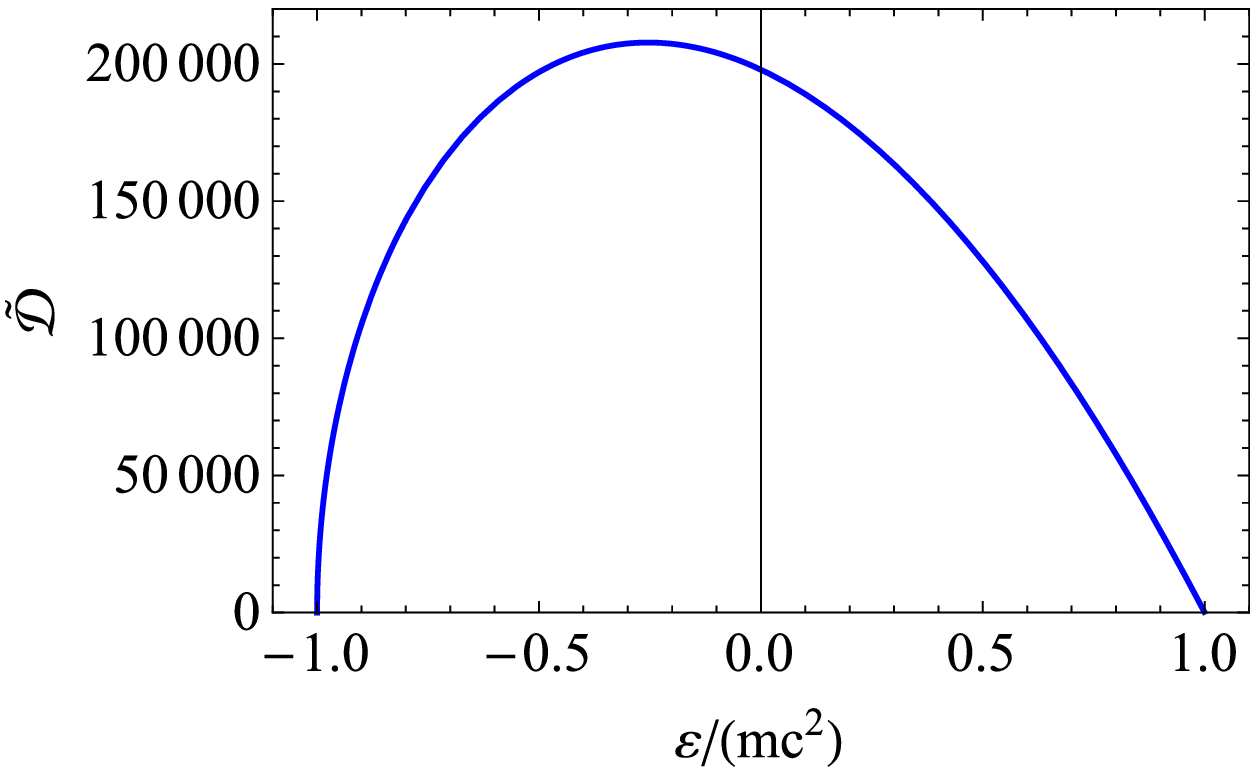}
   \label{Fig-det-non-zero}}
 \subfigure[]
 {\includegraphics[width=0.9 \linewidth,angle=0]
   {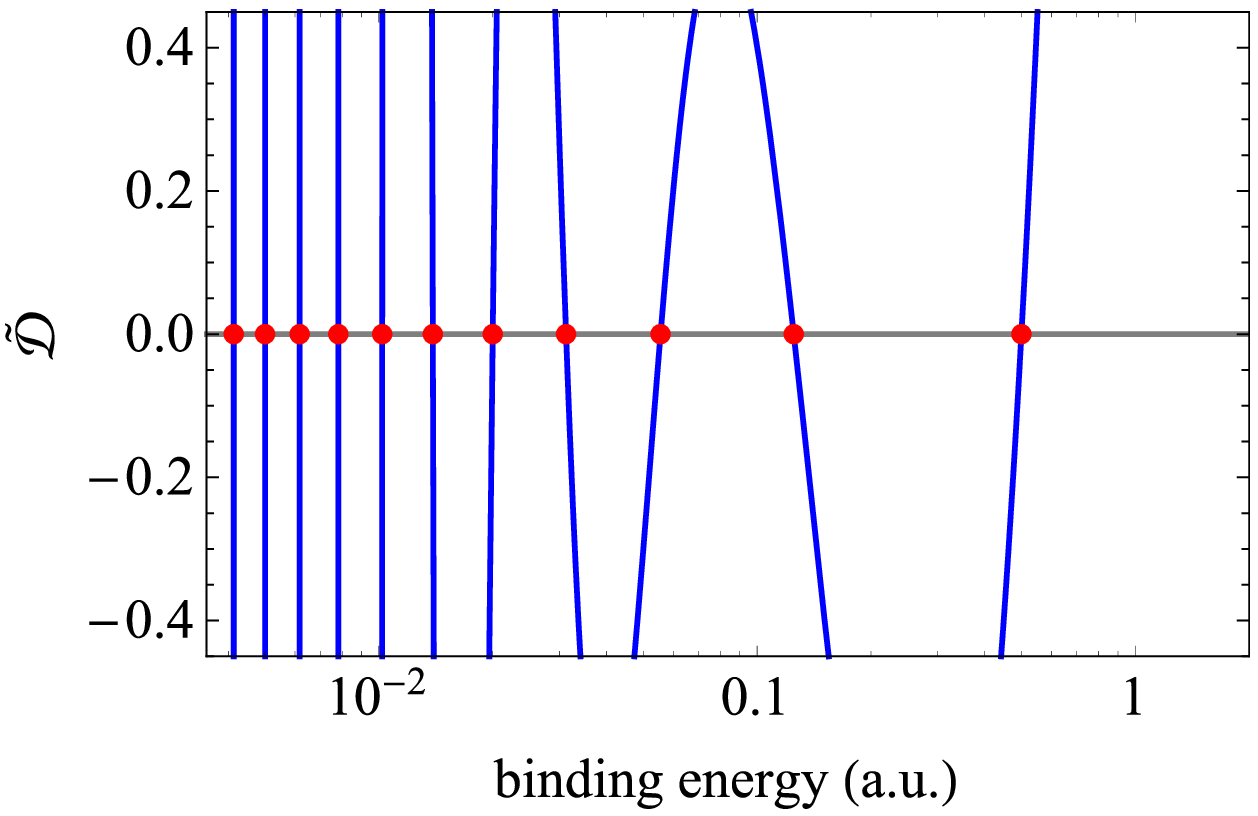}
   \label{Fig-det-zero}}
\caption{\footnotesize 
  (a) The function $\tilde{\mathfrak{D}}(\veps)$ defined in Eq.~(\ref{eq:tilde-D}) versus
  the energy $\veps$ for $|\veps | < 0.999 \, m_e c^2$.
  (b) Log-linear plot of $\tilde{\mathfrak{D}}(\veps)$ in Eq.~(\ref{eq:tilde-D})
  as a function of the binding energy $\epsilon = m c^2 - \veps$ (blue curve)
  for $\veps$ close to $m c^2$  (the range includes the 11 lowest $s$  ${}^{1}$H states). 
  The red dots show the Dirac binding energies $\epsilon_{n}^{(0)}$ of the hydrogen atom 
  with zero proton radius (i.e., the Dirac energies).  Here $\tilde{\mathfrak{D}}(\veps)$ is 
  dimensionless, and the binding energy in (b) is in Hartree.
  }
\label{Fig-det-zero-nonzero}
\end{figure}

The functions $f_{\pm}(r)$ and $g_{\pm}(r)$ satisfy
the boundary conditions  specified in Eq.~(\ref{eq:det}) at $r = r_N$.
Clearly, $f_{\pm}(r_N)$ and $g_{\pm}(r_N)$ in Eqs.~(\ref{subeqs:fg-res-plus}) and
(\ref{subeqs:fg-res-minus}), and hence $\mathfrak{D}$ in Eq.~(\ref{eq:det}) 
depend on $\veps$.  Solving the equation $\mathfrak{D}(\veps) = 0$ for $\veps$ yields
the energies of the hydrogen atom.  These are tabulated in Table \ref{Table:energies}.
Figure~\ref{Fig-det-zero-nonzero}(a) plots the function
\begin{equation}
  \tilde{\mathfrak{D}}(\veps) =
  \mathfrak{L}(\veps) \mathfrak{D}(\veps),
  \label{eq:tilde-D}
\end{equation}
versus $\veps$ for $Z=1$, $r_N = 0.8414 \times 10^{-13}$~cm (the proton RMS radius \cite{proton_radius}) 
and $| \veps | < 0.9 \, m_e c^2$, where $\mathfrak{D}(\veps)$ is given in Eq.~(\ref{eq:det}), and
$$
  \mathfrak{L}(\veps) = \frac{m_e^2 c^4 - \veps^2}{m_e^2 c^4}.
$$
$\mathfrak{L}(\veps)$ is positive for $|\veps| < m_e c^2$. 
Both $\tilde{\mathfrak{D}}(\veps)$ and $\mathfrak{L}(\veps)$ are dimensionless.
The wave functions in Eqs.~(\ref{subeqs:fg-res-plus})
and (\ref{subeqs:fg-res-minus}) are dimensionless
and will be normalized in what follows [by multiplying by the
factor $\mathcal{N}_{n}$ given by Eq.~(\ref{eq:normalization}),  
where the dimensions of $\mathcal{N}_{n}$ is
$[ \mathcal{N}_{n} ] = \mathrm{Length}^{-3/2}$].
$\tilde{\mathfrak{D}}(\veps)$ is positive, and there are no bound states in the interval 
$-m_e c^2 < \veps < 0.99 \, m_e c^2$.
Figure~\ref{Fig-det-zero} shows $\tilde{\mathfrak{D}}(\veps)$ for
$\veps$ close to and below $m_e c^2$.
$\tilde{\mathfrak{D}}(\veps) = 0$ when $\veps$ is very close
to the Dirac energies $\veps_{n}^{(0)}$ of
the hydrogen atom given by Eq.~(\ref{eq:energy-hydrgen})
with $Z = A = 1$, where the subscript $(0)$ indicates the
point charge nucleus limit.

\begin{table}
  \begin{tabular}{|c|c|c|c|}
     \hline
     $n$ & 1 & 2 & 3
     \\
     \hline
     $\epsilon_{n}$ &
     $0.499734640577$ &
     $0.1249340759790$ &
     $0.0555260711867$
     \\
     \hline
     $\epsilon_{n}^{(0)}$ &
     $0.499734640745$ &
     $0.1249340760015$ &
     $0.0555260711927$
     \\
     \hline
  \end{tabular}
  \caption{\footnotesize 
  Hydrogen binding energies $\epsilon_{n}(r_N)$ in Hartree
  calculated for model (a) using Eq.~(\ref{eq:det}),  and the hydrogen atom Dirac energies 
  $\epsilon_{n}^{(0)}$ for $r_N = 0$ (i.e., for point-like proton) in Eq.~(\ref{eq:energy-hydrgen}).
  $r_N$  is given in Table \ref{Table-atomic-nuclear-radii}.}
  \label{Table:energies}
\end{table}

The binding energies $\epsilon_{n}(r_N)$ (with $n = 1, 2, 3$
and $r_N$ being the nuclear radius of ${}^{1}$H)
calculated numerically from Eq.~(\ref{eq:det}) and
$\epsilon_{n}^{(0)} = m_e c^2 - \veps_{n}^{(0)}$, with $\veps_{n}^{(0)}$ given
in Eq.~(\ref{eq:energy-hydrgen}),  are listed in 
Table~\ref{Table:energies}.  It is seen that
$\big| \epsilon_{1}(r_N) - \epsilon_{1}^{(0)} \big| ~\approx~1.680 \times 10^{-10}$
Hartree, which is clearly very much smaller than $\epsilon_{1}^{(0)}$.  
The Dirac energies $\veps_{n}(r_N)$ are pushed up by the repulsive potential 
$W_a(r) = [V_C(r_N) - V_C(r)]\Theta(r_N - r)$ [see Eq.~(\ref{eq:W})].
Note the following inequalities:
$$
  \big|
      \epsilon_{n + 1}(r_N) -
      \epsilon_{n + 1}^{(0)}
  \big|
  ~\ll~
  \big|
      \epsilon_{n}(r_N) -
      \epsilon_{n}^{(0)}
  \big|,
$$
where $n = 1, 2, 3, \ldots$

\subsubsection{Wave functions}
  \label{subsubsec:wave-functions}

We now consider normalization of the wave functions $g_n$ and $f_n$.
Here $g_n(r)$ is given by Eqs.~(\ref{eq:g-res-plus}) for $r > r_N$
and by Eq.~(\ref{eq:g-res-minus}) for $r < r_N$, and
$f_n(r)$ is given by Eq.~(\ref{eq:f-res-plus}) for $r > r_N$
and by Eq.~(\ref{eq:f-res-minus}) for $r < r_N$.
The normalization constants $\mathcal{N}_{n}$
are chosen such that
\begin{equation}
  \mathcal{N}_{n}^{2} 
  \int\limits_{0}^{\infty}
  [g_{n}^{2}(r) + f_{n}^{2}(r)] r^2 dr = 1.
  \label{eq:normalization}
\end{equation}

\begin{figure}
\centering
 \subfigure[]
 {\includegraphics[width=0.9 \linewidth,angle=0]
   {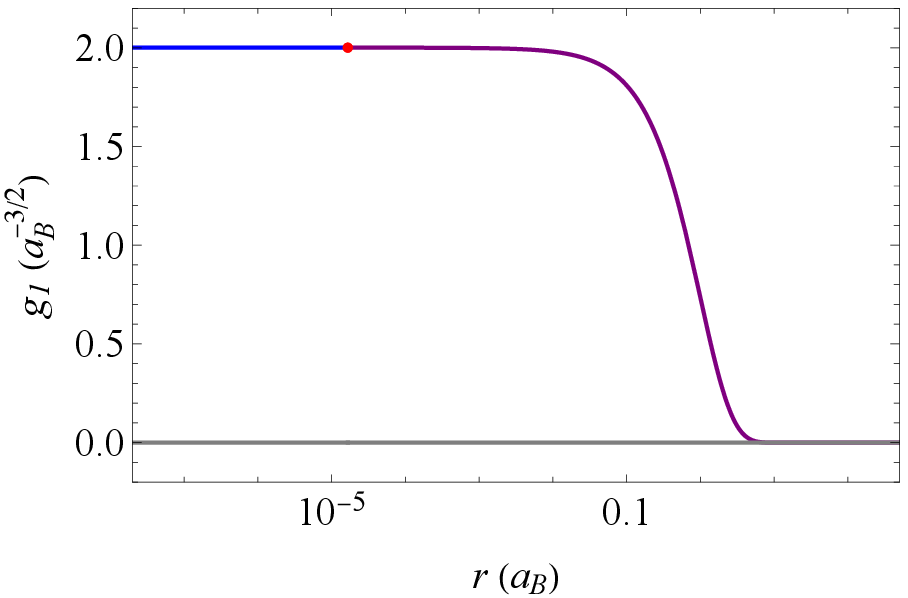}
   \label{Fig-g-1s}}
 \subfigure[]
 {\includegraphics[width=0.9 \linewidth,angle=0]
   {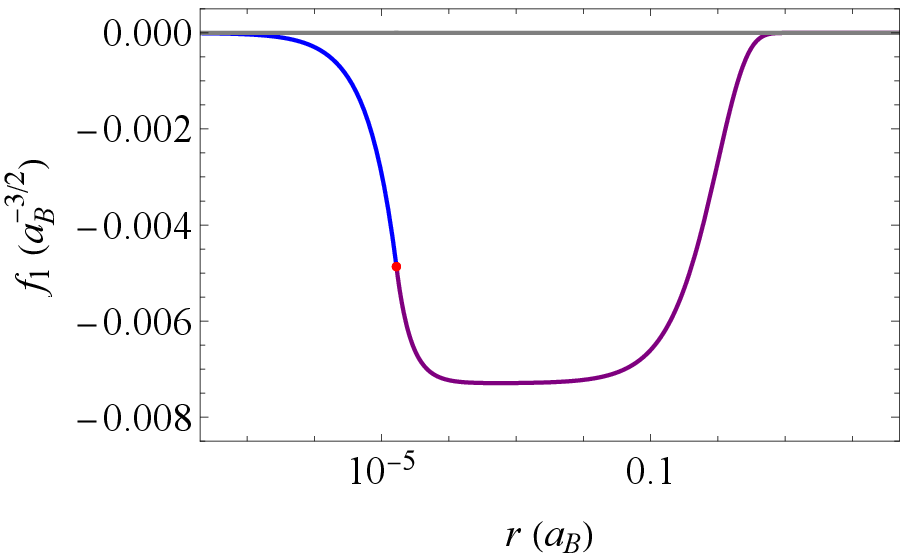}
   \label{Fig-f-1s}}
\caption{\footnotesize 
  Log-linear plot of (a) $\mathcal{N} g_1(r)$ and (b) $\mathcal{N} f_1(r)$
  for the $1s$ state of hydrogen atom with the energy $\veps_1$ (in Hartree).
  Here $g_1(r)$ is given by Eq.~(\ref{eq:g-res-plus}) for $r > r_N$
  (purple) and by Eq.~(\ref{eq:g-res-minus}) for $r < r_N$ (blue).
  $f_1(r)$ is given by Eq.~(\ref{eq:f-res-plus}) for $r > r_N$
  (purple) and by
  Eq.~(\ref{eq:f-res-minus}) for $r < r_N$ (blue).
  The normalization constant $\mathcal{N}$
  is found from Eq.~(\ref{eq:normalization}).
  The red dots show $g_1(r_N)$ and $f_1(r_N)$.
  }
\label{Fig-g-f-1s}
\end{figure}

Figure \ref{Fig-g-f-1s} plots $\mathcal{N}_1 (g_1(r), f_1(r))$ for the $1s$ state.
The functions $( g_1(r), f_1(r) )$ and their derivatives
$( g'_1(r), f'_1(r) )$ are continuous at $r = r_N$, where
$r_N = 1.658 \times 10^{-5} a_B$.
$g_1(r)$ has no nodes, has a maximum at $r = 0$, and vanishes
as $r \to \infty$.  $f_1(r)$ vanishes at $r = 0$ and is negative for finite $r$;
it reaches its minimum and tends to zero as $r \to \infty$.

\begin{figure}
\centering
 \subfigure[]
 {\includegraphics[width=0.9 \linewidth,angle=0]
   {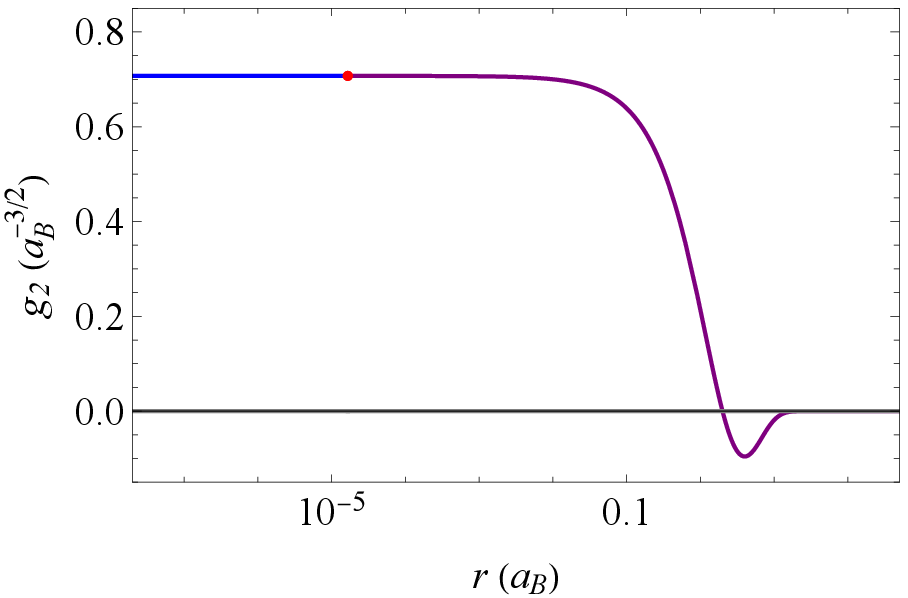}
   \label{Fig-g-2s}}
 \subfigure[]
 {\includegraphics[width=0.9 \linewidth,angle=0]
   {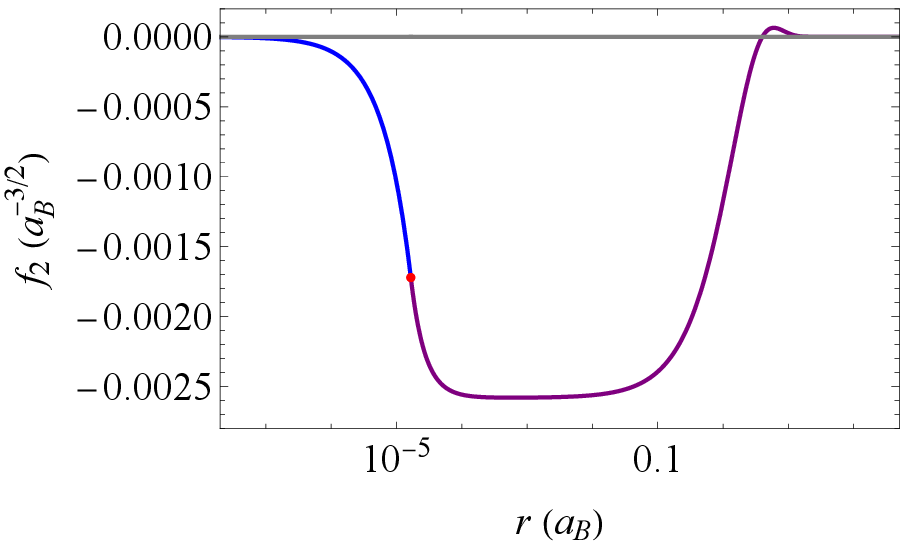}
   \label{Fig-f-2s}}
\caption{\footnotesize 
  The log-linear plot of
  (a) $\mathcal{N}_{2} g_2(r)$ and (b) $\mathcal{N}_{2} f_2(r)$
  for the $2s$ state of hydrogen atom with the energy $\veps_2$
  (in Hartree).
  Here $g_2(r)$ is given by Eq.~(\ref{eq:g-res-plus}) for $r > r_N$
  (purple) and
  by Eq.~(\ref{eq:g-res-minus}) for $r < r_N$ (blue).
  $f_2(r)$ is given by Eq.~(\ref{eq:f-res-plus}) for $r > r_N$
  (purple) and by
  Eq.~(\ref{eq:f-res-minus}) for $r < r_N$ (blue).
  The normalization constant $\mathcal{N}_{2}$
  is found from Eq.~(\ref{eq:normalization}).
  The red dots show the wave functions at $r = r_N$.}
\label{Fig-g-f-2s}
\end{figure}

Figure \ref{Fig-g-f-2s} is a log-linear plot for the $2s$ normalized wave
function $\mathcal{N}_{2} ( g_2(r), f_2(r))$.
The functions $( g_2(r), f_2(r) )$ and their derivatives
$( g'_2(r), f'_2(r) )$ are continuous at $r = r_N$.
$g_2(r)$ has a maximum at $r = 0$, and one node. It vanishes
at $r \gg a_B$.  $f_2(r)$ has two nodes: one node at $r = 0$, and
another node at finite $r$; it vanishes as $r \to \infty$.

\subsubsection{$1s$ and $2s$ state energies for various isotopes}
  \label{subsubsec_model-a-analyt}

We now solve the determinantal equation (\ref{eq:det}) 
for model (a) (constant potential inside the nucleus)
and find the FNS corrections to the energies of the ground
$1s$ state and the first excited $2s$ state for various isotopes.

\begin{table}
\caption{\footnotesize
  FNS corrections $\veps_{1,a}$ and $\veps_{1,b}$ (in Hartree) to the energy of the ground $1s$ state
  of H\&HLI.  The numbers in the brackets indicate the uncertainties $\delta \veps_{1,a}$
  and $\delta \veps_{1,b}$  in the calculated ground-state energy due to
  the uncertainty in the nuclear radius $r_N$ in Table~\ref{Table-atomic-nuclear-radii}
  [see Eqs.~(\ref{eq:delta-epsilon}) and (\ref{eq:delta-epsilon-b})]. The concise notation used
  for errors is explained in Ref.~\cite{errors}.}
\begin{center}
  \begin{tabular}{|c|c|c|}
    \hline
    Isotope ($Z,A$) &
    $\veps_{1,a} ( \delta \veps_{1,a} )$ &
    $\veps_{1,b} ( \delta \veps_{1,b} )$
    \\
    \hline
    H(1, 1) &
    $1.6837 (76) E^{-10}$ &
    $1.6837 (76) E^{-10}$
    \\
    \hline
    H(1, 2) &
    $1.07773 (75) E^{-9}$ &
    $1.07773 (75) E^{-9}$
    \\
    \hline
    H(1, 3) &
    $7.37 (30) E^{-10}$ &
    $7.37 (30) E^{-10}$
    \\
    \hline
    He(2, 3) &
    $1.4746 (45) E^{-8}$ &
    $1.4746 (45) E^{-8}$
    \\
    \hline
    He(2, 4) &
    $1.0711 (36) E^{-8}$ &
    $1.0711 (36) E^{-8}$
    \\
    \hline
    K(19, 39) &
    $4.1987 (46) E^{-4}$ &
    $4.1974 (46) E^{-4}$
    \\
    \hline
    K(19, 40) &
    $4.2064 (68) E^{-4}$ &
    $4.2051 (68) E^{-4}$
    \\
    \hline
    K(19, 41) &
    $4.240 (13) E^{-4}$ &
    $4.238 (13) E^{-4}$
    \\
    \hline
    Rb(37, 85) &
    $1.2547 (14) E^{-2}$ &
    $1.2534 (14) E^{-2}$
    \\
    \hline
    Rb(37, 87) &
    $1.2520 (12) E^{-2}$ &
    $12507 (12) E^{-2}$
    \\
    \hline
    Cs(55, 133) &
    $0.13135 (23)$ &
    $0.13107 (23)$
    \\
    \hline
    Cs(55, 135) &
    $0.13148 (23)$ &
    $0.13120 (23)$
    \\
    \hline
    Pb(82, 204) &
    $2.4687 (10)$ &
    $2.4586 (10)$
    \\
    \hline
    Pb(82, 206) &
    $2.4758 (10)$ &
    $2.4657 (10)$
    \\
    \hline
    Pb(82, 207) &
    $2.4787 (10)$ &
    $2.4686 (10)$
    \\
    \hline
    Pb(82, 208) &
    $2.48367 (93)$ &
    $2.47351 (93)$
    \\
    \hline
    Bi(83, 209) &
    $2.7559 (20)$ &
    $2.7444 (20)$
    \\
    \hline
    U(92, 235) &
    $7.3062 (75)$ &
    $7.2718  (75)$
    \\
    \hline
    U(92, 236) &
    $7.3234 (69)$ &
    $7.2889 (69)$
    \\
    \hline
    U(92, 238) &
    $7.3490 (60)$ &
    $7.3143 (60)$
    \\
    \hline
  \end{tabular}
\end{center}
\label{Table-energy-constant}
\end{table}

Tables~\ref{Table-energy-constant} and \ref{Table-transitions}
show the FNS corrections
\begin{equation}    \label{eq:correction_surface}
  \veps_{n,a} = \veps_n - \veps_{n}^{(0)},
  \quad
  n = 1, 2 .
\end{equation}
Here $\veps_{1}$ and  $\veps_{2}$ are calculated numerically using Eq.~(\ref{eq:det}),
and $\veps_{1}^{(0)}$ and $\veps_{2}^{(0)}$ are given in analytic form 
by Eq.~(\ref{eq:energy-hydrgen}).
$\veps_{1,a}$ and $\veps_{2,a}$ are functions of the nuclear radius
$r_a = r_N$, where $r_N$ is tabulated in Table~\ref{Table-atomic-nuclear-radii}.
The uncertainty in $r_N$ gives rise to uncertainty $\delta \veps_{n,a}$ in
$\veps_{n,a}$, which can be estimated as,
\begin{equation}
  \delta \veps_{n,a} =
  \bigg| \frac{\partial \veps_{n,a}}{\partial r_N} \bigg| \, \delta r_N.
  \label{eq:delta-epsilon}
\end{equation}
The FNS correction $\veps_{1,a}$  to the ground-state energy,
and the uncertainty $\delta \veps_{1,a}$  are tabulated in Table~\ref{Table-energy-constant},
and the FNS correction $\veps_{2,a}$ to the excited-state energy,
and the uncertainty $\delta \veps_{2,a}$  are tabulated in Table~\ref{Table-transitions}.
The corrections $\varepsilon_{1,a}$ and  $\varepsilon_{2,a}$ satisfy the inequalities 
$\varepsilon_{1,a} \gg \varepsilon_{2,a} > 0$.
Note that, for isotopes of atoms with a given $Z$, $\veps_{1,a}$ and $\veps_{2,a}$ 
do not always increase with $A$.  This is because the nuclear radii $r_N$ in
Table~\ref{Table-atomic-nuclear-radii} for some isotopes are not
monotonic with $A$ (see $r_N$ for He and Rb).
The repulsive potential $W_a(r)$ in Eq.~(\ref{eq:W}) increases
with the nuclear radius $r_N$, and $r_N$ increases with $Z$, 
see Table~\ref{Table-atomic-nuclear-radii},
hence the energy correction $\veps_a$ in
Table~\ref{Table-energy-constant} increases with $Z$.

\begin{table}
\caption{\footnotesize 
  FNS corrections $\veps_{2,a}$ and $\veps_{2,b}$ (in Hartree) to
  the energy of the excited $2s$ state of H\&HLI.
  The numbers in brackets indicate the uncertainties $\delta \veps_{2,a}$
  and $\delta \veps_{2,b}$  in the calculated ground-state energy due to
  the uncertainty in the nuclear radius $r_N$ given in Table~\ref{Table-atomic-nuclear-radii}
  [see Eqs.~(\ref{eq:delta-epsilon}) and (\ref{eq:delta-epsilon-b})].}
\begin{center}
  \begin{tabular}{|c|c|c|}
    \hline
    Isotope ($Z,A$) &
    $\veps_{2,a} ( \delta \veps_{2,a} )$ &
    $\veps_{2,b} ( \delta \veps_{2,b} )$
    \\
    \hline
    H(1, 1) &
    $2.1047 (95) E^{-11}$ &
    $2.1047 (95) E^{-11}$
    \\
    \hline
    H(1, 2) &
    $1.34722 (94) E^{-10}$ &
    $1.34722 (94) E^{-10}$
    \\
    \hline
    H(1, 3) &
    $9.21 (38) E^{-11}$ &
    $9.21 (38) E^{-11}$
    \\
    \hline
    He(2, 3) &
    $1.8436 (56) E^{-9}$ &
    $1.8436 (56) E^{-9}$
    \\
    \hline
    He(2, 4) &
    $1.3391 (45) E^{-9}$ &
    $1.3391 (45) E^{-9}$
    \\
    \hline
    K(19, 39) &
    $5.3381 (58) E^{-5}$ &
    $5.3365 (58) E^{-5}$
    \\
    \hline
    K(19, 40) &
    $5.3480 (86) E^{-5}$ &
    $5.3463 (86) E^{-5}$
    \\
    \hline
    K(19, 41) &
    $5.390 (17) E^{-5}$ &
    $5389 (17) E^{-5}$
    \\
    \hline
    Rb(37, 85) &
    $1.6732 (18) E^{-3}$ &
    $1.6714 (18) E^{-3}$
    \\
    \hline
    Rb(37, 87) &
    $1.6696 (16) E^{-3}$ &
    $1.6678 (16) E^{-3}$
    \\
    \hline
    Cs(55, 133) &
    $1.8969 (33) E^{-2}$ &
    $1.8928 (33) E^{-2}$
    \\
    \hline
    Cs(55, 135) &
    $1.8988 (34) E^{-2}$ &
    $1.8947 (34) E^{-2}$
    \\
    \hline
    Pb(82, 204) &
    $0.42868 (17)$ &
    $0.42692 (17)$
    \\
    \hline
    Pb(82, 206) &
    $0.42991 (17)$ &
    $0.42814 (17)$
    \\
    \hline
    Pb(82, 207) &
    $0.43041 (17)$ &
    $0.42864 (17)$
    \\
    \hline
    Pb(82, 208) &
    $0.43127 (16)$ &
    $0.42950 (16)$
    \\
    \hline
    Bi(83, 209) &
    $0.48259 (36)$ &
    $0.48057 (36)$
    \\
    \hline
    U(92, 235) &
    $1.3880 (14)$ &
    $1.3814 (14)$
    \\
    \hline
    U(92, 236) &
    $1.3912 (13)$ &
    $1.3846 (13)$
    \\
    \hline
    U(92, 238) &
    $1.3961 (11)$ &
    $1.3895 (11)$
    \\
    \hline
  \end{tabular}
\end{center}
\label{Table-transitions}
\end{table}

Table \ref{Table-transitions} shows the calculated FNS corrections $\veps_{2,a}$ and $\veps_{2,b}$.
The FNS corrections to the $1s$-$2s$ hydrogen transition frequency $\Delta \omega_{1s,2s}^{(a)}$ and
$\Delta \omega_{1s,2s}^{(b)}$ are defined as,
\begin{equation}    \label{eq:Delta-1s-2s}
   \Delta \omega_{1s,2s}^{(a)} = \frac{1}{\hbar} \, \big( \veps_{2,a} - \veps_{1,a} \big), 
   \quad
   \Delta \omega_{1s,2s}^{(b)} = \frac{1}{\hbar} \big( \veps_{2,b} - \veps_{1,b} \big).
\end{equation}
Note that the method to accurately measure the isotope shifts of atomic 
transitions reported in Ref.~\cite{Manovitz_19} can be used
to experimentally probe the isotope shifts calculated here.

Equation (\ref{eq:energy-hydrgen}) for a point-like nucleus
and the numerical calculations with the nuclear charge distribution
for models (a) and (b) take into account recoil effects by replacing
the electron mass $m_e$ by the reduced mass  $m \equiv m(Z,A)$ defined in Eq.~(\ref{eq:reduced-mass}).
This naive expression can be used as a starting
point for calculating the recoil corrections to the electron energy levels.
An explicit expression for the $s$-state energies with the reduced mass
dependence is derived from quantum electrodynamics in
Ref.~\cite{Eides-hydrogen-atom},
\begin{eqnarray}
  \veps_{n}^{\mathrm{(Dirac)}} &=& m_e c^2 +
  Z^2 h c R_{\infty} \times \nonumber \\
  && \Big[ \frac{2  \mathcal{G}(n)}{1 + \mu_{ep}} -
    \frac{\mu_{ep} \, Z^2 \alpha^2 \, \mathcal{G}^{2}(n)} {\big( 1 + \mu_{ep} \big)^{3}}
  \Big] .
  \label{eq:energy-point-recoil}
\end{eqnarray}
The dimensionless function $\mathcal{G}(n)$ is given in Eq.~(\ref{eq:F_n}).
The first two terms on the right hand side of Eq.~(\ref{eq:energy-point-recoil})
give the energy of the $s$ state in Eq.~(\ref{eq:energy-hydrgen}),
but the last term in this equation, which is proportional to $\mathcal{G}^{2}(n)$
is proportional to $\mu_{ep} (Z \alpha)^{4}$.  This term vanishes in the limit
of infinitely large nuclear mass, and hence is absent in Eq.~(\ref{eq:energy-hydrgen}).
For hydrogen, this correction term to the energy is
$$  
  h c R_{\infty}
  \frac{\mu_{e} \, \alpha^2 \, \mathcal{G}^{2}(n)}{\big( 1 + \mu_{ep} \big)^{3}}
  \approx  \frac{7.2 \times 10^{-9}}{n^2} ~ \epsilon_H.
$$

The Dirac energy in Eq.~(\ref{eq:energy-point-recoil}) can be expanded in a power series 
in $Z^2 \alpha^2$, as in Eq.~(3.4) of Ref.~\cite{Eides-hydrogen-atom}:
\begin{eqnarray}
  \tilde\veps_{n}^{\mathrm{(Dirac)}} &=&
  m_e c^2 -
  \frac{h c R_{\infty}}{1 + \mu_{ep}} \,
  \bigg\{
    \frac{Z^2}{n^2} +
    \frac{Z^4 \alpha^2}{n^3} \,
    \Big(
      1 -
      \frac{3}{4 n}
  \nonumber \\ && +
      \frac{\mu_{ep}}{(1 + \mu_{ep})^{2}} \, \frac{1}{4 n}
    \Big)
  \bigg\} .
  \label{eq:energy-ns-Dirac-recoil-34}
\end{eqnarray}
Equation (\ref{eq:energy-ns-Dirac-recoil-34}) is a power series expansion of 
Eq.~(\ref{eq:energy-point-recoil}) in $\alpha^2$, up to $\alpha^4$
(recall that a factor of $\alpha^2$ is contained in $R_\infty$).
The $1s$ energy obtained using (\ref{eq:energy-ns-Dirac-recoil-34}) differs from that
obtained using (\ref{eq:energy-point-recoil}) by about a MHz.

\subsection{Dirac equation (\ref{subeqs:radial}) solutions for model (b)}
\label{subsec:Dirac-finite-radius-model-b}

Here we find the wave functions and binding energies
of H\&HLI with nuclear charge distributed
homogeneously inside a sphere of radius $r_b$ defined in
Eq.~(\ref{eq:r_a-r_b-vs-r_N}).  The potential $V_b(r)$ in Eq.~(\ref{Eq:Vb}) 
is harmonic inside the nucleus.  In this case an analytic solution is unknown.
We rewrite Eq.~(\ref{subeqs:radial}) with $\kappa = -1$ as,
\begin{subequations}
  \label{subeqs:radial-s-states}
\begin{eqnarray}
  &&
  \frac{\hbar c \, r^2 g'_{-}(r)}{m c^2 - V_b(r) + \veps} =
  r^2 f_{b,-}(r),
  \label{eq1:radial-s-states}
  \\
  &&
  \big( r^2 f_{b,-}(r) \big)' =
  \frac{m c^2 + V_b(r) - \veps}{\hbar c} \,
  r^2 g_{b,-}(r).
  \label{eq2:radial-s-states}
\end{eqnarray}
\end{subequations}
Differentiating Eq.~(\ref{eq1:radial-s-states}) and using Eq.~(\ref{eq2:radial-s-states}) 
we get the following second-order differential equation for $g_{b,-}(r)$:
\begin{eqnarray}
  \frac{1}{r^2} \,
  \frac{d}{d r}
  \bigg(
       \frac{r^2}{m c^2 - V_b(r) + \veps} \,
       \frac{d g_{b,-}(r)}{d r}
  \bigg)
  \nonumber \\ =
  \frac{1}{\hbar^2 c^2} \,
  \big[ m c^2 + V_b(r) - \veps \big]
  g_{b,-}(r).
  \label{eq:for-g-2nd-order}
\end{eqnarray}
We solve Eq.~(\ref{eq:for-g-2nd-order}) numerically and find
the radial wave function $g_{b,-}(r)$ for $r < r_b$.  For this purpose, 
we need the binding energy of the atom (which we do not know).
In order to find the energy and the wave function, we
apply the iteration method described below.
The radial wave function $f_{b,-}(r)$ can be found from
the equation,
\begin{equation}
  f_{b,-}(r) =
  \frac{\hbar c \, g'_{b,-}(r)}{m c^2 - V_b(r) + \veps}.
  \label{eq:f-vs-g}
\end{equation}
The ground-state wave functions satisfy the boundary conditions
at $r = 0$,
\begin{equation}
  f_{b,-}(0) = 0,
  \quad
  g_{b,-}(0) = 1.
  \label{eq:bundary-conditions-r=0}
\end{equation}
The radial wave functions $g_{+}(r)$ and $f_{+}(r)$ with $r > r_b$
are given in Eq.~(\ref{subeqs:fg-res-plus}), and
the radial wave functions $g_b(r), f_b(r)$ satisfy the boundary
conditions at $r = r_b$,
\begin{equation}
  A_{-} g_{b,-}(r_b) = A_{+} g_{+}(r_b),
  \quad
  A_{-} f_{b,-}(r_b) = A_{+} f_{+}(r_b).
  \label{eq:boundary-conditions-b}
\end{equation}
Here $A_{\pm}$ are (unknown) normalization constants.
The set of equations (\ref{eq:boundary-conditions-b})
has nontrivial solutions when the determinant
$\mathfrak{D}(\veps)$ vanishes, where
\begin{equation}  \label{eq:det-b}
  \mathfrak{D}(\veps) =
  g_{b,-}(r_b) f_{+}(r_b) -
  g_{+}(r_b) f_{-}(r_b) .
\end{equation}
Appendix   \ref{append:iteration} explains the iteration method for finding the ground state 
and excited $s$ state energies using Eq.~(\ref{eq:det-b}).

The uncertainty $\delta \veps_{n,b}$ in the model (b) energy $\veps_{n,b}$ of the 
$n s$ state is estimated as,
\begin{equation}
  \delta \veps_{n,b} =
  \int\limits_{0}^{r_b}
  \delta V_b(r)
  \Big[
      \big| g_{n, -}(r) \big|^{2} +
      \big| f_{n, -}(r) \big|^{2}
  \Big]
  r^2 dr,
  \label{eq:delta-epsilon-b}
\end{equation}
where $\delta V_b(r)$ is given in Eq.~(\ref{eq:delta_V_b}), and
$n = 1$ for the ground state.

The FNS correction $\veps_{1,b}$  to the ground-state energy,
and the uncertainty $\delta\veps_{1,b}$ are tabulated
in Table~\ref{Table-energy-constant},
and the FNS correction $\veps_{2,b}$  to the excited-state energy,
and the uncertainty $\delta\veps_{2,b}$ are tabulated
in Table~\ref{Table-transitions}.
Note that for light isotopes, i.e., for H, He and K,
$\veps_{n,a}$ and $\veps_{n,b}$ [$n = 1,2$] and their uncertainties
$\delta\veps_{n, a}$ and $\delta\veps_{n,b}$ satisfy the relations
$$
  \delta\veps_{n,a} \approx \delta\veps_{n,b} > \veps_{n, a} - \veps_{n, b},
$$
whereas for the heavy isotopes,
$$
  \delta \veps_{n,a} \approx \delta\veps_{n,b} < \veps_{n, a} - \veps_{n, b}.
$$
Therefore, the comparison of the calculated and measured ground-state binding energies
of heavy HLI (such as Rb, Cs, Pb, Bi and U) can be used to determine the 
nuclear charge distribution.  The ground-state energy of the hydrogen atom, and of light 
HLI (such as He and K) are almost the same for models (a) and (b), hence,
measuring the ground-state binding energies would not be an accurate
method for determining the nuclear charge distributions.

The numerical solution technique developed in this subsection can be
generalized to the case of other charge distributions inside
the nucleus, e.g., for the TPFM charge distribution 
\cite{Shabaev-JPhysB-1993, Shabaev-PRA-1997} introduced in Eq.~(\ref{eq:Fermi-distribution}) below.

We calculated the FNS corrections analytically for model (a) and numerically for model (b).
The results were tabulated in Table~\ref{Table-energy-constant} for the ground $1s$ state of H\&HLI,
and in Table~\ref{Table-transitions} for the first excited $2s$ state.
For comparison, we will calculate in Sec.~\ref{sec:corrections} below
the FNS correction using perturbation theory.
We shall show that the perturbation theory is a good approximation for
the FNS corrections for H and light HLI, 
compare Tables~\ref{Table-energy-corrections} and \ref{Table-energy-constant}.
For heavy HLI, the perturbation theory results in Table~\ref{Table-energy-corrections}
are dissimilar to the results in Table~\ref{Table-energy-constant}, i.e., the perturbation
theory is poor.  References \cite{proton_radius} and \cite{CODATA-2002} (2018 CODATA and
2002 CODATA) and many other papers use perturbation theory to calculate FNS corrections for 
hydrogen and light HLIs.  These references expand the FNS corrections in a power series
in $r_N$.  In contrast, we numerically calculate the perturbation theory.

\subsection{Finite-nuclear size corrections in 2018 CODATA}
  \label{subsec:FNS-CODATA-2018}

Finite-nuclear-size and nuclear-polarizability corrections are ordered
by powers in $\alpha$, following Ref.~\cite{proton_radius}:
\begin{equation}   \label{eq:FNS-series}
  \veps_{n}^{(2018)} = \sum_{i = 4}^{\infty} \veps_{n}^{(i)} ,
\end{equation}
where $n$ is the principal quantum number, and index $i$ indicates
the power of $Z \alpha$ in the expansion.
The first and lowest-order contributions to $s$-state energies are
\begin{equation}   \label{eq:FNS-alpha-4}
  \veps_{n}^{(4)} =
  \frac{2}{3} \, m_e c^2 \, \frac{(Z \alpha)^{4}}{n^3} \,
  \bigg( \frac{m}{m_e} \bigg)^{3} \,
  \bigg( \frac{r_N}{\lambdabar_C} \bigg)^{2} ,
\end{equation}
where $\lambdabar_C = \hbar/(m_e c)$ is the reduced Compton wavelength.
The numerical value of $\lambdabar_C$ is given in Ref.~\cite{proton_radius} as,
$\lambdabar_C = 3.8615926796 (12) \times 10^{-12}$ m.
Equation~(\ref{eq:FNS-alpha-4}) is a non-relativistic approximation to
the FNS correction to the $s$-state energy of H\&HLI.
Reference~\cite{proton_radius} gives relativistic corrections $\veps_{n}^{(i)}$
with $i \geq 5$ just for hydrogen and deuterium atoms, but not for tritium
atoms or for hydrogen-like ions.

\begin{table}
\caption{\footnotesize 
  FNS corrections $\veps_{1}^{(4)}$ (in Hartree) to the energy of the ground $1s$ state
  of H\&HLI.  The numbers in the brackets indicate the uncertainties $\delta \veps_{1}^{(4)}$
  in the calculated ground-state energy due to the uncertainty in the nuclear radius $r_N$
  in Table~\ref{Table-atomic-nuclear-radii}.
  $\Delta \veps_{1, \nu}$ with $\nu = a, b$ is given in Eq.~(\ref{eq:Delta_n_ab}). }
\begin{center}
  \begin{tabular}{|c|c|c|c|}
    \hline
    Isotope ($Z, A$) &
    $\veps_{1}^{(4)} ( \delta \veps_{1}^{(4)} )$ &
    $\Delta \veps_{1, a}^{(4)}$ &
    $\Delta \veps_{1, b}^{(4)}$
    \\
    \hline
    H(1, 1) &
    $1.6827 (76) E^{-10}$ &
    $10 E^{-13}$ &
    $10 E^{-13}$
    \\
    \hline
    H(1, 2) &
    $1.07719 (75) E^{-9}$ &
    $5.4 E^{-13}$ &
    $5.4 E^{-13}$
    \\
    \hline
    H(1, 3) &
    $7.36 (30) E^{-10}$ &
    $4 E^{-13}$ &
    $4 E^{-13}$
    \\
    \hline
    He(2, 3) &
    $1.4716 (45) E^{-8}$ &
    $3.0 E^{-11}$ &
    $3.0 E^{-11}$
    \\
    \hline
    He(2, 4) &
    $1.0689 (36) E^{-8}$ &
    $2.2 E^{-11}$ &
    $2.2 E^{-11}$
    \\
    \hline
    K(19, 39) &
    $3.6604 (40) E^{-4}$ &
    $5.383 E^{-5}$ &
    $5.370 E^{-5}$
    \\
    \hline
    K(19, 40) &
    $36673 (60) E^{-4}$ &
    $5.392 E^{-5}$ &
    $5.379 E^{-5}$
    \\
    \hline
    K(19, 41) &
    $3.697 (12) E^{-4}$ &
    $5.43 E^{-5}$ &
    $5.42 E^{-5}$
    \\
    \hline
    Rb(37, 85) &
    $7.884 (9) E^{-3}$ &
    $4.663 E^{-3}$ &
    $4.650 E^{-3}$
    \\
    \hline
    Rb(37, 87) &
    $7.866 (8) E^{-3}$ &
    $4.653 E^{-3}$ &
    $4.640 E^{-3}$
    \\
    \hline
    Cs(55, 133) &
    $5.028 (10) E^{-2}$ &
    $8.108 E^{-2}$ &
    $8.079 E^{-2}$
    \\
    \hline
    Cs(55, 135) &
    $5.033 (10) E^{-2}$ &
    $8.115 E^{-2}$ &
    $8.087 E^{-2}$
    \\
    \hline
    Pb(82, 204) &
    $0.32327 (17)$ &
    $2.1455$ &
    $2.1354$
    \\
    \hline
    Pb(82, 206) &
    $0.32444 (17)$ &
    $2.1514$ &
    $2.1412$
    \\
    \hline
    Pb(82, 207) &
    $0.32493 (17)$ &
    $2.1538$ &
    $2.1437$
    \\
    \hline
    Pb(82, 208) &
    $0.32574 (16)$ &
    $2.15793$ &
    $2.14777$
    \\
    \hline
    Bi(83, 209) &
    $0.34440 (32)$ &
    $2.4115$ &
    $2.4000$
    \\
    \hline
    U(92, 235) &
    $0.58042 (82)$ &
    $6.7258$ &
    $6.6913$
    \\
    \hline
    U(92, 236) &
    $0.58229 (76)$ &
    $6.7411$ &
    $6.7066$
    \\
    \hline
    U(92, 238) &
    $0.58509 (66)$ &
    $6.7639$ &
    $6.7293$
    \\
    \hline
\end{tabular}
\end{center}
\label{Table-1s-FNS-CPDATA-2018}
\end{table}

For H\&HLIs the FNS correction $\veps_{n}^{(4)}$ to the $n s$-state energy and
its uncertainty $\delta \veps_{n}^{(4)}$ due to the uncertainty of the nuclear
RMS charge radius are tabulated in Table~\ref{Table-1s-FNS-CPDATA-2018}
for the ground $n = 1$ state, and in Table~\ref{Table-2s-FNS-CPDATA-2018}
for the first excited $n = 2$ state.  These tables also
show $\Delta \varepsilon^{(4)}_{n,a}$ and $\Delta \varepsilon^{(4)}_{n,b}$ where
\begin{equation}   \label{eq:Delta_n_ab}
\Delta \varepsilon_{n, \nu}^{(4)} = \veps_{n}^{(4)} - \veps_{n, \nu},
\end{equation}
and $\nu = a, b$ for the models (a) and (b) of the nuclear charge distribution.
We will now compare $\veps_{1}^{(4)} \approx \veps_{1}^{(2018)}$ with
$\veps_{1,a}$ and $\veps_{1,b}$ in Table~\ref{Table-energy-constant},
and $\veps_{2}^{(4)} \approx \veps_{2}^{(2018)}$ with
$\veps_{2,a}$ and $\veps_{2,b}$ in  Table~\ref{Table-transitions}
for H\&HLIs.
Moreover, for hydrogen and light HLIs, we compare $\Delta \veps_{n, \nu}^{(4)}$
with the uncertainty $\delta \veps_{n, \nu}$ of the FNS correction in Tablea~\ref{Table-energy-constant}
and \ref{Table-transitions}.
For ${}^{A}$H atoms and He ions, $\Delta \veps_{n, \nu}^{(4)} < \delta \veps_{n, \nu}$,
and therefore the non-relativistic approximation in Eq.~(\ref{eq:FNS-alpha-4})
is a good approximation to the FNS correction to the $1s$ state energy ($n = 1$)
and to the $2s$ state energy ($n = 2$).
For heavier HLI, $\Delta \veps_{n, \nu}^{(4)} > \delta \veps_{n, \nu}$, and
the non-relativistic approximation fails.

\begin{table}
\caption{\footnotesize 
  FNS corrections $\veps_{2}^{(4)}$ (in Hartree) to the energy of the first excited $2s$ state
  of H\&HLI.  The numbers in the brackets indicate the uncertainties $\delta \veps_{2}^{(4)}$
  in the calculated ground-state energy due to the uncertainty in the nuclear radius $r_N$
  in Table~\ref{Table-atomic-nuclear-radii}.
  $\Delta \veps_{2, \nu}$ with $\nu = a, b$ is given in Eq.~(\ref{eq:Delta_n_ab}). }
\begin{center}
  \begin{tabular}{|c|c|c|c|}
    \hline
    Isotope ($Z, A$) &
    $\veps_{2}^{(4)} (\delta \veps_{2}^{(4)} )$ &
    $\Delta \veps_{2, a}^{(4)}$ &
    $\Delta \veps_{2, b}^{(4)}$
    \\
    \hline
    H(1, 1) &
    $2.1034 (75) E^{-11}$ &
    $1.3 E^{-14}$ &
    $1.3 E^{-14}$
    \\
    \hline
    H(1, 2) &
    $1.34648 (94) E^{-10}$ &
    $7.4 E^{-14}$ &
    $7.4 E^{-14}$
    \\
    \hline
    H(1, 3) &
    $9.20 (38) E^{-11}$ &
    $5 E^{-14}$ &
    $5E^{-14}$
    \\
    \hline
    He(2, 3) &
    $1.8396 (56) E^{-9}$ &
    $4.1 E^{-12}$ &
    $4.1 E^{-12}$
    \\
    \hline
    He(2, 4) &
    $1.3361 (45) E^{-9}$ &
    $3.0 E^{-12}$ &
    $3.0 E^{-12}$
    \\
    \hline
    K(19, 39) &
    $4.5755 (51) E^{-5}$ &
    $7.626 E^{-6}$ &
    $7.610 E^{-6}$
    \\
    \hline
    K(19, 40) &
    $4.5841 (75) E^{-5}$ &
    $7.639 E^{-6}$ &
    $7.623 E^{-6}$
    \\
    \hline
    K(19, 41) &
    $4.621 (15) E^{-5}$ &
    $7.70 E^{-6}$ &
    $7.68 E^{-6}$
    \\
    \hline
    Rb(37, 85) &
    $9.855 (11) E^{-4}$ &
    $6.877 E^{-4}$ &
    $6.859 E^{-4}$ 
    \\
    \hline
    Rb(37, 87) &
    $9.833 (10) E^{-4}$ &
    $6.863 E^{-4}$ &
    $6.846 E^{-4}$
    \\
    \hline
    Cs(55, 133) &
    $6.285 (12) E^{-3}$ &
    $1.2684 E^{-2}$ &
    $1.2643 E^{-2}$
    \\
    \hline
    Cs(55, 135) &
    $6292 (12) E^{-3}$ &
    $1.2696 E^{-2}$ &
    $1.2655 E^{-2}$
    \\
    \hline
    Pb(82, 204) &
    $4.0409 (21) E^{-2}$ &
    $0.38827$ &
    $0.38651$
    \\
    \hline
    Pb(82, 206) &
    $4.0555 (21) E^{-2}$ &
    $0.38935$ &
    $0.38758$
    \\
    \hline
    Pb(82, 207) &
    $4.0616 (21) E^{-2}$ &
    $0.38980$ &
    $0.38803$
    \\
    \hline
    Pb(82, 208) &
    $4.0718  (19) E^{-2}$ &
    $0.39055$ &
    $0.38878$
    \\
    \hline
    Bi(83, 209) &
    $4.3050 (41) E^{-2}$ &
    $0.43954$ &
    $0.43752$
    \\
    \hline
    U(92, 235) &
    $7.255 (10) E^{-2}$ &
    $1.3154$ &
    $1.3088$
    \\
    \hline
    U(92, 236) &
    $7.2787 (95) E^{-2}$ &
    $1.3184$ &
    $1.3118$
    \\
    \hline
    U(92, 238) &
    $7.3136 (82) E^{-2}$ &
    $1.3229$ &
    $1.3163$
    \\
    \hline
\end{tabular}
\end{center}
\label{Table-2s-FNS-CPDATA-2018}
\end{table}

\section{Perturbation Theory Corrections due to FNS}
  \label{sec:corrections}

In this section we apply perturbation theory for models (a) and (b)  
and find the corrections to the ground-state energies of H\&HLI (assuming the 
potentials $W_\nu(r)$ and $W_{ba}(r)$ are  small).

To be specific, we will concentrate on the ground electronic state, $n = 1$, and therefore 
the subscript $n$ on energies and wave functions will often be omitted below. 
Since an analytic solution to the Dirac equation for a harmonic potential is not known,
we use perturbation theory in order to obtain FNS corrections for model (b),
but we also numerically solve the Dirac equation (see previous section).

Perturbation theory for models (a) and (b) with weak $W_\nu(r)$ is detailed in Sec.~\ref{subsec:perturbation_nu}. 
It starts from the Hamiltonian,
$H_{\nu} = -i \hbar c \boldsymbol\alpha \cdot \nabla + V_{\nu}(r) + \beta m c^2$,
rewriten as,
\begin{equation}    \label{eq:H=H_0+W_nu}
  H_{\nu} = H_0 + V_\nu(r) - V_C(r) \equiv H_0 + W_{\nu}(r),
\end{equation}
where $H_0 = -i \hbar c \boldsymbol\alpha \cdot \nabla + V_C(r) + \beta m c^2$
is the Dirac Hamiltonian for a point-like nucleus with $V_C(r) = -Z e^2/r$.

Perturbation theory for model (b) alone with weak $W_{ba}(r)$ is detailed in Sec.~\ref{subsec:perturbation_ba}. 
It starts from the Hamiltonoian $H_b$ in the form,
\begin{equation}   \label{eq:H_b=H_a+W_ba}
  H_b = H_a + W_{ba}(r),
\end{equation}
where $H_a$ is given in Eq.~(\ref{eq:H=H_0+W_nu}), and $W_{ba}(r) = V_b(r) - V_a(r)$
is given in Eq.~(\ref{Eq:W_ba}).

\subsection{Perturbation theory with $W_{\nu}(r)$}
  \label{subsec:perturbation_nu}

The first-order correction to the energy is,
\begin{equation}
  \veps^{(1)}_{n, \nu} =
  \mathcal{N}^{2}
  \int\limits_{0}^{r_\nu}
  W_\nu(r) \,
  \big[ g^2(r) + f^2(r) \big]
  r^2 dr,
  \label{eq:energy-1st-order-def}
\end{equation}
where $f(r)$ and $g(r)$ are defined in Eq.~(\ref{subeqs:f-g-ground-point}).  Note that
the singularities at $r = 0$ of $f(r)$ and $g(r)$ are integrable \cite{Landau-Lifshitz-4-hydrogen} 
and $W_{\nu}(r)$ is given by Eq.~(\ref{eq:W}). Explicitly,
\begin{eqnarray}    \label{eq:energy-1st-order-Q1-Q2}
  \veps^{(1)}_{n, \nu} &=&
  \frac{\hbar^3 c^3 \mathcal{N}^{2}}{8 \lambda^3}
  \int\limits_{0}^{\rho_\nu}
  W_\nu \bigg( \frac{\hbar c \rho}{2 \lambda} \bigg)
  \Big\{
      m_e c^2 \,
      \big[ Q_{1}^{2} (\rho) + Q_{2}^{2}(\rho) \big]
  \nonumber \\ && +
      2 \veps \,
      Q_1(\rho) Q_2(\rho)
  \Big\}
  \rho^{2 \gamma} e^{-\rho} d \rho,
\end{eqnarray}
where
\begin{eqnarray}  \label{eq:normalization-Q1-Q2}
  \mathcal{N}^{-2} &=&
  \frac{\hbar^3 c^3}{8 \lambda^3}
  \int\limits_{0}^{\infty}
  \Big\{
      m_e c^2 \,
      \big[ Q_{1}^{2} (\rho) + Q_{2}^{2}(\rho) \big]
  \nonumber \\ && +
      2 \veps \,
      Q_1(\rho) Q_2(\rho)
  \Big\}
  \rho^{2 \gamma} e^{-\rho} d \rho,
\end{eqnarray}
and the dimensionless nuclear radius is 
\begin{equation}    \label{eq:rho-n-def}
  \rho_\nu = \frac{2 \lambda r_\nu}{\hbar c} = \frac{2 Z r_\nu}{a_B}.
\end{equation}
Equation~(\ref{eq:rho-lambda-gamma-def}) was used to obtain the second equality in Eq.~(\ref{eq:rho-n-def}).
The nuclear radii $r_N$ are tabulated in Table~\ref{Table-atomic-nuclear-radii}
for a number of elements and isotopes.  Let us restore the $n$ dependence in 
$\veps^{(1)}_{n,\nu}$.  The FNS corrections $\veps^{(1)}_{1,\nu}$ to the ground-state
energy are given in Table~\ref{Table-energy-corrections}.  The dependence of 
$\veps^{(1)}_{1,\nu}$ on $A$ and $Z$ follows the behavior of the nuclear radius $r_N(A,Z)$;
because $r_N$ increases with $Z$, so does $\veps^{(1)}_{1,\nu}$.  
Dependence of $\veps^{(1)}_{1,\nu}$ on $A$ is less simple.  For most of the elements in 
Table~\ref{Table-atomic-nuclear-radii}, $r_N$ increases with $A$, and so does $\veps^{(1)}_{1,\nu}$, 
[see, e.g., $\veps^{(1)}_{1,a}$ and $\veps^{(1)}_{1,b}$ for isotopes of K, Cs, Pb and U].
However, for some elements, $r_N$ decreases with $A$, and so does $\veps^{(1)}_{1,\nu}$,
[see, e.g., $\veps^{(1)}_{1,a}$ and $\veps^{(1)}_{1,b}$ for isotopes of He and Rb].
Moreover, for some elements, $r_N$ is not a monotonic function of $A$,
hence $\veps^{(1)}_{1,\nu}$ is also not a monotonic function of $A$.
For example, the nuclear radii $r_N({}^{A}\text{H})$ for the isotopes
${}^{A}\text{H}$ satisfy the inequalities,
$$
  r_N \big( {}^{1}\text{H} \big) <
  r_N \big( {}^{3}\text{H} \big) <
  r_N \big( {}^{2}\text{H} \big).
$$
Thus $\veps^{(1)}_{1,\nu}({}^{A}\text{H})$ satisfy
$$
  \veps^{(1)}_{1,\nu} \big( {}^{1}\text{H} \big) <
  \veps^{(1)}_{1,\nu} \big( {}^{3}\text{H} \big) <
  \veps^{(1)}_{1,\nu} \big( {}^{2}\text{H} \big).
$$
Comparing $\veps^{(1)}_{1,a}$ in Table~\ref{Table-energy-corrections}
with $\veps_a$ in Table~\ref{Table-energy-constant} shows that
for light isotopes, perturbation theory predictions are
very close to the non-perturbative numerical calculations,
but for the heavy isotopes, the differences between $\veps_{1,a}$
and $\varepsilon^{(1)}_{1,a}$ are large (see the difference between 
the values of $\veps_{1,a}$ in Table \ref{Table-energy-constant} 
and $\veps^{(1)}_{1,a}$ in Table \ref{Table-energy-corrections}).

\begin{table}[htp]
\caption{\footnotesize 
  First-order perturbation theory correction due to FNS in Eq.~(\ref{eq:energy-1st-order-Q1-Q2}), 
  for $\veps^{(1)}_{1,\nu}$ [with $\nu = a, b$].
  The nuclear radii $r_a$ and $r_b$ are given in Eq.~(\ref{eq:r_a-r_b-vs-r_N}),
  and the RMS nuclear charge radii $r_N$ are listed in
  Table~\ref{Table-atomic-nuclear-radii}.}
\begin{center}
\begin{tabular}{|c|c|c|}  \hline
  Isotope ($Z, A$) &
  $\veps^{(1)}_{1,a}$ (Hartree) &
  $\veps^{(1)}_{1,b}$ (Hartree)
  \\
  \hline
  H (1, 1) &
  $1.683743 E^{-10}$ &
  $1.683736 E^{-10}$
  \\
  \hline
  H (1, 2) &
  $1.077785 E^{-9}$ &
  $1.077778 E^{-9}$
  \\
  \hline
  H (1, 3) &
  $7.36714 E^{-10}$ &
  $7.36710 E^{-10}$
  \\
  \hline
  He (2, 3) &
  $1.47484 E^{-8}$ &
  $1.47482 E^{-8}$
  \\
  \hline
  He (2, 4) &
  $1.07128 E^{-8}$ &
  $1.07126 E^{-8}$
  \\
  \hline
  K (19, 39) &
  $4.24722 E^{-4}$ &
  $4.24232 E^{-4}$
  \\
  \hline
  K (19, 40) &
  $4.25506 E^{-4}$ &
  $4.25015 E^{-4}$
  \\
  \hline
  K (19, 41) &
  $4.28869 E^{-4}$ &
  $4.28374 E^{-4}$
  \\
  \hline
  Rb (37, 85) &
  $1.31164 E^{-2}$ &
  $1.30612 E^{-2}$
  \\
  \hline
  Rb (37, 87) &
  $1.30882 E^{-2}$ &
  $1.30331 E^{-2}$
  \\
  \hline
  Cs (55, 133) &
  $0.145819$ &
  $0.144503$
  \\
  \hline
  Cs (55, 135) &
  $0.145964$ &
  $0.144646$
  \\
  \hline
  Pb (82, 204) &
  $3.25838$ &
  $3.19659$
  \\
  \hline
  Pb (82, 206) &
  $3.26777$ &
  $3.20580$
  \\
  \hline
  Pb (82, 207) &
  $3.27167$ &
  $3.20962$
  \\
  \hline
  Pb (82, 208) &
  $3.27822$ &
  $3.21605$
  \\
  \hline
  Bi(83, 209) &
  $3.67177$ &
  $3.60062$
  \\
  \hline
  U (92, 235) &
  $10.7543$ &
  $10.5056$
  \\
  \hline
  U (92, 236) &
  $10.7799$ &
  $10.5305$
  \\
  \hline
  U (92, 238) &
  $10.8180$ &
  $10.5677$
  \\
  \hline
\end{tabular}
\end{center}
\label{Table-energy-corrections}
\end{table}

In the non-relativistic approximation, the perturbation theory
of the FNS correction, $\veps^{(1)}_{n, \nu}$,
takes the form of the simple equation
\cite{CODATA-2002,proton_radius}
\begin{equation}  \label{eq:EFNS}
  \veps^{(1)}_{n, \nu} \approx
  \frac{2 m_e c^2}{3} \,
  \frac{(Z \alpha)^{4}}{n^3} \,
  \frac{m^3}{m_e^3} \,
  \frac{r_N^{2}}{\lambdabar_C^{2}}.
\end{equation}
We verified that this is an excellent approximation for hydrogen.
The perturbation theory result in
Eq.~(\ref{eq:EFNS}) is obtained from Eq.~(\ref{eq:energy-1st-order-Q1-Q2})
by using the approximations: $Q_1(\rho) \approx Q_1(0)$,
$Q_2(\rho) \approx Q_2(0)$ and $\gamma \approx 1$.
The first and second approximations are based on the inequalities
$0 < \rho < \rho_{\nu} \ll 1$ [see Eq.~(\ref{eq:energy-1st-order-Q1-Q2})]
which are valid for H and all HLIs.
The third approximation is based on the inequality $Z \alpha \ll 1$ which is satisfied
for H and light HLIs.
For the heavy HLIs, $Z \alpha$ is not a small parameter, and Eq.~(\ref{eq:EFNS})
should be modified. The FNS correction to the ground state energy of all HLIs is
\begin{subequations}
  \label{subeqs:EFNS-heavy-ab}
\begin{eqnarray}
  \veps^{(1)}_{1, a} &\approx&
  \frac{2 m_e c^2}{\gamma (2 \gamma + 1)} \,
  \frac{(Z \alpha)^{4}}{n^3} \,
  \frac{m^3}{m_e^3} \,
  \frac{r_N^{2 \gamma}}{\lambdabar_C^{2 \gamma}},
  \label{eq:EFNS-rheavy-a}
  \\
  \veps^{(1)}_{1, b} &\approx&
  \frac{5^{\gamma} \, 3^{1 - \gamma} \, 2 \, m_e c^2}{\gamma (2 \gamma + 1) (2 \gamma + 3)} \,
  \frac{(Z \alpha)^{4}}{n^3} \,
  \frac{m^3}{m_e^3} \,
  \frac{r_N^{2 \gamma}}{\lambdabar_C^{2 \gamma}}.
  \label{eq:EFNS-rheavy-b}
\end{eqnarray}
\end{subequations}

Note that the FNS correction to the $1s$ state energy of H and light HLIs in Eq.~(\ref{eq:EFNS})
depends just on the nuclear RMS charge radius $r_N$, and not on the details of
the nuclear charge distribution.
However, for heavy HLIs with high $Z$, the FNS correction to the $s$ state energy
is sensitive to the details of the nuclear charge distribution, and Eqs.~(\ref{eq:EFNS-rheavy-a})
and (\ref{eq:EFNS-rheavy-b}) give different FNS corrections for the models (a) and (b).
Therefore, it is important to develop and use more accurate models
for the nuclear charge distribution in order to obtain better comparisons
with the experimental energies of the $s$ state and the frequencies
of the $1s$-$2s$ and $1s$-$3s$ transitions.

\subsection{Perturbation theory with $W_{ba}(r)$}
  \label{subsec:perturbation_ba}

The first order correction $\varepsilon_{ba}^{(1)}$ is given by,
\begin{eqnarray}
  \varepsilon_{ba}^{(1)} &=&
  \frac{4 \pi A_{-}^{2}}{\mathcal{N}_a}
  \int\limits_{0}^{r_a}
  W_{ba}(r)
  \Big[ g_{-}^{2}(r) + f_{-}^{2}(r) \Big] r^2 dr +
  \nonumber \\ &&
  \frac{4 \pi A_{+}^{2}}{\mathcal{N}_a}
  \int\limits_{r_a}^{r_b}
  W_{ba}(r)
  \Big[ g_{+}^{2}(r) + f_{+}^{2}(r) \Big] r^2 dr,
  \label{Eq:epsilon_ba^1st}
\end{eqnarray}
where $g_{\pm}(r)$ and $f_{\pm}(r)$ are the radial wave functions
for the model (a).  The functions
$g_{+}(r)$ and $f_{+}(r)$ are given in Eq.~(\ref{subeqs:fg-res-plus}), and
$g_{-}(r)$ and $f_{-}(r)$ are given in Eq.~(\ref{subeqs:fg-res-minus}).
The constants $A_{\pm}$ are found from Eq.~(\ref{eq:boundary-conditions}),
and the normalization constant $\mathcal{N}_a$ is,
\begin{eqnarray}
  \mathcal{N}_a &=&
  4 \pi A_{-}^{2}
  \int\limits_{0}^{r_a}
  \Big[ g_{-}^{2}(r) + f_{-}^{2}(r) \Big] r^2 dr +
  \nonumber \\ &&
  4 \pi A_{+}^{2}
  \int\limits_{r_a}^{\infty}
  \Big[ g_{+}^{2}(r) + f_{+}^{2}(r) \Big] r^2 dr.
  \label{Eq:normalization}
\end{eqnarray}
In order to calculate $\veps_{ba}^{(1)}$,
it is useful to rewrite Eq.~(\ref{Eq:epsilon_ba^1st})
employing the electron density $\varrho_a(r)$ given by,
\begin{equation}
  \varrho_a(r) =
  \left\{
    \begin{array}{cc}
      \displaystyle
      \frac{4 \pi A_{-}^{2}}{\mathcal{N}} \,
      \Big[ g_{-}^{2}(r) + f_{-}^{2}(r) \Big],
      &
      r < r_a,
      \\
      \displaystyle
      \frac{4 \pi A_{+}^{2}}{\mathcal{N}} \,
      \Big[ g_{+}^{2}(r) + f_{+}^{2}(r) \Big],
      &
      r > r_a,
    \end{array}
  \right.
  \label{Eq:electron-density}
\end{equation}
hence Eq.~(\ref{Eq:epsilon_ba^1st}) can be written as
\begin{equation}   \label{Eq:epsilon_ba^1st-density}
  \varepsilon_{ba}^{(1)} =
  \int\limits_{0}^{r_b} W_{ba}(r) \varrho_a(r) r^2 dr.
\end{equation}
The two-component wave function in
Eq.~(\ref{eq:spinor-vs-spherical-harmonics}) is continuous
at $r = r_a$, i.e.,
$$
  A_{-} \, \big( g_{-}(r_a), \, f_{-}(r_a) \big)
  =
  A_{+} \, \big( g_{+}(r_a), \, f_{+}(r_a) \big),
$$
and thus the electron density $\varrho_a(r)$ in Eq.~(\ref{Eq:electron-density})
is also continuous.
Moreover, $\varrho_a(r)$ is smooth within the interval $0 < r < r_b$,
i.e., for any $r \in (0, r_b)$, the following inequality is satisfied:
$\big| \varrho(r) - \varrho(r_a) \big| \ll \varrho(r_a)$.
Note that, although one may expect that Eq.~(\ref{Eq:epsilon_ba^1st-density}) can be
approximated as
\begin{equation}   \label{Eq:epsilon_ba^1st-approx-wrong}
  \varepsilon_{ba}^{(1)} \approx
  \varrho(r_a)
  \int\limits_{0}^{r_b}
  W_{ba}(r)
  r^2 dr,
\end{equation}
it can be easily verified that
$\int\limits_{0}^{r_b} W_{ba}(r) r^2 dr = 0$,
hence the approximation in Eq.~(\ref{Eq:epsilon_ba^1st-approx-wrong})
is not valid.

\begin{table}
\caption{\footnotesize 
  Perturbation theory corrections $\veps_{ba}^{(1)}$ [in Hartree] to the ground
  state of H\&HLI systems [see Eq.~(\ref{Eq:epsilon_ba^1st-density-fluctuation})].
  The dimensionless parameter $\delta_{ba}$ is defined in Eq.~(\ref{eq:delta_ba}).
  The energy correction $\veps_{1,ba}$ and its uncertainty $\delta\veps_{1,ba}$
  due to the uncertainty in the nuclear RMS nuclear charge radius 
  [in units of Hartree] are calculated non-perturbatively from Eqs.~(\ref{eq:veps_ba}) and
  (\ref{eq:delta-veps_ba}).}
\begin{center}
  \begin{tabular}{|c|c|c|c|}
    \hline
    Isotope ($Z,A$) &
    $\veps_{ba}^{(1)}$ &
    $\delta_{ba}$ &
    $\veps_{1,ba} ( \delta\veps_{1,ba})$
    \\
    \hline
    H (1, 1) &
    $-2.39058 E^{-16}$ &
    $-4.78371 E^{-16}$ &
    $-3.26 (15) E^{-16}$
    \\
    \hline
    H (1, 2) &
    $-2.37631 E^{-15}$ &
    $-4.75385 E^{-15}$ &
    $-2.48(3) E^{-15}$
    \\
    \hline
    H (1, 3) &
    $-1.91704 E^{-15}$ &
    $-3.83472 E^{-15}$ &
    $-1.57(11) E^{-15}$
    \\
    \hline
    He (2, 3) &
    $-8.14903 E^{-14}$ &
    $-4.07504 E^{-14}$ &
    $-8.59 (4) E^{-14}$
    \\
    \hline
    He (2, 4) &
    $-5.4718 E^{-14}$ &
    $-2.73613 E^{-14}$ &
    $-5.80(3) E^{-14}$
    \\
    \hline
    K (19, 39) &
    $-1.20571 E^{-7}$ &
    $-6.64768 E^{-10}$ &
    $-1.2837 (15) E^{-7}$
    \\
    \hline
    K (19, 40) &
    $-1.20812 E^{-7}$ &
    $-6.66093 E^{-10}$ &
    $-1.286 (2) E^{-7}$
    \\
    \hline
    K (19, 41) &
    $-1.21844 E^{-7}$ &
    $-6.71783 E^{-10}$ &
    $-1.297 (4) E^{-7}$
    \\
    \hline
    Rb (37, 85) &
    $-1.24451 E^{-5}$ &
    $-1.78437 E^{-8}$ &
    $-1.3283 (15) E^{-5}$ 
    \\
    \hline
    Rb (37, 87) &
    $-1.24169 E^{-5}$ &
    $-1.78034 E^{-8}$ &
    $-1.3254 (13) E^{-5}$
    \\
    \hline
    Cs (55, 133) &
    $-2.65761 E^{-4}$ &
    $-1.68324 E^{-7}$ &
    $-2.842 (5) E^{-4}$
    \\
    \hline
    Cs (55, 135) &
    $-2.66035 E^{-4}$ &
    $-1.68497 E^{-7}$ &
    $-2.845 (5) E^{-4}$
    \\
    \hline
    Pb (82, 204) &
    $-9.41203 E^{-3}$ &
    $-2.52128 E^{-6}$ &
    $-1.0096 (4) E^{-2}$
    \\
    \hline
    Pb (82, 206) &
    $-9.43977 E^{-3}$ &
    $-2.52871 E^{-6}$ &
    $-1.0125 (4) E^{-2}$
    \\
    \hline
    Pb (82, 207) &
    $-9.45126 E^{-3}$ &
    $-2.53179 E^{-6}$ &
    $-1.0138 (4) E^{-2}$
    \\
    \hline
    Pb (82, 208) &
    $-947062 E^{-3}$ &
    $-2.53698 E^{-6}$ &
    $-1.0159 (4) E^{-2}$
    \\
    \hline
    Bi (83, 209) &
    $-1.06803 E^{-2}$ &
    $-2.78397 E^{-6}$ &
    $-1.1458 (9) E^{-2}$
    \\
    \hline
    U (92, 235) &
    $-3.20879 E^{-2}$ &
    $-6.60084 E^{-6}$ &
    $-3.447 (4) E^{-2}$
    \\
    \hline
    U (92, 236) &
    $-3.21652 E^{-2}$ &
    $-6.61674 E^{-6}$ &
    $-3.455 (3) E^{-2}$
    \\
    \hline
    U (92, 238) &
    $-3.22805 E^{-2}$ &
    $-6.64045 E^{-6}$ &
    $-3.467 (3) E^{-2}$
    \\
    \hline
  \end{tabular}
\end{center}
\label{Table-corrections-ba}
\end{table}%

The correction $\veps_{ba}^{(1)}$ in Eq.~(\ref{Eq:epsilon_ba^1st-density})
can be written as
\begin{equation}
  \varepsilon_{ba}^{(1)} =
  \int\limits_{0}^{r_b}
  W_{ba}(r) \,
  \big[ \varrho(r) - \varrho(0) \big] \,
  r^2 dr.
  \label{Eq:epsilon_ba^1st-density-fluctuation}
\end{equation}
The correction to the ground state energy of model (b) is given in
Eq.~(\ref{Eq:epsilon_ba^1st-density-fluctuation}) 
and tabulated in Table~\ref{Table-corrections-ba}, which also tabulates the dimensionless 
corrections to the ground-state binding energy,
\begin{equation}
  \delta_{ba} = \frac{\varepsilon_{ba}^{(1)}}{m c^2 - \veps_{1}^{(0)}},
  \label{eq:delta_ba}
\end{equation}
where $\veps_{1}^{(0)}$ is given in Eq.~(\ref{eq:energy-hydrgen-ground}).
For comparison, Table~\ref{Table-corrections-ba} shows
$\veps_{1,ba}$ and its uncertainty $\delta \veps_{1,ba}$ calculated
non-perturbatively from the equations,
\begin{eqnarray}
  \veps_{1,ba} &=&
  \veps_{1,b} -
  \veps_{1,a},
  \label{eq:veps_ba}
  \\
  \delta\veps_{1,ba} &=&
  \bigg| \frac{\partial \veps_{1,ba}}{\partial r_N} \bigg| \,
  \delta r_N
  \nonumber \\ &=&
  \Big| \delta\veps_{1,b} - \delta\veps_{1,a} \Big|,
  \label{eq:delta-veps_ba}
\end{eqnarray}
where $\delta \veps_{1,a}$ and $\delta \veps_{1,b}$ are given by
\begin{equation*}
  \delta\veps_{1,a} =
  \bigg| \frac{\partial \veps_{1,a}}{\partial r_N} \bigg| \,
  \delta r_N,
  \quad
  \delta\veps_{1,b} =
  \bigg| \frac{\partial \veps_{1,b}}{\partial r_N} \bigg| \,
  \delta r_N.
\end{equation*}
$\veps_{1,a}$, $\delta\veps_{1,a}$ and
$\veps_{1,b}$, $\delta\veps_{1,b}$ are tabulated in
Table~\ref{Table-energy-constant}.
Note that the first-order perturbation correction $\veps_{ba}^{(1)}$
is close to the non-perturbative correction $\veps_{1,ba}$ for
all the isotopes.  In other words, we conclude that the first-order perturbation 
correction $\veps_{ba}^{(1)}$ gives accurate results for both
light and heavy isotopes.

In summary, for light nuclei, perturbation theory gives results very 
close to the non-perturbative numerical calculations for 
the difference between $\veps_{1,a}$ and $\varepsilon^{(1)}_{1,a}$, 
but for heavy isotopes the difference is large (see the large differences  for the heavy isotopes
between the values of $\veps_{1,a}$ in Table \ref{Table-energy-constant} analytically
calculated using model (a) and $\veps^{(1)}_{1,a}$ calculated by perturbation theory using
model (a) in Table \ref{Table-energy-corrections}). 

\section{Two-parameter Fermi model for the nuclear charge distribution} \label{Fermi_model}

In model (c) the distribution of charge within the nucleus is taken to be the 
TPFM \cite{deVries_87}. 
References~\cite{Shabaev-JPhysB-1993, Shabaev-PRA-1997, Sunnergren_98}
studied the FNS corrections to the hyperfine splitting
of the ground state of H\&HLI using a TPFM to parameterize 
the distribution of the nuclear charge.  The TPFM is more realistic than either
model (a) or (b), at least for heavier nuclei; but it is unclear how realistic
it is for light nuclei. The TPFM charge distribution is,
\begin{equation}
  \label{eq:Fermi-distribution}
  \varrho_F(r) =
  \frac{Z e}{8 \pi a_F^3 \big| \mathrm{Li}_{3}(-e^{c_F/a_F}) \big|} \,
  \bigg[ \exp \bigg( \frac{r - c_F}{a_F} \bigg) + 1 \Bigg]^{-1},
\end{equation}
where $\mathrm{Li}_{n}(\bullet)$ is the polylogarithm function 
\cite{Abramowitz_Stegun} of order $n$.  Here $c_F$ is the half-density 
nuclear radius, and $a_F$ is the distribution thickness parameter.
The charge density is normalized as follows:
\begin{equation*}
  4 \pi \int\limits_{0}^{\infty} \varrho_F(r) r^2 dr = Z e.
\end{equation*}
The nuclear charge radius is $r_N = \sqrt{ \langle r^2 \rangle_N}$ is
obtained by the expression,
\begin{eqnarray} \label{rN_Fermi}
  r_{N}^2 \equiv \langle r^2 \rangle_{N} &=&
  \frac{4 \pi}{Z e} \int\limits_{0}^{\infty} \varrho_F(r) r^4 dr
  \nonumber \\ &=&
  12 a_{F}^{2} \,
  \frac{\mathrm{Li}_{5}(-e^{c_F/a_F})}{\mathrm{Li}_{3}(-e^{c_F/a_F})}.
\end{eqnarray}
Note that the arguments of the polylogarithm functions contain an exponential function
with argument $c_F/a_F$. We find $c_F$ as a function of $r_N$ and $a_F$ using Eq.~(\ref{rN_Fermi}).
The uncertainty $\delta c_F$ in $c_F$ due to the uncertainties $\delta r_N$ and $\delta a_F$
in $r_N$ and $a_F$ is,
\begin{equation}
  \delta c_F = \sqrt{ \bigg( \frac{\partial c_F}{\partial r_N} \, \delta r_N \bigg)^{2} +
  \bigg( \frac{\partial c_F}{\partial a_F} \, \delta a_F \bigg)^{2} }.
  \label{eq:delta-c_F}
\end{equation}
The nuclear charge radii $r_N$ and their uncertainties $\delta r_N$
for many nuclei are published in Ref.~\cite{deVries_87},
and the parameters $a_F$ and the uncertainties $\delta a_F$ in $a_F$
are published in Ref.~\cite{nuclear-radii-ADNDT-2013}.
As $a_F \to 0$, the charge distribution inside the nucleus becomes,
\begin{equation*}
  \lim_{a_F \to 0}
  \varrho_F(r)
  =
  \frac{3Ze}{4 \pi r_{b}^{3}} \, \Theta(r_b - r),
  \quad  \mathrm{and \, \, \,}
  r_b = \lim_{a_F \to 0} c_F .
\end{equation*}
Figure~\ref{Fig_Fermi_2_parameter_model} shows the charge density distribution versus $r/r_N$ 
for three values of the width parameter, $a_F = 0.02 \, r_N, 0.1 \, r_N, 0.15 \, r_N$ (red, green
and blue, respectively).  Clearly, as $a_F \to 0$, the charge distribution approaches the limit given 
above.  Moreover, in the limit as $a_F \to 0$, $c_F \to \sqrt{5/3} \, r_N$.  The dashed black curve in
Fig.~\ref{Fig_Fermi_2_parameter_model} shows the charge density in this limit. 

The TPFM charge distribution yields the following potential (for simplicity 
we have used $c$ and $a$ for $a_F$ and $c_F$ in this equation):
\begin{eqnarray}
  V_F(r) =
  -\frac{Ze^2}{r} +
  \frac{4 \pi}{r} \,
  \int\limits_r^{\infty }
   \varrho _F (r') \,
    r' ( r' - r ) \, dr'  = \nonumber \\ 
    Z e^2 \left[
  \frac{\text{Li}_3\left(-e^{\frac{c}{a}-\frac{r}{a}}\right)
   }{r \, \text{Li}_3\left(-e^{\frac{c}{a}}\right)}+\frac{
   \text{Li}_2\left(-e^{\frac{c}{a}-\frac{r}{a}}\right)
   }{2 a \, \text{Li}_3\left(-e^{\frac{c}{a}}\right)}-\frac{1}{r} \right].
  \label{V_F}
\end{eqnarray}

\begin{table}
\caption{\footnotesize 
  The half-density nuclear radii $c_F$ (in Bohr), the
  distribution thickness parameter $a_F$ (in Bohr), and
  the uncertainties $\delta c_F$ and $\delta a_F$ in $c_F$ and $a_F$
  for a number of heavy isotopes.
  $a_F$ and $\delta a_F$ are taken from Ref.~\cite{deVries_87},
  $c_F$ is calculated from Eq.~(\ref{rN_Fermi}),
  $\delta c_F$ is calculated from Eq.~(\ref{eq:delta-c_F}),
  and $r_N$ and $\delta r_N$ are tabulated in
  Table~\ref{Table-atomic-nuclear-radii}.}
\begin{center}
  \begin{tabular}{|c|c|c|}
    \hline
    Isotope ($Z,A$) &
    $c_F$ ($\delta c_F$) &
    $a_F$ ($\delta a_F$)
    \\
    \hline
    Pb (82, 206) &
    $1.2449 (29) E^{-4}$ &
    $1.030 (15) E^{-5}$
    \\
    \hline
    Pb (82, 207) &
    $1.2456 (36) E^{-4}$ &
    $1.032 (19) E^{-5}$
    \\
    \hline
    Bi (83, 209) &
    $1.278 (12) E^{-4}$ &
    $8.84 (74) E^{-6}$
    \\
    \hline
    U (92, 238) &
    $1.3194 (61) E^{-4}$ &
    $1.143 (30) E^{-5}$
    \\
    \hline
  \end{tabular}
\end{center}
\label{Table-Fermi-distribution-radii}
\end{table}%

We numerically solve Dirac equation~(\ref{subeqs:radial})  for the heavy nuclei 
${}^{206}$Pb, ${}^{207}$Pb, ${}^{209}$Bi, and ${}^{238}$U to obtain radial wave functions
$g_n(r)$ and $f_n(r)$ and eigen-energies $\veps_n$ with potential (\ref{V_F}) with
$c_F$ and $a_F$ as tabulated in Table~\ref{Table-Fermi-distribution-radii}.
The uncertainty $\delta \veps_n$ in the energy $\veps_n$ of the $n s$ state
can be estimated as follows,
\begin{equation}    \label{eq:uncertainty-energy-Fermi}
  \delta \veps_n =
  \sqrt{ \Big( \delta \veps_{n}^{(r)}  \Big)^{2} + \Big( \delta \veps_{n}^{(a)}  \Big)^{2} },
\end{equation}
where $\delta \veps_{n}^{(r)}$ is the uncertainty in $\veps_n$ due to the uncertainty 
in $r_N$, and $\delta \veps_{n}^{(a)}$ is the uncertainty in $\veps_n$ due to
the uncertainty in $a_F$.  $\delta \veps_{n}^{(r)}$ and $\delta \veps_{n}^{(a)}$ can be approximated as,
\begin{eqnarray*}
  \delta \veps_{n}^{(r)} &\approx&
  \frac{1}{2}
  \bigg|
       \veps_n \Big( c_F(r_N + \delta r_N, a_F), a_F \Big)
  \nonumber \\ && -
       \veps_n \Big( c_F(r_N - \delta r_N, a_F), a_F \Big)
  \bigg|,
  \\
  \delta \veps_{n}^{(a)} &\approx&
  \frac{1}{2}
  \bigg|
       \veps_n \Big( c_F(r_N, a_F + \delta a_F), a_F + \delta a_F \Big)
  \nonumber \\ && -
       \veps_n \Big( c_F(r_N, a_F - \delta a_F), a_F - \delta a_F \Big)
  \bigg|,
\end{eqnarray*}
where $\veps_n ( c_F(r_N \pm \delta r_N, a_F), a_F )$ is the $n s$ state energy 
for the nuclear RMS charge radius $r_N \pm \delta r_N$, and
$\veps_n ( c_F(r_N, a_F \pm \delta a_F), a_F \pm \delta a_F )$ is the $n s$ state energy 
for the distribution thickness parameter $a_F \pm \delta a_F$.
The FNS energy correction $\veps_{n,c}$ is
\begin{equation}
  \veps_{n,c} = \veps_n - \veps_{n}^{(0)},
  \label{eq:FNC-energy-Fermi}
\end{equation}
where $\veps_{n}^{(0)}$ is the energy of the $n s$ state for a point-like nucleus.
$\veps_{n,c}$ and $\delta \veps_n$ are tabulated in
Table~\ref{Table-energy-correction-Fermi} for heavy isotopes.
Comparing $\veps_{n, c}$ with $\veps_{n, a}$ and $\veps_{n, b}$ in
Tables~\ref{Table-energy-constant} and \ref{Table-transitions},
one sees that $\veps_{n, a} > \veps_{n, b} > \veps_{n, c} > 0$
and $\veps_{n, a} - \veps_{n, b} > \veps_{n, b} - \veps_{n, c}$.
For ${}^{206}$Pb, ${}^{207}$Pb and ${}^{238}$U,
the uncertainty $\delta \veps_n$ in $\veps_n$ is the same
for the models (a), (b) and (c).  This is because
$\delta \veps_{n}^{(a)} \ll \veps_{n}^{(r)}$,
see Eq.~(\ref{eq:uncertainty-energy-Fermi}).
For ${}^{209}$Bi, $\delta \veps_{n}^{(a)}$ and $\veps_{n}^{(r)}$ 
are of the same order of magnitude, and the uncertainty 
$\delta \veps_n$ for the model (c) is larger than the uncertainty for models (a) and (b).  

\begin{table}
\caption{\footnotesize 
  The FNS corrections $\veps_{n,c}$ to the energies
  $\veps_n$ of the $n s$ states in Eq.~(\ref{eq:FNC-energy-Fermi}),
  and the uncertainties $\delta \veps_n$ in $\veps_n$ in
  Eq.~(\ref{eq:uncertainty-energy-Fermi}).}
\begin{center}
  \begin{tabular}{|c|c|c|}
    \hline
    Isotope ($Z,A$) &
    $\veps_{1, c}$ ($\delta \veps_1$) &
    $\veps_{2, c}$ ($\delta \veps_2$)
    \\
    \hline
    Pb (82, 206) &
    $2.4605 (10)$ &
    $0.42724 (17)$
    \\
    \hline
    Pb (82, 207) &
    $2.4634 (10)$ &
    $0.42774 (17)$
    \\
    \hline
    Bi (83, 209) &
    $2.740 (29)$ &
    $0.4798 (50)$
    \\
    \hline
    U (92, 238) &
    $7.2952 (60)$ &
    $1.3858 (11)$
    \\
    \hline
  \end{tabular}
\end{center}
\label{Table-energy-correction-Fermi}
\end{table}%

\section{Ground state hyperfine splitting}
  \label{sec:hyperfine}

Within the Dirac equation formalism, the hyperfine interaction between the magnetic moments 
of the electron (located at space point ${\bf r}$) and the nucleus (located at the origin) can be 
expressed  in terms of the vector potential $\mathbf{A}(\mbfr)$.  The interaction Hamiltonian is
\begin{equation}   \label{eq:H-mag-def}
  H_{\mathrm{mag}} =
  [c] \, e \, \boldsymbol\alpha \cdot \mathbf{A}(\mbfr),
\end{equation}
where the factor $[c]$ is present in SI units but is unity in Gaussian units,
and $\mathbf{A}(\mbfr)$ is the vector potential generated by the magnetic moment 
of the nucleus with magnetic moment density $\boldsymbol\mu(\mbfr')$,
\begin{equation} \label{Eq:vector_potential}
  \mathbf{A}_{\nu}(\mbfr) =
  \left[{\frac{{\mu_0 }} {{4\pi }}} \right]
  \int
  \frac{{\boldsymbol \mu_{\nu}(\mbfr')}\times (\mbfr - \mbfr')}{|\mbfr - \mbfr'|^3} \,
  d^3 \mbfr' ,
\end{equation}
where integration is over all space.  The subscript $\nu$ specifies the models for the 
magnetic moment distribution: $\nu=0,a, b, c$ are the point-like, uniform surface, 
uniform volume, and TPFM.
Here the factor $\left[ {\frac{{\mu_0 }} {{4\pi }}} \right]$ is present in SI units 
but is unity in Gaussian units.
We restrict ourselves by considering isotropic models where the magnetic moment
density $\boldsymbol\mu(r')$ depends only on the distance $r' = |\mbfr'|$ from
the center of mass of the nucleus.  
For the point-like distribution, $\nu=0$, 
\begin{equation}   \label{eq:magnetic-moment-point}
  \boldsymbol\mu_0(r) =
  g_I \mu_N \mathbf{I} \,
  \delta^3(\mbfr),
\end{equation}
where $\mu_N$ is the nuclear magneton, $\mu_N = e \hbar/(2 m_p c)$ in Gaussian units,
$m_p$ is the proton mass,  $g_I$ is the nuclear $g$-factor, $\mathbf{I}$ is the 
nuclear spin operator, and $\delta^3(\mbfr)$ is the three dimensional Dirac delta function.  
For $\nu = a$,
\begin{equation}   \label{eq:magnetic-moment-a}
  \boldsymbol\mu_a(r) =
  \frac{g_I \mu_N}{4 \pi r_{a}^{2}} \,
  \mathbf{I} \,
  \delta (r - r_a ),
\end{equation}
where $\delta(\bullet)$ is the Dirac $\delta$ function.  For $\nu = b$,
\begin{equation}   \label{eq:magnetic-moment-b}
  \boldsymbol\mu_b(r) =
  \frac{3 g_I \mu_N}{4 \pi r_{b}^{3}} \,
  \mathbf{I} \,
  \Theta (r_b - r).
\end{equation}
For the Fermi model, $\nu = c$,
\begin{equation}   \label{eq:magnetic-moment-Fermi}
  \boldsymbol\mu_c(r) =
  \frac{\vartheta_0}{e^{(r-c_F)/a_F} + 1} \,
  g_I \mu_N \mathbf{I}.
\end{equation}
The factor $\vartheta_0$ is determined from the normalization
condition,
\begin{equation*}
  4 \pi \vartheta_0
  \int\limits_{0}^{\infty}
  \frac{r^2 dr}{e^{(r-c_F)/a_F} + 1}
  = 1.
\end{equation*}
This yields,
\begin{equation*}
  \vartheta_0 =
  -\frac{1}{8 \pi a_{F}^{3} \mathrm{Li}_{3}(-e^{c_F / a_F})}.
\end{equation*}

\begin{figure}
\centering 
 \includegraphics[width=0.8 \linewidth,angle=0]
   {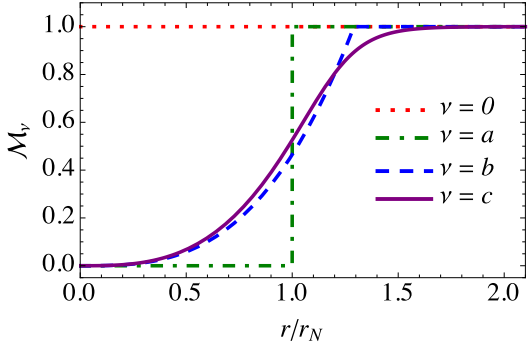}
\caption{\footnotesize 
  $\mathcal{M}_{\nu}(r)$ versus $r$ for ${}^{207}$Pb nucleus
  for different magnetic moment distribution models:
  The point-like nucleus, $\nu = 0$ (red dotted curve),
  distributed homogeneously on a sphere of radius $r_a$, $\nu = a$
  (green dot-dashed curve), distributed homogeneously
  inside a sphere of radius $r_b$, $\nu = b$ (dashed blue curve),
  and the TPFM in Eq.~(\ref{eq:M_nu-def}),
  $\nu = c$ (solid purple curve).}
\label{Fig-M_nu-vs-r}
\end{figure}

Substituting $\boldsymbol\mu_\nu(r')$ into Eq.~(\ref{Eq:vector_potential}), we get,
\begin{equation} \label{Eq:vector_potential-isotropic-1}
  \mathbf{A}_{\nu}(\mbfr) =
  \left[{\frac{{\mu_0 }} {{4\pi }}} \right]
  \frac{g_I \mu_N}{r^2} \,
  \big[ \mathbf{I} \times \mbfe_r \big] \,
  \mathcal{M}_{\nu}(r),
\end{equation}
where
\begin{eqnarray}
  \mathcal{M}_{0}(r) &=& 1,
  \label{eq:M_point}
  \\
  \mathcal{M}_{a}(r) &=&
  \Theta ( r - r_a),
  \label{eq:M_a}
  \\
  \mathcal{M}_{b}(r) &=&
  \frac{r^3}{r_{b}^{3}} \,
  \Theta (r_b - r) +
  \Theta (r - r_b).
  \label{eq:M_b} \\
    \mathcal{M}_{c}(r) &=&
  4 \pi \vartheta_0
  \int\limits_{0}^{r}
  \frac{{r'}^2 dr'}{e^{(r' - c_F)/a_F} + 1}
  \nonumber \\ &=&
  \frac{1}{ \mathrm{Li}_{3} \big( -e^{c_F/a_F} \big) } \,
  \Bigg[\frac{r^2}{2a^2} \ln \big(1+e^{(c_F - r)/a_F} \big) 
  \nonumber \\ && +
       \mathrm{Li}_{3} \big( -e^{c_F/a_F} \big) -
       \mathrm{Li}_{3} \big( -e^{(c_F - r)/a_F} \big)
  \nonumber \\ && -
       \frac{r}{a_F} \mathrm{Li}_{2} \big( -e^{(c_F-r)/a_F} \big) \Bigg].
  \label{eq:M_nu-def}
\end{eqnarray}
The function $\mathcal{M}_{\nu}(r)$ is plotted versus $r$ in Fig.~\ref{Fig-M_nu-vs-r}
for the four magnetic moment distribution models.
Substituting Eq.~(\ref{Eq:vector_potential-isotropic-1}) into
Eq.~(\ref{eq:H-mag-def}), we obtain
\begin{equation}
  H_{\mathrm{hf}}^{(\nu)} =
  \left[ \frac{\mu_0 c}{4 \pi} \right]
  \frac{e g_I \mu_N}{r^2} \,
  \mathcal{M}_{\nu}(r) \,
  \Big(
      \mathbf{I}
      \cdot
      \big[ \mbfe_r \times \boldsymbol\alpha \big]
  \Big).
  \label{eq:Hmag-1}
\end{equation}

Assuming that the hyperfine splitting of the ground state is much
smaller than the excitation energy of the $2p$ state, we can restrict
ourselves to first-order perturbation theory,
\begin{equation}
  \epsilon_{\mathrm{hf}}^{(\nu)}(F) =
  \big\langle
      \Psi_{F, m_F}^{(\nu)}
  \big|
      H_{\mathrm{hf}}^{(\nu)}
  \big|
      \Psi_{F, m_F}^{(\nu)}
  \big\rangle ,
  \label{eq:energy-hyperfine-1st}
\end{equation}
where the ground-state wave function $| \Psi_{F, m_F} \rangle$ is
\begin{eqnarray}   \label{eq:WF-ground-F-m_F}
\langle {\bf r}  \big|
      \Psi_{F, m_F}^{(\nu)}
  \big\rangle
  &=&
  \sum_{m_J, m_F}
  \big\langle
       \frac{1}{2}, m_J; \,
       I, m_I
  \big|
       F, m_F
  \big\rangle
  \nonumber \\ && \times
  \big\langle {\bf r} \big|
       \nu; \,
       \frac{1}{2}, m_J; \,
       I, m_I
  \big\rangle,
\end{eqnarray}
$| \nu; \frac{1}{2}, m_J; I, m_I \rangle$ is the atomic angular ket
with total electronic angular momentum vector ${\bf J}$ ($J = 1/2$) and 
projection $m_J=\pm \tfrac{1}{2}$, nuclear spin ${\bf I}$, 
and projection $m_I$, and
\begin{equation*}
  \big\langle
       \mbfr
  \big|
       \nu; \,
       \frac{1}{2}, m_J; \,
       I, m_I
  \big\rangle
  =
  \chi_{I, m_I} \,
  \psi_{m_J}^{(\nu)}(\mbfr),
\end{equation*}
$\chi_{I, m_I}$ is a nuclear spin wave function, and
$\psi_{m_J}^{(\nu)}(\mbfr)$ is a spatial wave function,
\begin{equation}   \label{eq:WF-ground-J-m_J}
  \psi_{m_J}^{(\nu)}(\mbfr) =
  \left(
    \begin{array}{c}
      g_{\nu}(r) \Omega_{\frac{1}{2} 0 m_J}
      \\
      i f_{\nu}(r) \Omega_{\frac{1}{2} 1 m_J}
    \end{array}
  \right).
\end{equation}

For $\nu = 0$, the radial wave functions $g_0(r)$ and $f_0(r)$
are given in Eq.~(\ref{subeqs:f-g-ground-point}).
For $\nu = a$, the radial wave functions $g_a(r)$ and $f_a(r)$
are
\begin{subequations}    \label{subeqs:g_a-f_a}
\begin{eqnarray}
  g_a(r) &=&
  g_{+}(r) \Theta(r - r_a) +
  g_{-}(r) \Theta(r_a - r),
  \label{eq:g_a}
  \\
  f_a(r) &=&
  f_{+}(r) \Theta(r - r_a) +
  f_{-}(r) \Theta(r_a - r),
  \label{eq:f_a}
\end{eqnarray}
\end{subequations}
where $g_{+}(r)$ and $f_{+}(r)$ are given in Eq.~(\ref{subeqs:f-g-vs-Q1-Q2}),
and $g_{-}(r)$ and $f_{-}(r)$ are given in Eq.~(\ref{subeqs:fg-res-minus}).
For $\nu = b$, the wave functions $( g_b(r), f_b(r) )$ are numerically calculated 
as described in Sec.~\ref{subsec:Dirac-finite-radius-model-b}, and for  $\nu = c$
the wave functions must be numerically calculated using the Dirac equation for
a potential which includes a two-parameter Fermi nuclear model distribution.

Substituting Eq.~(\ref{eq:WF-ground-F-m_F}) into
Eq.~(\ref{eq:energy-hyperfine-1st}), we find
$\epsilon_{\mathrm{hf}}^{(\nu)}(F)$,
\begin{equation}
  \epsilon_{\mathrm{hf}}^{(\nu)}(F) =
  \frac{\Delta_{\nu}}{2} \,
  \big[ F (F+1) - I (I+1) - S(S+1) \big],
  \label{eq:energy-hyperfine-1st-res}
\end{equation}
where $S = 1/2$.
The hyperfine splitting $\Delta_{\nu} = %
\epsilon_{\mathrm{hf}}^{(\nu)}(I + 1/2) - %
\epsilon_{\mathrm{hf}}^{(\nu)}(I - 1/2)$ is
\begin{eqnarray}
  \Delta_{\nu} &=&
  -\left[ \frac{\mu_0 c}{4 \pi} \right] 
   \Big( I + \frac{1}{2} \Big)\,
  \frac{8 e g_I \mu_N}{3}
  \nonumber \\ && \times
  \int\limits_{0}^{\infty}
  \mathcal{M}_{\nu}(r)
  g_{\nu}(r) f_{\nu}(r) dr.
  \label{eq:A-hyperfine-def}
\end{eqnarray}

The expression for the hyperfine splitting of the ground state of H\&HLI given in 
Ref.~\cite{Shabaev-JPhysB-1993} is equivalent to
\begin{eqnarray}
  \Delta &=&
  \frac{4 m c^2}{3} \,
  \frac{\alpha ( Z \alpha)^{3}}{\gamma(2 \gamma - 1)} \,
  g_I \, \frac{m_e}{m_p} \,
  \Big( I + \frac{1}{2} \Big)
  \nonumber \\ && \times
  \bigg[ (1 - \xi) (1 - \eta) + \zeta_{\mathrm{QED}} \bigg].
  \label{eq:hyperfine-Shabaev}
\end{eqnarray}
Note the use of the electron reduced mass $m$ in Eq.~(\ref{eq:hyperfine-Shabaev}),
as discussed in Refs.~\cite{reduced-mass1, reduced-mass2, reduced-mass3, Eides-hydrogen-atom}.
In the following we shall write Eq.~(\ref{eq:hyperfine-Shabaev}) in the form
\begin{equation}    \label{eq:hyperfine}
  \Delta = \Delta_0 \bigg[ (1 - \xi) (1 - \eta) + \zeta_{\mathrm{QED}} \bigg],
\end{equation}
where
\begin{equation}    \label{eq:A-hyperfine-00}
  \Delta_0 =
  \frac{4 m c^2}{3} \,
  \alpha \big( Z \alpha \big)^{3} \,
  g_I \, \frac{m_e}{m_p} \,
   \Big( I + \frac{1}{2} \Big) \,
  \frac{1}{\gamma(2 \gamma - 1)}.
\end{equation}
$\Delta_0$ is the well-known uncorrected relativistic hyperfine splitting 
\cite{Shabaev-JPhysB-1993}.  Here $\xi$ and $\eta$ are the FNS corrections to 
the hyperfine splitting due to the charge distribution and the magnetic moment 
distribution (Bohr-Weisskopf correction) respectively [see 
Eqs.~(\ref{eq:delta-epsilon_a-vs-G_0-G_plus-G_minus}), (\ref{eq:delta-epsilon_b-vs-G_0-G_plus-G_minus}) and (\ref{eq:xi_c-eta_c}), and the text below for models (a), (b) and (c)],
and the quantity $\zeta_{\mathrm{QED}}$ is the QED radiative correction to the hyperfine splitting.  
Relativistic QED radiative corrections are addressed in Ref.~\cite{Shabaev-JPhysB-1993}, where a 
quantity $\chi_{\mathrm{QED}}$ is defined.  We prefer to present the QED radiative correction in 
terms of the quantity $\zeta_{\mathrm{QED}} \equiv \gamma (2 \gamma - 1) \chi_{\mathrm{QED}}$ 
because $\zeta_{\mathrm{QED}}$ can be directly compared with $\xi$ and $\eta$, as is 
clear from Eq.~(\ref{eq:hyperfine-Shabaev}).  Note that we do not calculate the QED radiative 
corrections here but rather use values reported in the literature.
\ \\
\ \\
Below, we calculate hyperfine splitting for the four models $\nu = 0, a, b, c$ for the
{\em magnetic moment distribution}. We should point out that the nuclear magnetic moment distribution
can be very different from the nuclear charge distribution.  This
is particularly true for nuclei with one nucleon outside a closed
nuclear shell.  Appendix \ref{append:magnetic-moment-distribution}
discusses this case.  In our calculations we used the same distribution 
for both nuclear charge and nuclear magnetic moment.

\subsection{Point-like charge and magnetic moment}
  \label{subsec:model_00}

\begin{figure}
\vspace{0.5in}
\centering 
 \includegraphics[width = 0.9 \linewidth,angle=0]
   {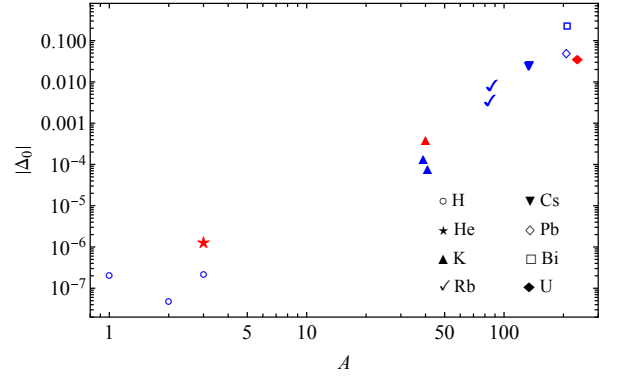}
\caption{\footnotesize 
  Absolute value  of the hyperfine splitting $|\Delta_0|$ 
  (in Hartree) in Eq.~(\ref{eq:A-hyperfine-00})
  for a point-like nucleus versus $A$.
  Positive values of the hyperfine splitting $\Delta_0$
  (positive $g_I$) are marked by blue symbols, and
  negative values of $\Delta_0$ (negative $g_I$)
  are shown in red.}
\label{Delta_0_vs_A}
\end{figure}

Numerical values of $\Delta_0$ are tabulated in Table~\ref{Table-hyperfine-point}
for a number of isotopes.  For some ions, $g_I$ is negative, and the hyperfine
interaction is ferromagnetic, hence $\Delta_0 < 0$.
Figure~\ref{Delta_0_vs_A} plots the magnitude of the uncorrected
hyperfine splitting versus $A$.

\begin{table}
\caption{\footnotesize 
  Nuclear $g$-factors $g_I$, nuclear spin $I$, and ground-state hyperfine splitting 
  $\Delta_0$ calculated using Eq.~(\ref{eq:A-hyperfine-00}) [in Hartree], for point-like nuclei.
  The nuclear $g$ factors for ${}^{1}$H, ${}^{2}$H, ${}^{3}$H and ${}^{3}$He nuclei
  are taken from 2018 CODATA, i.e., Ref.~\cite{proton_radius}, and those for 
  ${}^{39}$K, ${}^{40}$K, ${}^{41}$K, ${}^{85}$Rb, ${}^{87}$Rb,
  ${}^{133}$Cs are taken from Ref.~\cite{alkali-hyperfine-1977},
  those for ${}^{135}$Cs and ${}^{235}$U are taken from Ref.~\cite{ADNDT-1989},
  those for ${}^{207}$Pb and ${}^{209}$Bi are taken from Ref.~\cite{hyperfine-Pb-Bi-PRL-1998}.}
\begin{center}
\begin{tabular}{|c|c|c|c|}
  \hline
  Isotope ($Z, A$) &
  $g_I$ &
  $I$ &
  $\Delta_0$
  \\
  \hline
  H (1, 1) &
  $5.5856946893(16)$ &
  $1/2$ &
  $2.1589180648 (6) E^{-7}$
  \\
  \hline
  H (1, 2) &
  $0.8574382338(22)$ &
  $1$ &
  $4.972458067 (13) E^{-8}$
  \\
  \hline
  H (1, 3) &
  $5.957924931(12)$ &
  $1/2$ &
  $2.3036226460 (46) E^{-7}$
  \\
  \hline
  He (2, 3) &
  $-4.255250615 (50)$ &
  $1/2$ &
  $1.316544463 (15) E^{-8}$
  \\
  \hline
  K (19, 39) &
  $0.26061413 (22)$ &
  $3/2$ &
  $1.4234088 (12) E^{-4}$
  \\
  \hline
  K (19, 40) &
  $-0.324063 (62)$ &
  $4$ &
  $-3.98239 (77) E^{-4}$
  \\
  \hline
  K (19, 41) &
  $0.14304731 (15)$ &
  $3/2$ &
  $7.812889 (8) E^{-5}$
  \\
  \hline
  Rb (37, 85) &
  $0.5391679 (11)$ &
  $5/2$ &
  $3.554396 (7) E^{-3}$
  \\
  \hline
  Rb (37, 87) &
  $1.8272315 (18)$ &
  $3/2$ &
  $8.030529 (8) E^{-3}$
  \\
  \hline
  Cs (55, 133) &
  $0.732356746 (95)$ &
  $7/2$ &
  $2.4735911 (32) E^{-2}$
  \\
  \hline
  Cs (55, 135) &
  $0.78069 (6)$ &
  $7/2$ &
  $2.63683 (19) E^{-2}$
  \\
  \hline
  Pb (82, 207) &
  $1.16438 (4)$ &
  $1/2$ &
  $5.1434 (2) E^{-2}$
  \\
  \hline
  Bi (83, 209) &
  $0.91347 (4)$ &
  $9/2$ &
  $2.14590 (10) E^{-1}$
  \\
  \hline
  U (92, 235) &
  $-0.109 (9)$ &
  $7/2$ &
  $-3.66 (29) E^{-2}$
  \\
  \hline
\end{tabular}
\end{center}
\label{Table-hyperfine-point}
\end{table}%

\subsection{Charge and magnetic moment distributions for model (a)}
  \label{subsec:model_a}

Substituting Eqs.~(\ref{eq:M_a}) and (\ref{subeqs:g_a-f_a})
into Eq.~(\ref{eq:A-hyperfine-def}) with $\nu = a$, we find
\begin{eqnarray}
  \Delta_a &=&
  -\frac{4 m c^2}{3} \,
  \alpha \big( Z \alpha \big)^{3} \,
  g_I \, \frac{m_e}{m_p}
  \Big( I + \frac{1}{2} \Big)
  \nonumber \\ && \times
  a_{B}^{2}
  \int\limits_{r_a}^{\infty}
  g_{+}(r) f_{+}(r) dr.
  \label{eq:A-hyperfine-a}
\end{eqnarray}
As in Eq.~(\ref{eq:hyperfine}), it is convenient to write $\Delta_a$ in the form
\begin{equation}
  \Delta_a = \Delta_0 \big( 1 - \xi_a \big) \big( 1 - \eta_a \big),
  \label{eq:Delta_a0-vs-Delta_00}
\end{equation}
where $\Delta_0$ is given in Eq.~(\ref{eq:A-hyperfine-00}).
The corrections $\xi_a$ and $\eta_a$ are calculate within model (a).
Explicitly,
\begin{equation}   \label{eq:delta-epsilon_a-vs-G_0-G_plus-G_minus}
  \xi_a =
  \frac{\mathcal{G}_{0} - \mathcal{G}_{a}}{\mathcal{G}_{0}},
  \quad
  \eta_a =
  \frac{\mathcal{G}_{-}^{(a)}}{\mathcal{G}_{a}},
\end{equation}
where
\begin{subequations}  \label{subeqs:G_0_minus_plus}
\begin{eqnarray}
  \mathcal{G}_{0} &=&
  \frac{a_{B}^{2}}{Z^3 \alpha}
  \int\limits_{0}^{\infty}
  g_0(r) f_0(r) dr
  \nonumber \\ &=&
  -\frac{1}{\gamma (2 \gamma - 1)},
  \label{eq:G_0}
  \\
  \mathcal{G}_{-}^{(a)} &=&
  \frac{a_{B}^{2}}{Z^3 \alpha}
  \int\limits_{0}^{r_a}
  g_{a,-}(r) f_{a,-}(r) dr,
  \label{eq:G_minus}
  \\
  \mathcal{G}_{a} &=&
  \mathcal{G}_{-}^{(a)} +
  \frac{a_{B}^{2}}{Z^3 \alpha}
  \int\limits_{r_a}^{\infty}
  g_{+}(r) f_{+}(r) dr.
  \label{eq:G_a}
\end{eqnarray}
\end{subequations}

\bigskip

\begin{table}[ht]
\caption{\footnotesize
  The corrections $\xi_a$ and $\xi_b$ to the hyperfine splitting for models
  (a) [Eq.~(\ref{eq:delta-epsilon_a-vs-G_0-G_plus-G_minus})],
  and (b) [Eq.~(\ref{eq:delta-epsilon_b-vs-G_0-G_plus-G_minus})].}
\begin{center}
\begin{tabular}{|c|c|c|}
  \hline
  Isotope ($Z, A$) &
  $\xi_a$ &
  $\xi_b$
  \\
  \hline
  H (1, 1) &
  $1.1092 (2) E^{-3}$ &
  $1.1106 (2) E^{-3}$
  \\
  \hline
  H (1, 2) &
  $5.9912 (2) E^{-4}$ &
  $6.0261 (2) E^{-4}$
  \\
  \hline
  H (1, 3) &
  $4.087 (9) E^{-4}$ &
  $4.116 (10) E^{-4}$
  \\
  \hline
  He (2, 3) &
  $5.126 (2) E^{-4}$ &
  $5.0786 (10) E^{-4}$
  \\
  \hline
  K (19, 39) &
  $2.8513 (15) E^{-3}$ &
  $2.7643 (7) E^{-3}$
  \\
  \hline
  K (19, 40) &
  $2.853 (2) E^{-3}$ &
  $2.7661 (10) E^{-3}$
  \\
  \hline
  K (19, 41) &
  $2.864 (4) E^{-3}$ &
  $2.776 (2) E^{-3}$
  \\
  \hline
  Rb (37, 85) &
  $9.173 (5) E^{-3}$ &
  $8.911 (2) E^{-3}$
  \\
  \hline
  Rb (37, 87) &
  $9.163 (4) E^{-3}$ &
  $8.902 (2) E^{-3}$
  \\
  \hline
  Cs (55, 133) &
  $2.4526 (19) E^{-2}$ &
  $2.3913 (9) E^{-2}$
  \\
  \hline
  Cs (55, 135) &
  $2.454 (2) E^{-2}$ &
  $2.3924 (9) E^{-2}$
  \\
  \hline
  Pb (82, 207) &
  $0.107053 (17)$ &
  $0.105226 (7)$
  \\
  \hline
  Bi (83, 209) &
  $0.11338 (3)$ &
  $0.111483 (13)$
  \\
  \hline
  U (92, 235) &
  $0.19317 (6)$ &
  $0.19062 (3)$
  \\
  \hline
\end{tabular}
\end{center}
\label{Table-hyperfine-xi}
\end{table}%

\begin{table}[ht]
\caption{\footnotesize 
 The corrections $\eta_a$ and $\eta_b$ to the hyperfine splitting for models
  (a) [Eq.~(\ref{eq:delta-epsilon_a-vs-G_0-G_plus-G_minus})],
  and (b) [Eq.~(\ref{eq:delta-epsilon_b-vs-G_0-G_plus-G_minus})].}
\begin{center}
\begin{tabular}{|c|c|c|}
  \hline
  Isotope ($Z, A$) &
  $\eta_a$ &
  $\eta_b$
  \\
  \hline
  H (1, 1) &
  $1.060 (11) E^{-5}$ &
  $1.847 (19) E^{-5}$
  \\
  \hline
  H (1, 2) &
  $2.6812 (9) E^{-5}$ &
  $4.6729 (16) E^{-5}$
  \\
  \hline
  H (1, 3) &
  $2.22 (5) E^{-5}$ &
  $3.86 (8) E^{-5}$
  \\
  \hline
  He (2, 3) &
  $4.961 (8) E^{-5}$ &
  $8.646 (13) E^{-5}$
  \\
  \hline
  K (19, 39) &
  $9.083 (5) E^{-4}$ &
  $1.5807 (9) E^{-3}$
  \\
  \hline
  K (19, 40) &
  $9.092 (7) E^{-4}$ &
  $1.5822 (13) E^{-3}$
  \\
  \hline
  K (19, 41) &
  $9.127 (14) E^{-4}$ &
  $1.588 (2) E^{-3}$
  \\
  \hline
  Rb (37, 85) &
  $2.6911 (14) E^{-3}$ &
  $4.665 (3) E^{-3}$
  \\
  \hline
  Rb (37, 87) &
  $2.6883 (12) E^{-3}$ &
  $4.660 (2) E^{-3}$
  \\
  \hline
  Cs (55, 133) &
  $6.179 (5) E^{-3}$ &
  $1.0645 (9) E^{-2}$
  \\
  \hline
  Cs (55, 135) &
  $6.182 (5) E^{-3}$ &
  $1.0650 (9) E^{-2}$
  \\
  \hline
  Pb (82, 207) &
  $1.8471 (3) E^{-2}$ &
  $3.1386 (6) E^{-2}$
  \\
  \hline
  Bi (83, 209) &
  $1.9209 (6) E^{-2}$ &
  $3.2619 (11) E^{-2}$
  \\
  \hline
  U (92, 235) &
  $2.7368 (11) E^{-2}$ &
  $4.620 (2) E^{-2}$
  \\
  \hline
\end{tabular}
\end{center}
\label{Table-hyperfine-eta}
\end{table}%

The corrections $\xi_a$ and $\xi_b$ are tabulated in Table~\ref{Table-hyperfine-xi},
and $\eta_a$ and $\eta_b$ are tabulated in Table~\ref{Table-hyperfine-eta}.
Here the numbers in the parenthesis are the uncertainties. For example,
$\xi_a$ for the ${}^{1}$H isotope is
$$
  \xi_a =
  1.1092(2) E^{-3} \equiv
  (1.1092 \pm 0.0002) \times 10^{-3}.
$$
For all the isotopes, $| \xi_a - \xi_b |$ is larger than the uncertainties in
$\xi_a$ and $\xi_b$, and $| \eta_a - \eta_b |$ is much larger than
the uncertainties in $\eta_a$ and $\eta_b$.
Thus the ground-state hyperfine splitting is very sensitive to
the nuclear charge and magnetic moment distribution.

\subsection{Charge and magnetic moment distribution
  inside the nucleus for model (b)}  \label{subsec:model_b}

Substituting Eqs.~(\ref{eq:M_b}) and (\ref{subeqs:g_a-f_a})
into Eq.~(\ref{eq:A-hyperfine-def}) with $\nu = b$, we get
\begin{eqnarray}
  \Delta_b &=&
  -\frac{4 m c^2}{3} \,
  \alpha \big( Z \alpha \big)^{3} \,
  g_I \, \frac{m_e}{m_p}
  \Big( I + \frac{1}{2} \Big)
  \nonumber \\ && \times
  \bigg\{
       a_{B}^{2}
       \int\limits_{0}^{r_b}
       \frac{r^3}{r_{b}^{3}} \,
       g_{b,-}(r) f_{b,-}(r) dr
  \nonumber \\ && +
       a_{B}^{2}
       \int\limits_{r_b}^{\infty}
       g_{+}(r) f_{+}(r) dr
  \bigg\}.
  \label{eq:A-hyperfine-b}
\end{eqnarray}
We write $\Delta_b$ in the form
\begin{equation}   \label{eq:Delta_b-vs-Delta_0}
  \Delta_b = \Delta_0 \big( 1 - \xi_b \big) \big( 1 - \eta_b \big),
\end{equation}
where $\Delta_0$ is defined in Eq.~(\ref{eq:A-hyperfine-00}),
\begin{equation}   \label{eq:delta-epsilon_b-vs-G_0-G_plus-G_minus}
  \xi_b =
  \frac{\mathcal{G}_{0} - \mathcal{G}_{b}}{\mathcal{G}_{0}},
  \quad
  \eta_b =
  \frac{\mathcal{G}_{-}^{(b)} - \mathcal{F}_{-}^{(b)}}{\mathcal{G}_{b}},
\end{equation}
$\mathcal{G}_{0}$ is given in Eq.~(\ref{eq:G_0}), and
\begin{subequations}  \label{subeqs:G_minus_plus}
\begin{eqnarray}
  \mathcal{F}_{-}^{(b)} &=&
  \frac{a_{B}^{2}}{Z^3 \alpha}
  \int\limits_{0}^{r_b}
  \frac{r^3}{r_{b}^{3}} \,
  g_{b,-}(r) f_{b,-}(r) dr,
  \label{eq:F_b_minus}
  \\
  \mathcal{G}_{-}^{(b)} &=&
  \frac{a_{B}^{2}}{Z^3 \alpha}
  \int\limits_{0}^{r_b}
  g_{b,-}(r) f_{b,-}(r) dr,
  \label{eq:G_b_minus}
  \\
  \mathcal{G}_{b} &=&
  \mathcal{G}_{-}^{(b)} +
  \frac{a_{B}^{2}}{Z^3 \alpha}
  \int\limits_{r_b}^{\infty}
  g_{+}(r) f_{+}(r) dr.
  \label{eq:G_b}
\end{eqnarray}
\end{subequations}

\begin{figure}
\centering 
 \includegraphics[width=0.9 \linewidth,angle=0]
   {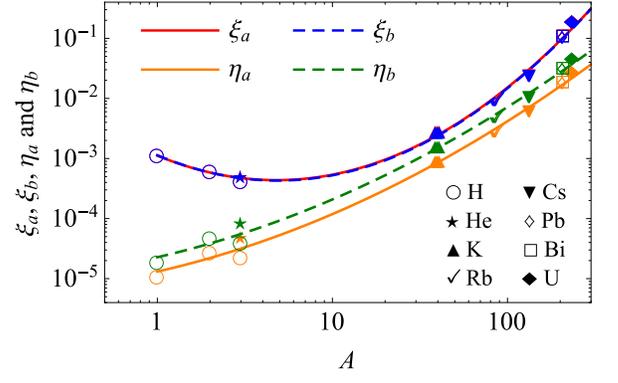}
\caption{\footnotesize 
  Corrections to the ground-state hyperfine splitting
  due to FNS distributions of the nuclear charge and
  magnetic moment as a function of the atomic mass number $A$.
  The corrections $\xi_a$ and $\eta_a$ [red and orange] are defined in
  Eq.~(\ref{eq:delta-epsilon_a-vs-G_0-G_plus-G_minus}) and the correction 
  $\xi_b$ and $\eta_b$ [blue and green] is defined in 
  Eq.~(\ref{eq:delta-epsilon_b-vs-G_0-G_plus-G_minus})
  [blue].  The solid curves that fit the data for $\xi_a$, $\eta_a$ and 
  $\xi_b$, $\eta_b$ are meant to guide the eye (see the text).\\
  Redo figure.}
\label{Fig-delta-epsilon-vs-A}
\end{figure}

\subsection{FNS corrections to hyperfine splitting  in models (a) and (b)}  
\label{subsec:hyperfine_corrections_a_b}

The FNS corrections to the hyperfine splitting 
$\xi_a$, $\xi_b$, $\eta_a$ and $\eta_b$ are tabulated in 
Tables~\ref{Table-hyperfine-xi} and \ref{Table-hyperfine-eta}
and are plotted in Fig.~\ref{Fig-delta-epsilon-vs-A} as functions of  $A$. 
Numerical analysis shows that $\ln \xi_a$, $\ln \eta_a$, $\ln \xi_b$ and $\ln \eta_b$
can be well-fit by
\begin{subequations}    \label{subeqs:delta-epsilon_a-epsilon_b-fit}
\begin{eqnarray}
  \ln \xi_a(A) &=&
  -6.79743 -
  1.20214 \, \ln A
  \nonumber \\ && +
  0.384078 \, \ln^2 A,
  \label{eq:xi_a-fit}
  \\
  \ln \eta_a(A) &=&
  -11.2423 +
  0.661711 \, \ln A
  \nonumber \\ && +
  0.128055 \, \ln^2 A,
  \label{eq:eta_a-fit}
  \\
  \ln \xi_b(A) &=&
  -6.78386 -
  1.22495 \, \ln A
  \nonumber \\ && +
  0.38719 \, \ln^2 A,
  \label{eq:xi_b-fit}
  \\
  \ln \eta_b(A) &=&
  -10.6924 +
  0.673483 \, \ln A
  \nonumber \\ && +
  0.125229 \, \ln^2 A.
  \label{eq:eta_b-fit}
\end{eqnarray}
\end{subequations}

FNS corrections to the hyperfine splitting of the ground state
of heavy HLI were calculated in Ref.~\cite{Shabaev-PRA-1997}.
The corrections $\xi_a$ for ${}^{207}$Pb and ${}^{209}$Bi in
Table~\ref{Table-hyperfine-xi} are not far from the values calculated in 
Ref.~\cite{Shabaev-PRA-1997}, $\xi(\mathrm{Pb}) = 0.1049$,
and $\xi(\mathrm{Bi}) = 0.1111$, the values of
$\eta_a$ for ${}^{209}$Bi in Table~\ref{Table-hyperfine-eta}
not far from the value reported in \cite{Shabaev-PRA-1997}, $\epsilon = 0.0118$,
and $\eta_b$ for ${}^{207}$Pb not too far away from $\epsilon = 0.0419$
in Ref.~\cite{Shabaev-PRA-1997}.

Within the assumption that $W_{ba}(r)$ in Eq.~(\ref{Eq:W_ba})
is small [i.e., $W_{ba}(r)$ is much weaker than $W_a(r)$ or $W_b(r)$
in Eq.~(\ref{eq:W}), see Fig.~\ref{Fig:V_W_surface-volume}], we showed 
that the correction $\xi$ is almost the same for both models (a) and (b), 
whereas models (a) and (b) predict very different values for the corrections 
$\eta_{\nu}$ due to the magnetic moment distribution.

\subsection{Charge and magnetic moment distribution
  inside the nucleus for model (c)}  \label{subsec:model_c}

For the TPFM, model (c),
\begin{equation}
  \label{eq:Delta_c}
  \Delta_c = \Delta_0 \big( 1 - \xi_c \big) \big(1 - \eta_c \big) ,
\end{equation}
where $\Delta_0$ is given in (\ref{eq:A-hyperfine-00}),
\begin{equation}   \label{eq:xi_c-eta_c}
  \xi_c = \frac{ \mathcal{G}_{0} - \mathcal{G}_{c} }{ \mathcal{G}_{0} },
  \quad
  \eta_c = \frac{ \mathcal{F}_{c} }{ \mathcal{G}_{c} },
\end{equation}
and
\begin{subequations}  \label{subeqs:G_c}
\begin{eqnarray}
  \mathcal{G}_{c} &=&
  \frac{a_{B}^{2}}{Z^3 \alpha}
  \int\limits_{0}^{\infty}
  g_c(r) f_c(r) dr,
  \label{eq:G_c}
  \\
  \mathcal{F}_{c} &=&
  \frac{a_{B}^{2}}{Z^3 \alpha}
  \int\limits_{0}^{\infty}
  \Big[ 1 - \mathcal{M}_{c}(r) \Big]
  g_c(r) f_c(r) dr.
  \label{eq:F_c}
\end{eqnarray}
\end{subequations}

\subsection{Relativistic QED radiative corrections for hyperfine splitting}  
\label{subsec:radiative_corr}

We now include the relativistic QED radiative corrections 
$\zeta_{\mathrm{QED}} = \gamma (2 \gamma - 1) \chi_{\mathrm{QED}}$
given in Eq.~(\ref{eq:hyperfine}).  The quantity $\chi_{\mathrm{QED}}$ is 
calculated in Refs.~\cite{Shabaev-PRA-1997, Sunnergren_98} for some elements.
The corrections $\zeta_{\mathrm{QED}}$ are tabulated in Table~\ref{Table-radiative}.

\begin{table}
\caption{Radiative QED correction $\zeta_{\mathrm{QED}} = \gamma (2 \gamma - 1) \chi_{\mathrm{QED}}$
  to the hyperfine splitting in Eq.~(\ref{eq:hyperfine}),
  where $\gamma$ is given in Eq.~(\ref{eq:rho-lambda-gamma-def}) and
  $\chi_{\mathrm{QED}}$ is calculated in Refs.~\cite{Shabaev-PRA-1997,
  Sunnergren_98} for  some elements. The fit 
  $\zeta_{\mathrm{QED}}^{\mathrm{(fit)}}$ is calculated using 
  Eq.~(\ref{eq:xi-fit}).}
\begin{center}
  \begin{tabular}{|c|c|c|}
    \hline
    $Z$ & $\zeta_{\mathrm{QED}}$ & $\zeta_{\mathrm{QED}}^{\mathrm{(fit)}}$
    \\
    \hline
    $1$ & $0.00105756$ & $0.00106387$
    \\
    \hline
    $2$ & & $0.000946973$
    \\
    \hline
    $3$ & $0.000835844$ & $0.000831409$
    \\
    \hline
    $5$ & $0.000609149$ & $0.000603679$
    \\
    \hline
    $7$ & $0.000382069$ & $0.00038002$
    \\
    \hline
    $10$ & $0.0000466108$ & $0.0000516092$
    \\
    \hline
    $19$ & & $-0.000884945$
    \\
    \hline
    $37$ & & $-0.00253464$
    \\
    \hline
    $49$ & $-0.00348472$ & $-0.00345706$
    \\
    \hline
    $53$ & $-0.00373754$ & $-0.00373092$
    \\
    \hline
    $55$ & & $-0.00386134$
    \\
    \hline
    $57$ & $-0.0039463$ & $-0.00398737$
    \\
    \hline
    $63$ & $-0.00427325$ & $-0.00433865$
    \\
    \hline
    $67$ & $-0.00467521$ & $-0.00455007$
    \\
    \hline
    $71$ & $-0.00461933$ & $-0.00474287$
    \\
    \hline
    $75$ & $-0.00501339$ & $-0.00491674$
    \\
    \hline
    $82$ & $-0.0051733$ & $-0.0051745$
    \\
    \hline
    $83$ & $-0.0051822$ & $-0.00520641$
    \\
    \hline
    92 & & $-0.00543714$
    \\
    \hline
  \end{tabular}
\end{center}
\label{Table-radiative}
\end{table}

\begin{figure}[htb]
\centering 
 \includegraphics[width=0.9 \linewidth,angle=0]
   {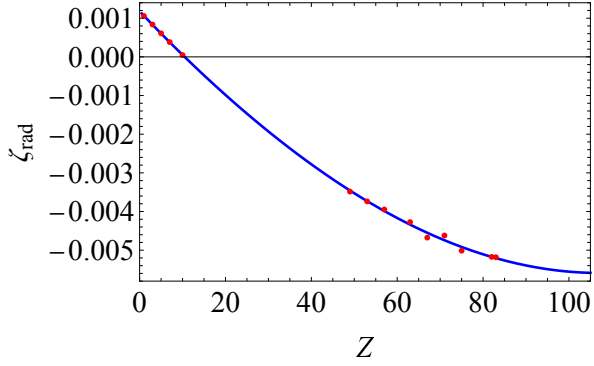}
\caption{\footnotesize 
  Radiative QED correction (red dots)
  $\zeta_{\mathrm{QED}} = \gamma (2 \gamma - 1) \chi_{\mathrm{QED}}$
  to the hyperfine splitting in Eq.~(\ref{eq:hyperfine}), versus nuclear charge $Z$
  where $\chi_{\mathrm{QED}}$ is calculated in Refs.~\cite{Shabaev-PRA-1997, Sunnergren_98}.
  The blue curve is the fit of $\zeta_{\mathrm{QED}}$ in Eq.~(\ref{eq:xi-fit}).}
\label{Fig-radiative-vs-Z}
\end{figure}

We fit $\zeta_{\mathrm{QED}}$ by the formula \cite{Sunnergren_98},
\begin{eqnarray}
  &&
  \zeta_{\mathrm{QED}}^{\mathrm{(fit)}}(Z) =
  \frac{\alpha}{\pi}
  \Big[
      A_0 -
      A_1 \, \alpha Z +
      A_2 \, \big(\alpha Z)^2
  \nonumber \\ && ~~~~~ +
      A_3 \, (\alpha Z)^2 \ln(\alpha Z) +
      A_4 \big( \alpha Z \ln(\alpha Z) \big)^2
  \Big].
  \label{eq:xi-fit}
\end{eqnarray}
The fit gives,
\begin{eqnarray*}
  A_0 &=& 0.509092
\\
  A_1 &=& 7.08715,
  \\
  A_2 &=& 4.43975,
    \\
  A_3 &=& 0.696079,
  \\
  A_4 &=& 0.448402,
\end{eqnarray*}
and the mean standard deviation $\sigma_{\mathrm{QED}}$ is
\begin{eqnarray*}
  \sigma_{\mathrm{QED}} &=&
  (\frac{1}{N_Z}
  \sum_{n_Z}^{N_Z}
  \big( \zeta_{\mathrm{QED}}(Z) - \zeta_{\mathrm{QED}}^{\mathrm{(fit)}}(Z) \big)^{2})^{1/2}
  \nonumber \\ &=& 0.000058397,
\end{eqnarray*}
where $\zeta_{\mathrm{QED}}$ was taken from Table~\ref{Table-radiative}, 
$N_Z = 14$ is the number of elements for which $\zeta_{\mathrm{QED}}$ is 
tabulated in the table, and $n_Z$ is a running index over the elements used.
The QED radiative correction $\zeta_{\mathrm{QED}}$ and its fit using 
Eq.~(\ref{eq:xi-fit}) are shown in Fig.~\ref{Fig-radiative-vs-Z}.
The fit (blue curve) is close to the red dots, hence we conclude that
the approximation in Eq.~(\ref{eq:xi-fit}) is satisfactory.  
The QED radiative corrections $\zeta_{\mathrm{QED}}^{\mathrm{(fit)}}(Z)$
in Eq.~(\ref{eq:xi-fit}) are tabulated in Table~\ref{Table-radiative} for H\&HLI.
For heavy isotopes,  $\zeta_{\mathrm{QED}}(Z)$ satisfies the inequalities:
$\zeta_{\mathrm{QED}} < 0$, and $\eta_a < |\zeta_{\mathrm{QED}}| < \eta_b \ll \xi$,
where $\xi_a$ and $\xi_b$ are tabulated in Table~\ref{Table-hyperfine-xi},
and $\eta_a$ and $\eta_b$ are tabulated in Table~\ref{Table-hyperfine-eta}.
Therefore, we conclude that the main contribution to the hyperfine splitting is the correction 
$\xi$ due to the finite nuclear charge distribution, which is much larger than the 
relativistic QED radiative corrections and the correction of the magnetic momentum
 distribution due to FNS.  For light elements (e.g., H and He), 
 $\zeta_{\mathrm{QED}} > 0$ and satisfies the inequalities, $\zeta_{\mathrm{QED}} > 
 \xi_a, \xi_b > \eta_a, \eta_b$, i.e., the QED radiative corrections to the ground-state 
 hyperfine splitting are larger than the FNS effects.  For heavy elements, $\zeta_{\mathrm{QED}} < 0$
and satisfies the inequalities, $|\zeta_{\mathrm{QED}}| < \eta_a, \eta_b < \xi_a, \xi_b$;
The FNS corrections to the hyperfine splitting due to the charge distribution are the largest
correction using either model (a) or (b).

\section{Estimates of the uncertainty of the hyperfine splitting in models $\text{(a)}$, $\text{(b)}$ and $\text{(c)}$}  \label{sec:uncertainty}

The ground-state hyperfine splittings $\Delta_a$, Eq.~(\ref{eq:Delta_a0-vs-Delta_00}),
and $\Delta_b$, Eq.~(\ref{eq:Delta_b-vs-Delta_0}), depend on the 
nuclear g factor $g_I$ and the RMS nuclear charge radius $r_N$.
Both $g_I$ and $r_N$ are measured in experiment with their uncertainties
$\delta g_I$ and $\delta r_N$, see Tables~\ref{Table-atomic-nuclear-radii} and 
\ref{Table-hyperfine-point}.  The uncertainties $\delta g_I$ and $\delta r_N$ give rise 
to uncertainties $\delta \Delta_a$ and $\delta \Delta_b$ that are given by
\begin{equation}
  \label{eq:delta-Delta-nu}
  \delta \Delta_{\nu} =
  \sqrt{ \bigg( \frac{\partial \Delta_{\nu}}{\partial g_I} \, \delta g_I \bigg)^{2} +
  \bigg( \frac{\partial \Delta_{\nu}}{\partial r_N} \, \delta r_N \bigg)^{2}} \,.
\end{equation}
Using Eq.~(\ref{eq:Delta_a0-vs-Delta_00}) and (\ref{eq:Delta_b-vs-Delta_0}),
we can write Eq.~(\ref{eq:delta-Delta-nu}) as,
\begin{equation}
  \label{eq:delta-Delta-vs-delta-g_I-delta-r_N}
  \delta \Delta_{\nu} = \Delta_{\nu} \,
  \sqrt{ \bigg( \frac{\delta g_I}{g_I} \bigg)^{2} +
  \bigg( \frac{\delta \xi_{\nu}}{1 - \xi_{\nu}} + 
  \frac{\delta \eta_{\nu}}{1 - \eta_{\nu}} \bigg)^{2} } \, .
\end{equation}

The ground-state hyperfine splitting $\Delta_c$ in Eq.~(\ref{eq:Delta_c})
depends on $g_I$, $r_N$ and the distribution thickness parameter $a_F$.
The parameters $g_I$, $r_N$ and $a_F$, and their uncertainties $\delta g_I$, 
$\delta r_N$ and $\delta a_F$, are measured in experiment, see 
Tables~\ref{Table-atomic-nuclear-radii}, \ref{Table-Fermi-distribution-radii}
and \ref{Table-hyperfine-point}.
The uncertainty $\delta \Delta_c$ in $\Delta_c$ duet to the uncertainties
$\delta g_I$, $\delta r_N$ and $\delta a_F$ is estimated as,
\begin{equation}
  \delta \Delta_c =
  \sqrt{ \bigg( \frac{\partial \Delta_c}{\partial g_I} \, \delta g_I \bigg)^{2} +
  \bigg( \frac{\partial \Delta_c}{\partial r_N} \, \delta r_N \bigg)^{2} +
  \bigg( \frac{\partial \Delta_c}{\partial a_F} \, \delta a_F \bigg)^{2}} \,.
  \label{eq:delta-Delta-c}
\end{equation}
Using Eq.~(\ref{eq:Delta_c}), we can rewrite Eq.~(\ref{eq:delta-Delta-c})
in the form,
\begin{eqnarray}
  \delta \Delta_c &=&
  \Bigg[
       \bigg( \frac{\delta g_I}{g_I} \bigg)^{2} +
       \bigg(
            \frac{\delta \xi_{c}^{(r)}}{1 - \xi_c} +
            \frac{\delta \eta_{c}^{(r)}}{1 - \eta_c}
       \bigg)^{2}
  \nonumber \\ && +
       \bigg(
            \frac{\delta \xi_{c}^{(a)}}{1 - \xi_c} +
            \frac{\delta \eta_{c}^{(a)}}{1 - \eta_c}
       \bigg)^{2}
  \Bigg]^{1/2},
  \label{eq:delta-Delta_c-vs-delta-g_I-delta-r_N-delta-a_F}
\end{eqnarray}
where $\delta \xi_{c}^{(r)}$ and $\delta \xi_{c}^{(a)}$ are the uncertainties
in $\xi_c$ due to the uncertainties $\delta r_N$ and $\delta a_F$ in
$r_N$ and $a_F$,
\begin{equation*}
  \delta \xi_{c}^{(r)} =
  \bigg| \frac{\partial \xi_c}{\partial r_N} \bigg| \,
  \delta r_N,
  \quad
  \delta \xi_{c}^{(a)} =
  \bigg| \frac{\partial \xi_c}{\partial a_F} \bigg| \,
  \delta a_F,
\end{equation*}
and $\delta \eta_{c}^{(r)}$ and $\delta \eta_{c}^{(a)}$ are the uncertainties
in $\xi_c$ due to the uncertainties $\delta r_N$ and $\delta a_F$ in
$r_N$ and $a_F$,
\begin{equation*}
  \delta \eta_{c}^{(r)} =
  \bigg| \frac{\partial \eta_c}{\partial r_N} \bigg| \,
  \delta r_N,
  \quad
  \delta \eta_{c}^{(a)} =
  \bigg| \frac{\partial \eta_c}{\partial a_F} \bigg| \,
  \delta a_F.
\end{equation*}

\section{Hyperfine splitting: Comparison with experiment}
  \label{sec:comparison}

We now compare the results of experiments on the ground state hyperfine splitting of
H\&HLI with the results of our calculations using model (a) and (b).
We present results of the calculations without and with relativistic QED radiative corrections.

The percent uncertainties of the theoretical results compared with the experimental values
are defined as,
\begin{eqnarray}
  \label{eq:hyperfine-comparison}
  \tau_{\nu} = \frac{\Delta_{\mathrm{exp}} - \tilde\Delta_{\nu}}{\Delta_{\mathrm{exp}}},
\end{eqnarray}
where $\nu = a, b, c$, and $\tilde\Delta_{\nu}$ is a hyperfine splitting with the 
QED radiative correction given by
\begin{equation}   \label{eq:Delta-tot}
  \tilde\Delta_{\nu} =
  \Delta_0 \,
  \Big[ \big( 1 - \xi_{\nu} \big) \big( 1 - \eta_{\nu} \big) + \zeta_{\mathrm{QED}} \Big]
  = \Delta_{\nu} + \Delta_0 \,  \zeta_{\mathrm{QED}}.
\end{equation}

The experimental hyperfine splitting $\Delta_{\mathrm{exp}}$ for hydrogen atoms 
and helium ions are taken from Refs.~\cite{Karshenboim-CanJPhys-2000, 
Ramsey-Phys-Rev-1972, Mathur-Phys-Rev-1967, Schluessler-Phys-Rev-1969, 
Karshenboim-H-He-HFS-PhysRevD-2002}, values for hydrogenic lead ions are taken from 
Ref.~\cite{Stohlker-presentation}, and values for hydrogenic bismuth ions are taken from 
Ref.~\cite{Ullmann-BiHFS-nature-comms-2017}. The quantities
$\tilde\Delta_a$, $\tilde\Delta_b$, $\Delta_{\mathrm{exp}}$, $\tau_a$ and $\tau_b$
are tabulated in Table~\ref{Table-hyperfine-experiment}.
For ${}^{1}$H atoms and ${}^{3}$He ions, $\tau_a < 0$ and $\tau_b > 0$, and
$\tilde\Delta_a < \Delta_{\mathrm{exp}} < \tilde\Delta_b$.
Moreover, $|\tau_a|$ and $\tau_b$ are smaller than the FNS corrections
$\eta_{\nu}$ and $\xi_{\nu}$ and the QED radiative correction $\zeta_{\mathrm{QED}}$.
For ${}^{2}$H and ${}^{3}$H atoms, both $\tau_a$ and $\tau_b$ are positive,
and $\tilde\Delta_a$ and $\tilde\Delta_b$ are below $\Delta_{\mathrm{exp}}$.
Moreover, $\tau_{a}$ and $\tau_b$ exceed $\eta_{\nu}$, but are below
$\xi_{\nu}$ and $\zeta_{\mathrm{QED}}$.
For ${}^{207}$Pb ions, $\tau_a < 0$ and $\tau_b > 0$, and
for ${}^{209}$Bi ions,  $\tau_a > 0$ and $\tau_b > 0$.
For both ${}^{207}$Pb and ${}^{209}$Bi, 
$| \tau_a |$ and $\tau_b$ are below $\eta_{\nu}$ and $\xi_{\nu}$,
but above $\zeta_{\mathrm{QED}}$.

\begin{widetext}

\begin{table}[ht]
\caption{\footnotesize 
  Experimental values of the hyperfine splitting $\Delta_{\mathrm{exp}}$ [in Hartree],
  as compared with the calculated results $\tilde\Delta_a$ and $\tilde\Delta_b$ in 
  Eq.~(\ref{eq:Delta-tot}) using models (a) and (b) with relativistic QED radiative corrections, 
  and the percent uncertainties ($\tilde\tau_a, \tilde\tau_b$) of theoretical results compared with 
  $\Delta_{\mathrm{exp}}$.  The numbers in brackets show uncertainty $\delta \Delta_{\mathrm{exp}}$ 
  in $\Delta_{\mathrm{exp}}$, and $\delta \Delta_{\nu}$ in $\Delta_{\nu}$ with $\nu = a, b$,
  see Eq.~(\ref{eq:delta-Delta-vs-delta-g_I-delta-r_N}).
  The experimental values for hydrogen atoms and helium ions are taken from
  Refs.~\cite{Karshenboim-CanJPhys-2000, Ramsey-Phys-Rev-1972,%
  Mathur-Phys-Rev-1967, Schluessler-Phys-Rev-1969, %
  Karshenboim-H-He-HFS-PhysRevD-2002}, values for hydrogenic
  lead ions are taken from Ref.~\cite{Stohlker-presentation} and values
  for hydrogenic bismuth ions are taken from Ref.~\cite{Ullmann-BiHFS-nature-comms-2017}.
  The percent uncertainties $\tau_{\nu}$ of the theoretical results compared with the
  experimental values [see Eq.~(\ref{eq:hyperfine-comparison})].  The model (c) results
  for heavy ions are not presented in this table; they are presented below.}
\begin{center}
\begin{tabular}{|c|c|c|c|c|c|c|c|}
  \hline
  Isotope ($Z, A$) &
  $\tilde\Delta_a$ &
  $\delta \Delta_a/\Delta_a$ &
  $\tilde\Delta_b$ &
  $\delta \Delta_b/\Delta_b$ &
  $\Delta_{\mathrm{exp}}$ &
  $\tau_a$ &
  $\tau_b$
  \\
  \hline
  H (1, 1) &
  $2.1587837 (7) E^{-7}$ &
  $3 E^{-7}$ &
  $2.1587637 (8) E^{-7}$ &
  $4 E^{-7}$ &
  $2.158775054569 (2) E^{-7}$ &
  $-4.0 E^{-6}$ &
  $5.2 E^{-6}$
  \\
  \hline
  H (1, 2) &
  $4.97460440(14) E^{-8}$ &
  $3 E^{-8}$ &
  $4.97448807 (18) E^{-8}$ &
  $4 E^{-8}$ &
  $4.9756850998 (3) E^{-8}$ &
  $2.2 E^{-4}$ &
  $2.4 E^{-4}$
  \\
  \hline
  H (1, 3) &
  $2.305066 (3) E^{-7}$ &
  $1.4 E^{-6}$ &
  $2.305022 (4) E^{-7}$ &
  $1.8 E^{-6}$ &
  $2.30512816232(1) E^{-7}$ &
  $2.7 E^{-5}$ &
  $4.6 E^{-5}$
  \\
  \hline
  He (2, 3) &
  $-1.3170511 (4) E^{-6}$ &
  $3 E^{-7}$ &
  $-1.3170088 (3) E^{-6}$ &
  $2 E^{-7}$ &
  $-1.317031330040 (2) E^{-6}$ &
  $-1.5 E^{-5}$ &
  $1.7 E^{-5}$
  \\
  \hline
  K(19, 39) &
  $1.416801 (3) E^{-4}$ &
  $2 E^{-6}$ &
  $1.415971 (3) E^{-4}$ &
  $1.8 E^{-6}$ &
  &
  &
  \\
  \hline
  K(19, 40) &
  $-3.9639 (8) E^{-4}$ &
  $1.9 E^{-4}$ &
  $-3.9616 (8) E^{-4}$ &
  $1.9 E^{-4}$ &
  &
  &
  \\
  \hline
  K(19, 41) &
  $7.77649 (4) E^{-5}$ &
  $5 E^{-6}$ &
  $7.77191 (3) E^{-5}$ &
  $4 E^{-6}$ &
  &
  &
  \\
  \hline
  Rb(37, 85) &
  $3.50330 (2) E^{-3}$ &
  $7 E^{-6}$ &
  $3.497280 (19) E^{-3}$ &
  $5 E^{-6}$ &
  &
  &
  \\
  \hline
  Rb(37, 87) &
  $7.91520 (4) E^{-3}$ &
  $4 E^{-6}$ &
  $7.90160 (3) E^{-3}$ &
  $4 E^{-6}$ &
  &
  &
  \\
  \hline
  Cs(55, 133) &
  $2.38846 (6) E^{-2}$ &
  $2 E^{-5}$ &
  $2.37919 (4) E^{-2}$ &
  $1.8 E^{-5}$ &
  &
  &
  \\
  \hline
  Cs(55, 135) &
  $2.54604 (19) E^{-2}$ &
  $8 E^{-5}$ &
  $2.53615 (19) E^{-2}$ &
  $8 E^{-5}$ &
  &
  &
  \\
  \hline
  Pb (82, 207) &
  $4.5117 (19) E^{-2}$ &
  $4 E^{-5}$ &
  $4.46840 (17) E^{-2}$ &
  $4 E^{-5}$ &
  $4.4684 (7) E^{-2}$ &
  $-9.7 E^{-3}$ &
  $8.3 E^{-3}$
  \\
  \hline
  Bi (83, 209) &
  $1.85493 (12) E^{-1}$ &
  $1.2 E^{-5}$ &
  $1.83335 E^{-1}$ &
  $1.0 E^{-5}$ &
  $1.86871 (3) E^{-1}$ &
  $1.4 E^{-3}$ &
  $3.5 E^{-3}$
  \\
  \hline
  U(92, 235) &
  $-2.85 (22) E^{-2}$ &
  $8 E^{-2}$ &
  $-3.46 (27)$ &
  $8 E^{-2}$ &
  &
  &
  \\
  \hline
\end{tabular}
\end{center}
\label{Table-hyperfine-experiment}
\end{table}
\end{widetext}

We now compare our results for model (c) with experiment.
Calculations for the ${}^{207}$Pb ion yield:
\begin{equation*}
  \xi_c = 0.11105 (8),
  \quad
  \eta_c = 0.02086 (3).
\end{equation*}
Taking into account that $\Delta_0 = 5.1434 (2) \times 10^{-2}$, we find
$$
  \Delta_c = 4.4858 (2)\times 10^{-2},
$$
hence
$$
  \tau_c = \frac{\Delta_{\mathrm{exp}} - \Delta_c}{\Delta_{\mathrm{exp}}} =
  -0.0039.
$$
Comparing $\tau_c$ with $\tau_a = -0.0097$ and $\tau_b = 0.0083$
in Table~\ref{Table-hyperfine-experiment},
one can see that model (c) results are better for the ground-state
hyperfine splitting than for models (a) and (b). However,
$| \Delta_{\mathrm{exp}} - \Delta_c |> \delta \Delta_c$.

Calculations for ${}^{209}$Bi ion give:
\begin{equation*}
  \xi_c = 0.104649 (27),
  \quad
  \eta_c = 0.020142 (10).
\end{equation*}
Taking into account that $\Delta_0 = 2.14590(10) E^{-2}$, we get
$$
  \Delta_c = 0.185665 (15) E^{-2},
$$
such that
$$
  \tau_c =
  \frac{\Delta_{\mathrm{exp}} - \Delta_c}{\Delta_{\mathrm{exp}}} =
  0.0064526.
$$
Comparing $\tau_c$ with $\tau_a = 0.0014$ and $\tau_b = 0.0035$
in Table~\ref{Table-hyperfine-experiment}, one sees that model (a)
gives somewhat better results for the ground-state hyperfine splitting than
models (b) and (c) for ${}^{209}$Bi.

\section{Summary and Conclusions}
  \label{sec:conclude}

We calculated FNS corrections for the ground state ($1s$) and the first excited state ($2s$) of H\&HLI using the Dirac equation with three different models for the atomic nucleus charge distribution. For model (a) [nuclear charge on the surface of the nucleus], we present the analytic solution of the Dirac equation, whereas for model (b) [nuclear charge uniformly distributed inside the nucleus] and model (c) [the two-parameter Fermi model -- TPFM], we use both numerical calculations and first-order perturbation theory (the latter is not valid for large $Z$ nuclei).  The FNS corrections to the ground state energy and the first excited state energy are smaller than the electron-nucleus reduced mass correction and the relativistic QED radiative corrections for {\em light nuclei} (i.e., low $Z$ nuclei) \cite{{Eides-hydrogen-atom}}.  Nevertheless, it is important to obtain accurate FNS corrections so that a good comparison can be made with the extremely high-accuracy experimental results for hydrogen \cite{Parthey, Matveev} and deuterium \cite{Pachucki_94, Jentschura}.  A detailed study of the hydrogen $1s$-$2s$ transition has been reported in Ref.~\cite{Kuzmenko-letter}. For heavy nuclei, the electron-nucleus reduced mass correction is small because $m_e/M(Z,A)$ is small, hence the FNS correction is larger than the electron-nucleus reduced mass correction.  Furthermore, for heavy nuclei, $Z \alpha$ is not a small parameter, hence higher order relativistic QED radiative corrections, which can be neglected for hydrogen and light nuclei, are required, and therefore it is not clear whether FNS corrections are more important than relativistic QED radiative corrections since the latter have not been calculated.

We find that for H atoms and light HLI, the difference of the FNS corrections obtained using 
model (a) and model (b), $\veps_{n,a} - \veps_{n,b}$, are small; the difference
between model (a) and model (b) energies $\veps_n$ for $n = 1$ (the $1s$ state) 
and for $n = 2$ (the $2s$ state) is smaller than the uncertainties $\delta \veps_{n,a}$ and 
$\delta \veps_{n,b}$ in $\veps_{n,a}$ and $\veps_{n,b}$, see Tables~\ref{Table-energy-constant}
and \ref{Table-transitions}.  Hence, measurement of the $1s$-$2s$ transition frequency cannot be used to 
determine details of the nuclear charge distribution for low $Z$ nuclei
[the perturbation theory expression in Eq.~(\ref{eq:EFNS}) is a good approximation
for low $Z$ nuclei, and perturbation theory expression does not depend on the details of the charge distribution,
but only on $r_N$].  However, for heavy HLIs, the difference, $\veps_{n,a} - \veps_{n,b}$, is larger 
than the uncertainties $\delta \veps_{n,a}$ and $\delta \veps_{n,b}$ in $\veps_{n,a}$ and $\veps_{n,b}$.
Thus measuring the $1s$-$2s$ transition frequency for heavy nuclei HLIs can be used to determine 
not only the nuclear charge radii but the details of the nuclear charge distribution.
Note that QED radiative corrections are available only for $Z \alpha \ll 1$. Hence, there is a 
problem applying the QED corrections for heavy nuclei.

Experimental results for the $1s$-$3s$ transition frequency for hydrogen and deuterium have also been reported, see e.g., Ref.~\cite{Grinin_20}.  We intend to calculate this transition frequency to compare with these experiments in future work.

We have also calculated the FNS corrections for the ground-state
{\em hyperfine splitting} of H\&HLI using models (a), (b) and (c).
After calculating the hyperfine splitting for point-like nuclei (model $\nu = 0$),
we calculate the corrections $\xi_{\nu}$ (where $\nu = a, b, c$) due to
the nuclear charge distribution, and the corrections $\eta_{\nu}$
due to the magnetic moment distribution.  The percent uncertainties 
$\tau_{\nu}$ of the theoretical results are compared with the experimental values 
[see Eq.~(\ref{eq:hyperfine-comparison})].  Both $|\tau_a|$ and $|\tau_b|$ are smaller 
than the correction $\xi_{\nu}$ to the hyperfine splitting. For ${}^{1}$H atoms and ${}^{3}$He ions, 
$|\tau_{\nu}| < \eta_{\nu}$.  For ${}^{2}$H, ${}^{3}$H atoms, and ${}^{207}$Pb and ${}^{209}$Bi HLIs,
$|\tau_{\nu}| > \eta_{\nu}$.  We calculate also the ground-state hyperfine splitting 
for ${}^{207}$Pb and ${}^{209}$Bi HLIs with model (c) for the nuclear charge and 
magnetic moment distribution.
For the ${}^{207}$Pb HLI, model (c) gives a somewhat better result for
the ground-state hyperfine splitting than models (a) and (b).
For the ${}^{209}$Bi HLI, model (a) gives a somewhat better result for
the ground-state hyperfine splitting than models (b) and (c).

We show that $\xi_a > \eta_a$ for all the isotopes (see tables \ref{Table-hyperfine-xi}
and \ref{Table-hyperfine-eta}), i.e., the nuclear charge
distribution correction to the ground-state hyperfine splitting is larger than
the magnetic moment distribution correction.
On the other hand, the difference $| \xi_a - \xi_b |$ is much larger than the uncertainties
$\delta\xi_a$ and $\delta\xi_b$ in $\xi_a$ and $\xi_b$, and
the difference $| \eta_a - \eta_b |$ is much larger than the uncertainties
$\delta\eta_a$ and $\delta\eta_b$ in $\eta_a$ and $\eta_b$.  Hence,
it is crucial to test more accurate models of the nuclear charge and magnetic moment distributions in order to obtain results closer to experimental hyperfine splittings. The work that has been invested into calculating QED radiative corrections, can only bear fruit if accurate FNS corrections are developed.  Given the fact that the FNS effects for hyperfine splittings are as big or bigger than the QED radiative corrections for all the nuclei studied, how can one obtain more accurate nuclear charge and magnetic moment distributions in order to improve the FNS corrections?  Several approaches are possible. (1) One can use more elaborate models for the charge and magnetic moment distributions.  For example, one can use the three-parameter Fermi model, which has an additional term, $(1 + w_F \, r^2/c_F^2)$, multiplying the TPFM \cite{{deVries_87}} or the double three-parameter Fermi model \cite{Abdulghany}.  (2) Direct numerical fits to nuclear charge and magnetic moment distributions can be employed using data from electron scattering from nuclei \cite{Uberall}.  (3) The charge and magnetic moment distributions are in general not equal, see e.g., Appendix~\ref{append:magnetic-moment-distribution}, hence distributions should be determined for each from experiment.

\bigskip

\noindent{\bf Acknowledgement}: We would like to thank Shalom Shlomo for useful correspondence.

\bigskip

\appendix

\section{Nuclear charge radius}  \label{append:nuclear-charge-radius}

The nuclear charge radius $r_N = \sqrt{ \langle r^2 \rangle_{N}}$ is
defined as a RMS charge radius which is a measure of
the normalized charge distribution $\rho_N(\mbfr)$,
\begin{equation}
  \langle r^2 \rangle_{N} =
  \int r^2 \rho_N(r) d^3\mbfr,
  \label{eq:nuclear-charge-radius-def}
\end{equation}
where $\rho_N(\mbfr)$ is normalized by the condition
\begin{equation*}
  \int \rho_N(\mbfr) d^3\mbfr = 1.
\end{equation*}
We now apply Eq.~(\ref{eq:nuclear-charge-radius-def})
and calculate the nuclear radius for models (a) and (b).

\underline{\textbf{Model (a)}}:
When the charge is distributed uniformly on a sphere of
radius $r_a$, the normalized proton distribution
$\rho_N(\mbfr) \equiv \rho_a(r)$ is,
\begin{equation}
  \rho_a(r) = \frac{Z e}{4 \pi r_{a}^{2}} \, \delta ( r - r_a),
  \label{eq:charge-density-a}
\end{equation}
Substituting Eq.~(\ref{eq:charge-density-a}) into
Eq.~(\ref{eq:nuclear-charge-radius-def}), we get
\begin{equation}
  r_a = r_N.
  \label{eq:r_a=r_N}
\end{equation}

\underline{\textbf{Model (b)}}:
When the charge is distributed uniformly inside a sphere of
radius $r_b$, the normalized proton distribution
$\rho_N(\mbfr) \equiv \rho_b(r)$ is,
\begin{equation}
  \rho_b(r) = \frac{3 Z e}{4 \pi r_{b}^{3}} \, \Theta ( r_b - r),
  \label{eq:charge-density-b}
\end{equation}
where $\Theta( r_b - r)$ is equal to $1$ for $r < r_b$, $\frac{1}{2}$ 
for $r = r_b$ and $0$ for $r > r_b$.
Substituting Eq.~(\ref{eq:charge-density-b}) into
Eq.~(\ref{eq:nuclear-charge-radius-def}), we get
$r_N = r_b \sqrt{3/5}$, or
\begin{equation}
  r_b = r_N \, \sqrt{ \frac{5}{3} }.
  \label{eq:r_b=r_N-sqrt}
\end{equation}

\section{Matrix elements of $\mbfe_r \times \boldsymbol\alpha$}
  \label{append:er_alpha_matrix_elemets}

The vector operator $\mbfe_r \times \boldsymbol\alpha$ can be written as,
\begin{eqnarray*}
  \mbfe_r \times \boldsymbol\alpha &=&
  \left(
    \begin{array}{cc}
      0 & \mbfe_r \times \boldsymbol\sigma
      \\
      \mbfe_r \times \boldsymbol\sigma & 0
    \end{array}
  \right),
\end{eqnarray*}
where the 2$\times$2 dimensional matrices
$\mbfe_r \times \boldsymbol\sigma$ are given by
\begin{eqnarray*}
  \mbfe_r \times \boldsymbol\sigma &=&
  \left(
    \begin{array}{cc}
      \sin\theta \sin\phi & i \cos\theta
      \\
      -i \cos\theta & -\sin\theta \sin\phi
    \end{array}
  \right)
  \mbfe_x +
  \nonumber \\ &&
  \left(
    \begin{array}{cc}
      -\sin\theta \cos\phi & \cos\theta
      \\
      \cos\theta & \sin\theta \cos\phi
    \end{array}
  \right)
  \mbfe_y +
  \nonumber \\ &&
  \left(
    \begin{array}{cc}
      0 & -i e^{-i \phi}
      \\
      i e^{i \phi} & 0
    \end{array}
  \right)
  \mbfe_z.
\end{eqnarray*}
The matrix elements of $\mbfe_r \times \boldsymbol\alpha$
with the wave functions $\psi_{m_J}(\mbfr)$ in
Eq.~(\ref{eq:WF-ground-J-m_J}) take the form,
\begin{eqnarray}
  &&
  \int
  \psi_{m_J}^{\dag}(\mbfr) \,
  \big[ \mbfe_r \times \boldsymbol\alpha \big] \,
  \psi_{m'_J}(\mbfr)
  \sin\theta d\theta d\phi
  \times \nonumber \\ && ~~~ =
  g(r) f(r)
  \int\limits_{0}^{\pi} \sin\theta d\theta
  \int\limits_{0}^{2\pi} d\phi
  \times \nonumber \\ && ~~~~~ ~
  \Big\{
      i \, \Omega_{\frac{1}{2} 0 m_J}^{\dag}
      \big[ \mbfe_r \times \boldsymbol\sigma \big]
      \Omega_{\frac{1}{2} 1 m'_J} +
      \mathrm{c.c.}
  \Big\} \nonumber \\ && ~~~ =
  -\frac{8}{3} \,
  g(r) f(r) \,
  \mathbf{J}_{m_J, m'_J},
  \label{eq:e-alpha-matrix-elements}
\end{eqnarray}
where $\mathbf{J}_{m_J, m'_J}$ are matrix elements of
the orbital angular momentum operator $\mathbf{J}$.

\section{Iteration method for determining $s$ state energies using Eq.~(\ref{eq:det-b})}
  \label{append:iteration}
 
We shall now explain the iteration method, using 
the case of the ${}^{3}$He ion for specificity.
The ground-state energy of the ${}^{3}$He ion for model (a) is
$\veps_{1}^{(0)} = 1.47460 \times 10^{-8} \epsilon_H$
(see Table~\ref{Table-energy-constant}), where
$\veps_{1}^{(0)}$ is the ground-state energy of the He ion with
the point-like nucleus given in Eq.~(\ref{eq:energy-hydrgen-ground}).
The energy using model (b) is close to the energy for model (a), 
see Table~\ref{Table-corrections-ba} below.  Thus we take 
$\mathcal{E}_{1} = \veps_{1}^{(0)} - 4 \times 10^{-8} \epsilon_H$
as a first iteration for the energy.
Solving Eq.~(\ref{eq:for-g-2nd-order}) numerically
for $\mathcal{E}$, we find $g_{b,-}(r), f_{b,-}(r)$ for $r < r_b$.
Substituting $g_{b,-}(r_b), f_{b,-}(r_b)$, and
$g_{+}(r_b), f_{+}(r_b)$ with $\mathcal{E}_1$
[see in Eq.~(\ref{subeqs:fg-res-plus})]
into Eq.~(\ref{eq:det-b}), we get
$\mathfrak{D}_1 \equiv \mathfrak{D}(\mathcal{E}_1) = -5.16387 \times 10^{-5}$.
The second iteration is $\mathcal{E}_{2} = \veps_{1}^{(0)} - 1.4 \times 10^{-8} \epsilon_H$.
Solving Eq.~(\ref{eq:for-g-2nd-order}) numerically
for $\mathcal{E}_2$, we find $g_{b,-}(r), f_{b,-}(r)$ for $r < r_b$.
Substituting $g_{b,-}(r_b), f_{b,-}(r_b)$, and
$g_{+}(r_b), f_{+}(r_b)$ with $\mathcal{E}_2$
[see in Eq.~(\ref{subeqs:fg-res-plus})]
into Eq.~(\ref{eq:det-b}), we get
$\mathfrak{D}_2 \equiv \mathfrak{D}(\mathcal{E}_2) = 1.52535 \times 10^{-6}$.
In order to find the third iteration for the ground-state energy $\varepsilon$,
we plot a graph of $\mathfrak{D}(\varepsilon)$ versus $\varepsilon$, 
with the two points, $(\mathcal{E}_1, \mathfrak{D}_{1})$,
and $(\mathcal{E}_2, \mathfrak{D}_{2})$.
We then connect these points by a line, and find the crossing
point $\mathcal{E}_3$ of this line with the $x$ axis,
$$
  \mathcal{E}_3 =
  \frac{\mathcal{E}_1 \mathfrak{D}_2 - \mathcal{E}_2 \mathfrak{D}_1}
       {\mathfrak{D}_2 - \mathfrak{D}_1} =
  \veps_{1}^{(0)} -  1.474597673 \times 10^{-8} \, \epsilon_H.
$$
Solving Eq.~(\ref{eq:for-g-2nd-order}) numerically
for $\mathcal{E}_3$, we find $g_{b,-}(r), f_{b,-}(r)$ for $r < r_b$.
Substituting $g_{b,-}(r_b), f_{b,-}(r_b)$, and
$g_{+}(r_b), f_{+}(r_b)$ with $\mathcal{E}_3$
[see in Eq.~(\ref{subeqs:fg-res-plus})]
into Eq.~(\ref{eq:det-b}), we get
$\mathfrak{D}_3 \equiv \mathfrak{D}(\mathcal{E}_3) = 4.59235 \times 10^{-13}$.
The fourth iteration for the energy is found from the equation,
$$
  \mathcal{E}_4 =
  \frac{\mathcal{E}_2 \mathfrak{D}_3 - \mathcal{E}_3 \mathfrak{D}_2}
       {\mathfrak{D}_3 - \mathfrak{D}_2} =
  \veps_{1}^{(0)} -  1.474597699 \times 10^{-8} \, \epsilon_H,
$$
such that $\mathcal{E}_3 - \mathcal{E}_4 = 2.6 \times 10^{-16} \epsilon_H$.
Solving Eq.~(\ref{eq:for-g-2nd-order}) numerically
for $\mathcal{E}_4$, we find $g_{b,-}(r), f_{b,-}(r)$ for $r < r_b$.
Substituting $g_{b,-}(r_b), f_{b,-}(r_b)$, and
$g_{+}(r_b), f_{+}(r_b)$ with $\mathcal{E}_4$
[see in Eq.~(\ref{subeqs:fg-res-plus})]
into Eq.~(\ref{eq:det-b}), we get
$\mathfrak{D}_4 \equiv \mathfrak{D}(\mathcal{E}_4) = -9.41473 \times 10^{-13}$.
Since $\mathcal{E}_3 - \mathcal{E}_4$ and $\mathfrak{D}_4$ are very small, 
 $\mathcal{E}_4$ is practically equal to the  ${}^{3}$He ground state energy.
-----------------

\section{Nuclear magnetic moment distribution for special nuclei}
  \label{append:magnetic-moment-distribution}
  
This appendix discusses the magnetic moment distribution for
nuclei with one nucleon outside a closed nuclear shell. 

\subsection{Nuclear shell model}
  \label{append:subsec:shell}

According to the nuclear shell model \cite{shell-model-huperphysics, 
Landau-Lifshitz-3-shell}, each nucleon in a nucleus
moves in a self-consistent field due to the other nucleons.
The nuclear self-consistent potential decreases rapidly outside 
the volume bounded by the surface of the nucleus.
The quantum state of the nucleus is described by specifying
the states of the individual nucleons.
The self-consistent field is spherically symmetric, and the center
of symmetry is the center of mass of the nucleus.
Hence, the quantum states of the individual nucleons are parametrized
by a radial quantum number $n_r$,
angular momentum quantum number $L$,
orbital angular momentum $J = L \pm 1/2$,
and projection $M_J$ of the vector $\mathbf{J} = \mbfL + \mbfS$.
Nucleons fill shells according to the Pauli principle
such that the nucleon states are distributed among the groups
shown in Table~\ref{Table:groups-of-shells}
\cite{shell-model-huperphysics, Landau-Lifshitz-3-shell}.
For each group, the total number of proton or neutron occupation is 
shown.  Hence, the occupation of a group is completed
when the number $Z$ of protons or the number $N$ of neutrons
is equal to one of the numbers: 2, 8, 28, 50, 82, 126, $\ldots$.
These numbers are called \textit{magic numbers} \cite{Landau-Lifshitz-3-shell}.
Nuclei with both $Z$ and $N$ being magic numbers, are called
\textit{double magic} nuclei and have nuclear spin $I = 0$.

\begin{table}[ht]
\begin{center}
  \begin{tabular}{l c}
    shells & nucleons
    \\
    $1s_{1/2}$ & 2
    \\
    $1p_{3/2}$, $1p_{1/2}$ & 6
    \\
    $1d_{5/2}$, $2s_{1/2}$, $1d_{3/2}$, $1f_{7/2}$ & 20
    \\
    $2p_{3/2}$, $1f_{5/2}$, $2p_{1/2}$, $1g_{9/2}$ & 22
    \\
    $2d_{5/2}$, $1g_{7/2}$, $1h_{11/2}$, $2d_{3/2}$, $3s_{1/2}$ & 32
    \\
    $1h_{9/2}$, $2f_{7/2}$, $2i_{13/2}$, $2f_{5/2}$, $3p_{3/2}$, $3p_{1/2}$ & 44
\end{tabular}
\end{center}
\caption{Groups of shells and the number of nucleon completing
  each group.
  Here the quantum number $n_r$ before the letter is
  a radial quantum number, and
  the letters $s$, $p$, $d$, $f$, $g$, $h$, $i$, $\dots$
  refer to the angular momentum quantum number,
  $L_j = 0, 1, 2, 3, 4, 5, 6, \ldots$, where $j = p, n$ for protons
  and neutrons, and the subscript on the letters indicates
  the total orbital angular momentum of the nucleon,
  $J_j = L_j \pm 1/2$.}
\label{Table:groups-of-shells}
\end{table}%

A nucleon moves in an effective attractive potential $V_{\mathrm{nucl}}^{(j)}(r)$ 
formed by all the other nucleons.  The potential $V_{\mathrm{nucl}}^{(j)}(r)$ is 
given by  \cite{shell-model-huperphysics}
\begin{equation}
  \label{eq:V_nucl-model-b}
  V_{\mathrm{nucl}}^{(j)}(r) =
  -\frac{V_j}{e^{(r - R_0)/a_F} + 1} +
  U_{\mathrm{rep}}(r) \delta_{j,p},
\end{equation}
where the first term on the right hand side of Eq.~(\ref{eq:V_nucl-model-b})
is the Woods-Saxon attractive potential, and the second term,
$U_{\mathrm{rep}}(r)$, is the electrostatic repulsive interaction of
a proton with the other $Z-1$ protons in the nucleus.
The Kronecker delta indicates that the Coulomb interaction
is present for protons and absent for neutrons.
Here $R_0 = 1.25 \, A^{1/3}$, and $V_j$ is the depth of
the potential well for protons ($j = p$) and neutrons ($j = n$).
Note that $R_0 > c_F$, where $c_F$ is the half-density nuclear 
radius in the two-parameter Fermi nuclear charge distribution,
Eq.~(\ref{eq:Fermi-distribution}), and the difference $R_0 - c_F$ is due to 
the finite range of the nuclear force.  The potential depth $V_j$ is
\begin{eqnarray}
  V_j &=&
  U_0 +
  \eta_j U_1 \,
  \frac{N - Z}{A},
  \label{eq:V_nuclear-depth}
\end{eqnarray}
where $U_0 = 57~\text{MeV}$, $U_1 = 27~\text{MeV}$, $Z$ is the number 
of protons in the nucleus, $N$ is the number of neutrons in the nucleus, and
$A = Z + N$ is the total number of nucleons in the nucleus.
The second term in the right hand side of Eq.~(\ref{eq:V_nuclear-depth})
is the symmetry energy arising when $Z \neq N$, where $\eta_p = 1$
and $\eta_n = -1$ \cite{shell-model-huperphysics}.
The second term in the right hand side of Eq.~(\ref{eq:V_nucl-model-b}),
$U_{\mathrm{rep}}(r)$, is the electrostatic repulsion of a proton with
the other $Z-1$ protons in the nucleus. For the model (b) for the
charge distribution, $U_{\mathrm{rep}}(r)$ is
\begin{equation*}
  U_{\mathrm{rep}}(r) =
  \left\{
    \begin{array}{ccc}
      \displaystyle
      \frac{(Z-1) e^2}{2 r_b}
      \bigg( 3 - \frac{r^2}{r_{b}^{2}} \bigg),
      & \text{for} &
      r \leq r_b,
      \\
      \displaystyle
      \frac{(Z - 1) e^2}{r},
      & \text{for} &
      r > r_b.
    \end{array}
  \right.
  \label{eq:V-nuclear-Coulomb}
\end{equation*}

In addition, spin-orbit interaction of a nucleon in
the self-consistent potential is present, which is weaker than
the strong interaction and can be taken into account using
perturbation theory.

A nucleon wave function $\Psi_{n_r, L_j, M_L}(\mbfr)$ in a
shell with the radial quantum number $n_r$,
the angular momentum quantum number $L_j$ and
the projection $M_L$ of the angular momentum $\mathbf{L}_{j}$
on the $z$ axis is found from the Schr\"odinger equation,
\begin{equation}
  \label{eq:Schrodinger-nucleon}
  \left[-\frac{\hbar^2}{2 M_j} \,
  \nabla^2  + V_{\mathrm{nucl}}^{(j)}(r) \right] \Psi_{n_r, L_j, M_L}(\mbfr)
  = \epsilon \Psi_{n_r, L_j, M_L}(\mbfr) ,
\end{equation}
where $M_j$ is the nucleon mass.
$\Psi_{n_r, L_j, M_L}(\mbfr)$ can be written as,
$$
  \Psi_{n_r, L_j, M_L}(\mbfr) =
  \psi_{n_r, L_j}(r) Y_{L_j, M_L}(\theta, \phi),
$$
where $\psi_{n_r, L_j}(r)$ is a radial wave function and
$Y_{L_j, M_L}(\theta, \phi)$ is a spherical harmonic.
$\psi_{n_r, L_j}(r)$ is found from the equation,
\begin{eqnarray}
  &&-\frac{\hbar^2}{2 M_j r^2} \,
  \frac{d}{d r}
  \bigg(
       r^2 \, \frac{d \psi_{n_r,L}(r)}{d r}
  \bigg) +
  \frac{\hbar^2 L_j (L_j+1)}{2 M_j r^2} \, \psi_{n_r, L}(r)
  \nonumber \\ 
  && \; \; + V_{\mathrm{nucl}}^{(j)}(r) \psi_{n_r, L_j}(r)
  =
  \epsilon \psi_{n_r, L_j}(r),
  \quad
  \label{eq:Schrodinger-nucleon-radial}
\end{eqnarray}
where the potential $V_{\mathrm{nucl}}^{(j)}(r)$ is given
in Eq.~(\ref{eq:V_nucl-model-b}).

We shall now apply the shell model to describe 
the nuclear magnetic moment distribution for ${}^{207}$Pb and ${}^{209}$Bi isotopes.
For ${}^{207}$Pb, $R_0 = 7.394$~fm,
$c_F = 6.591$~fm, and the difference is
$R_0 - c_F = 0.803$~fm.
For ${}^{209}$Bi, $R_0 = 7.418$~fm,
$c_F = 6.765$~fm, and the difference is
$R_0 - c_F = 0.653$~fm.

\subsection{${}^{207}\text{Pb}$ isotope nuclear density}
  \label{appdend:subsec:nuclear-density-207Pb}

\begin{figure}[htb]
\centering
\subfigure[]
  {\includegraphics[width=0.9\linewidth,angle=0] {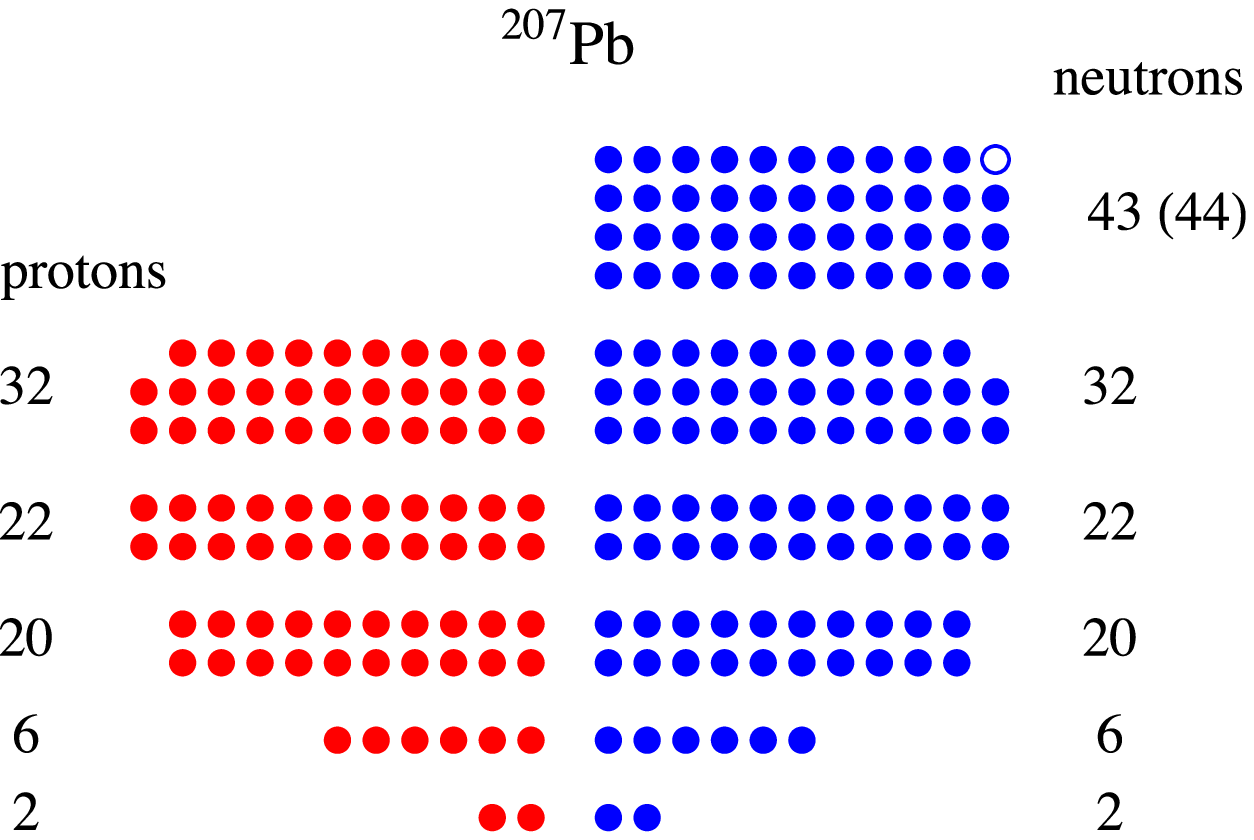}
   \label{Fig:shell-207Pb}}
\subfigure[]
  {\includegraphics[width=0.9\linewidth,angle=0] {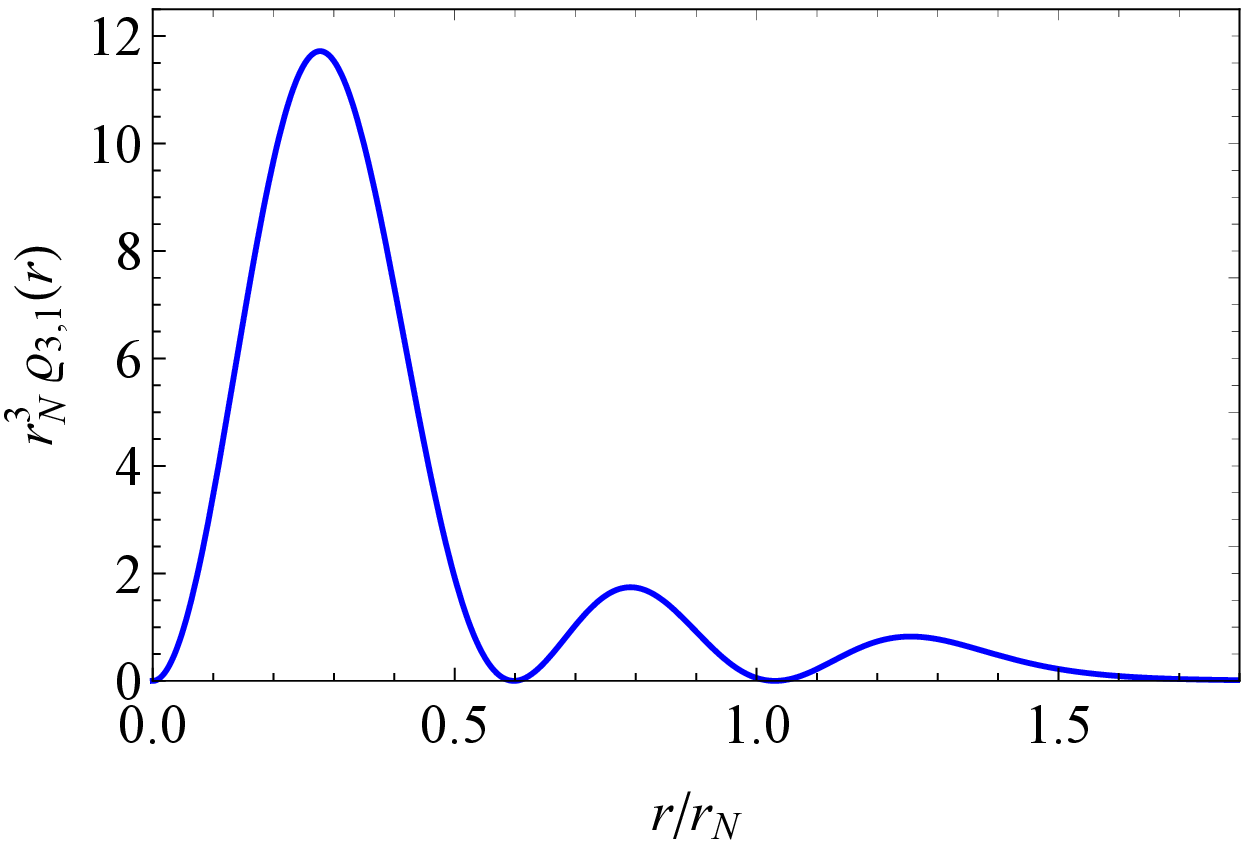}
   \label{Fig:magnetic-density-207Pb}}
\caption{\footnotesize  
  (a) Protons (red) and neutrons (blue) in the ${}^{207}$Pb nucleus
  fill the nuclear shells.
  The closed shells consist of 2, 6, 20, 22, 32 and 44 nucleons.
  The number of protons $Z = 82$ is a magic number, and the number
  of neutrons $N = 125$ is one less than the magic number 126.
  The empty circle shows the vacancy (a hole) in the outer shell. 
  The nuclear spin $I = 1/2$ is the orbital angular momentum
  of the hole on the $3p_{1/2}$ shell.
  (b) Nuclear magnetic moment density for ${}^{207}$Pb isotope.}
\label{Fig:shell-density-207Pb}
\end{figure}

The nucleus of ${}^{207}$Pb consists of $Z = 82$ protons
and $N = 125$ neutrons. Filling of the nuclear shells by
the nucleons is illustrated in Fig.~\ref{Fig:shell-207Pb}(a):
The number of protons is a magic number, i.e., the proton
close the lowest shells and form a singlet state.
The number of neutrons is one less than the magic number
126. The lowest 5 shells are closed by the neutrons, and
the outer shell is partially filled with 43 neutrons.
We consider the outer shell as a closed shell,
partially filled by 1 hole, and derive the wave function for
this hole. The nuclear spin $I$ is equal to the orbital angular
momentum of the hole.
The nuclear spin of ${}^{207}$Pb nucleus is $I = 1/2$.
The hole is on the $3p_{1/2}$ orbit with the radial
quantum number $n_r = 3$, angular momentum quantum number
$L = 1$, and the orbital angular momentum quantum number $I = 1/2$.

Wave function $\Psi_{3,1,M_L}(\mbfr)$ of a hole with the radial quantum
number $n_r = 3$, angular momentum number $L = 1$ and the projection
$M_L$ of $\mathbf{L}$ on the $z$ axis can be written as,
$$
  \Psi_{n_r,L, M_L}(\mbfr) = \psi_{n_r, L}(r) Y_{L, M_L}(\theta, \phi),
$$
where $\psi_{n_r, L}(r)$ is a radial wave function and
$Y_{L, M_L}(\theta, \phi)$ is a spherical harmonics.
$\psi_{n_r, L}(r)$ is found from the equation,
\begin{eqnarray}
  &&-\frac{\hbar^2}{2 M_n r^2} \,
  \frac{d}{d r}
  \bigg(
       r^2 \, \frac{d \psi_{n_r,L}(r)}{d r}
  \bigg) +
  \frac{\hbar^2 L (L+1)}{2 M_n r^2} \, \psi_{n_r, L}(r)
  \nonumber \\ 
  && \; \; +V_{\mathrm{nucl}}^{(n)}(r) \psi_{n_r, L}(r)
  =
  \epsilon \psi_{n_r, L}(r),
  \label{eq:Schrodinger-nucleus-207Pb}
\end{eqnarray}
where $M_n$ is the neutron mass, and the potential
$V_{\mathrm{nucl}}^{(n)}(r)$ is given
in Eq.~(\ref{eq:V_nucl-model-b}) with $j = n$.
We apply the {\it Mathematica} command NDEigensystem to
solve Eq.~(\ref{eq:Schrodinger-nucleon-radial})
and find radial wave functions $\psi_{n_r,1}(r)$ and energies 
of the $1p$, $2p$ and $3p$ states.

The magnetic moment density $\varrho_m(r)$ is
\begin{equation}
  \varrho_m(r) = \big| \psi_{3, 1}(r) \big|^{2}.
  \label{eq:magnetic-density-207Pb}
\end{equation}
The magnetic moment density $\varrho_m(r)$ in
Eq.~(\ref{eq:magnetic-density-207Pb}) is plotted in
Fig.~\ref{Fig:magnetic-density-207Pb}.
$\varrho_m(r)$ has three nodes, at $r = 0$, $0.59821 \, r_N$
and $1.03036 \, r_N$, and three maxima,
$\varrho_m(r_1) = 11.7192 \, r_{N}^{-3}$ at $r_1 = 0.277111 \, r_N$,
$\varrho_m(r_2) = 1.74229 \, r_{N}^{-3}$ at $r_2 = 0.79112 \, r_N$, and
$\varrho_m(r_3) = 0.824837\, r_{N}^{-3}$ at $r_3 = 1.25533 r_N$.
Note that the highest peak position $r_1$ in $\varrho(r)$
is far below $r_N$.

\subsection{${}^{209}\text{Bi}$ isotope nuclear density}
  \label{append:sec:nuclear-density-209Bi}

\begin{figure}[ht]
\centering
\subfigure[]
  {\includegraphics[width=0.9\linewidth,angle=0] {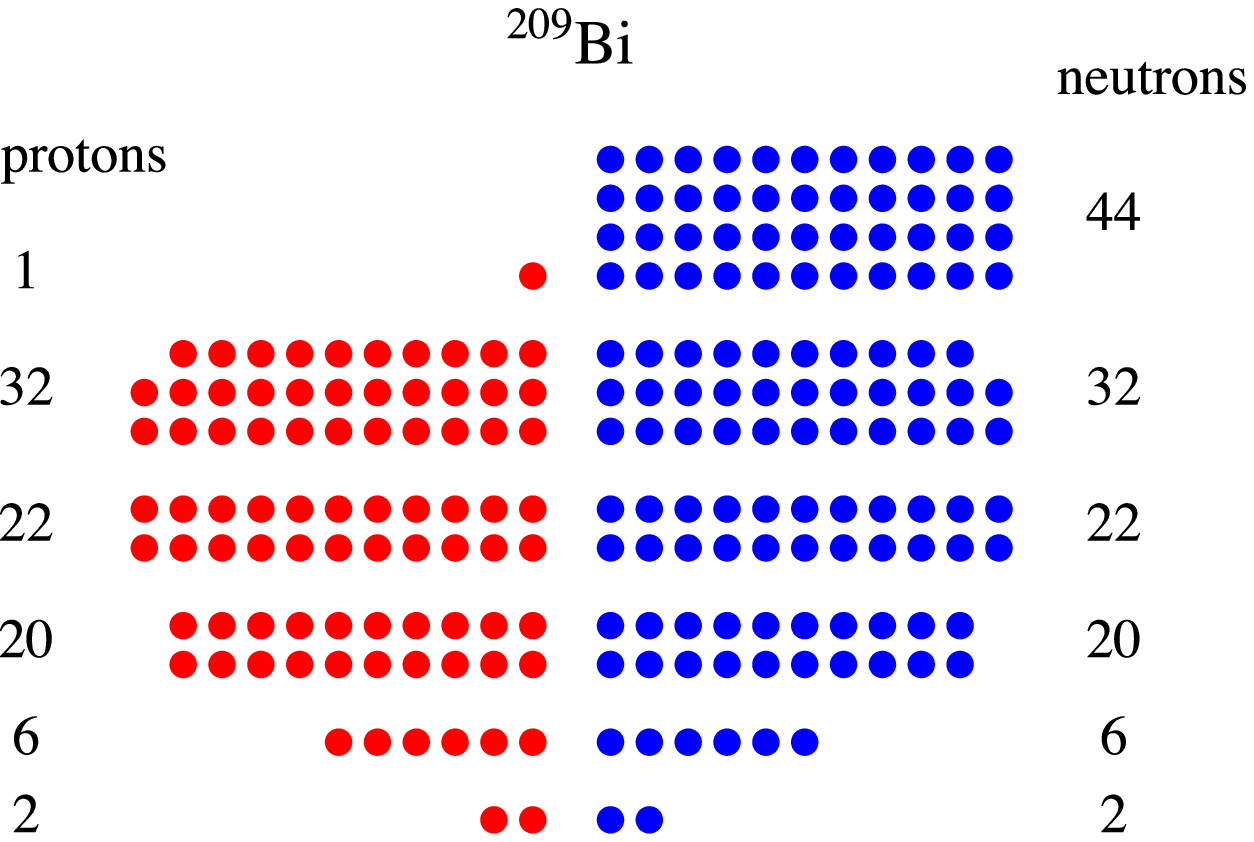}
   \label{Fig:shell-209Bi}}
\subfigure[]
  {\includegraphics[width=0.9\linewidth,angle=0] {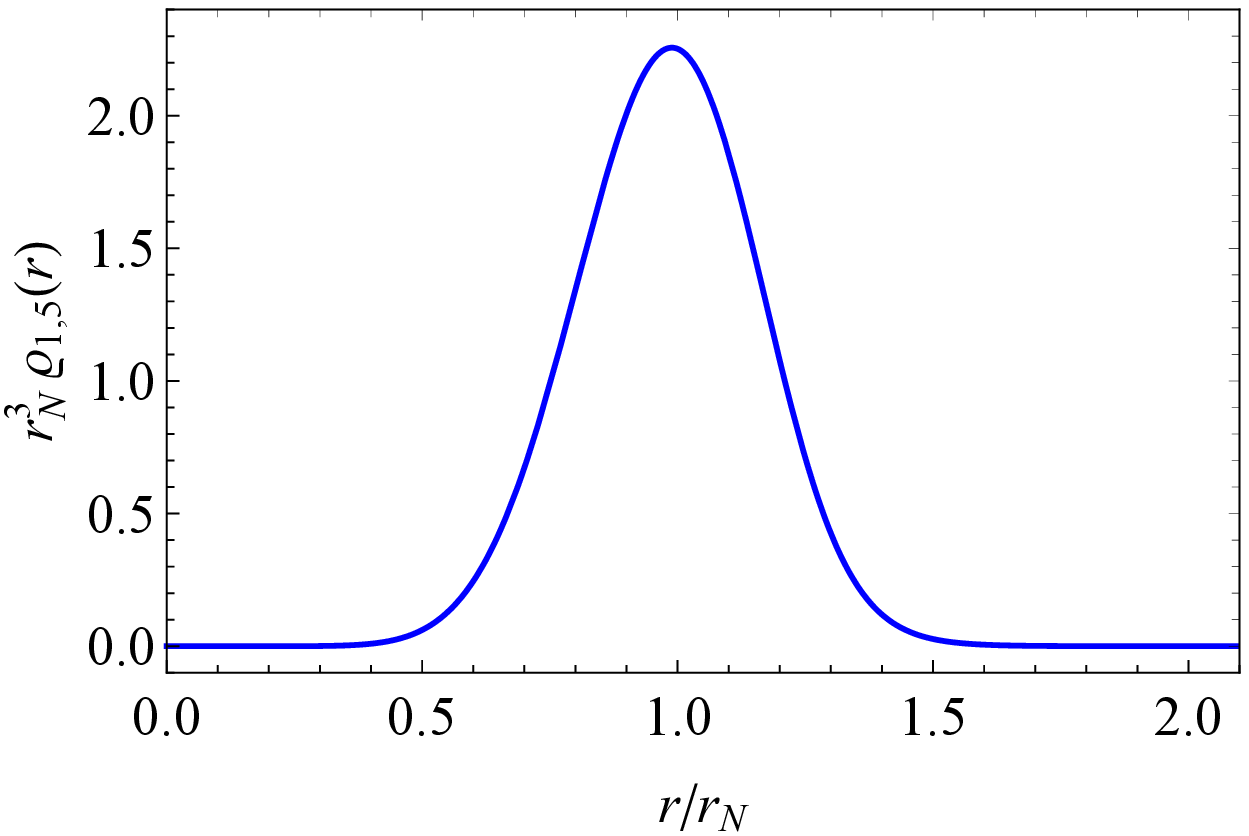}
   \label{Fig:magnetic-density-209Bi}}
\caption{\footnotesize  
  (a) Protons (red) and neutrons (blue) in the ${}^{209}$Bi nucleus
  fill the nuclear shells.
  The closed shells consist of 2, 6, 20, 22, 32, and 44 nucleons.
  The number of protons $Z = 83$ is one more than a magic number,
  and the number of neutrons $N = 126$ is a magic number.
  Nuclear spin $I = 9/2$ is equal to the orbital angular momentum
  of the proton occupying the outer $1h_{9/2}$ shell.
  (b) Nuclear magnetic moment density for ${}^{209}$Bi isotope.}
\label{Fig:shell-density-209Bi}
\end{figure}

The nucleus of ${}^{209}$Bi consists of $Z = 83$ protons
and $N = 126$ neutrons. Filling of the nuclear shells by
the nucleons is illustrated in Fig.~\ref{Fig:shell-density-209Bi}.
The number of neutrons is a magic number, i.e., the neutrons
close the lowest 6 groups of shells and form
a singlet state, see Table~\ref{Table:groups-of-shells}.
The number of protons is one more than the magic number
82. The lowest 5 groups of shells are closed by protons, and
the outer shell is partially filled by a proton.
The nuclear spin of ${}^{209}$Bi, $I = 9/2$, is equal to
the total orbital angular momentum $J_p = 9/2$ of
the proton on the $1h_{9/2}$ shell with the radial
quantum number $n_r = 1$, and the angular momentum
quantum number $L_p = 5$.

Using the {\it {\it Mathematica}} command NDEigensystem, we numerically
solve Eq.~(\ref{eq:Schrodinger-nucleon-radial}) to
find the wave function $\psi_{1,5}(r)$ and the eigen-energy of
the proton on the $1h$ shell.

The magnetic moment density $\varrho_m(r)$ of the ${}^{209}$Bi
nucleus is
\begin{equation}
  \varrho_m(r) = \big| \psi_{1, 5}(r) \big|^{2}.
  \label{eq:magnetic-density-209Bi}
\end{equation}
The magnetic moment density $\varrho_m(r)$ in
Eq.~(\ref{eq:magnetic-density-209Bi}) is plotted in
Fig.~\ref{Fig:magnetic-density-209Bi}.
$\varrho_m(r)$ has a node at $r = 0$,
increases with $r$, reaches its maximum,
 $\varrho_m(r_1) = 2.56639 \, r_{N}^{-3}$ at $r_1 = 0.988875 \, r_N$,
decreases for $r > r_1$ and vanishing as $r \to \infty$.
The half-width of the peak at half-maximum is $\Delta R_0 = 0.211582 \, r_N$,
i.e., $\Delta R_0 \ll r_N$.


\begin{thebibliography}{99}
\bibitem{Bohr_radius}
1 Bohr $\equiv \frac{\hbar^2}{m e^2} = 5.2917721 \times 10^{-11} \,\mathrm{m} \, \equiv a_0$, this quantity is called the Bohr radius.
\bibitem{comment-normalization}
  The wave functions in Eqs.~(\ref{subeqs:fg-res-plus})
  and (\ref{subeqs:fg-res-minus}) are dimensionless and
  not normalized.
  The normalized wave function has additional
  factor $\mathcal{N}$ given by Eq.~(\ref{eq:normalization}).
  The dimensionality of $\mathcal{N}_{n}$ is
  $[ \mathcal{N}_{n} ] = \mathrm{cm}^{-3/2}$.
\end{thebibliography}
\end{document}